\pdfoutput=1
\documentclass[12pt,a4paper]{article}
\usepackage{a4wide}
\usepackage{subcaption}
\usepackage[english]{babel}
\usepackage{amsmath}   
\usepackage{graphicx}  
\usepackage{dcolumn}   
\usepackage{bm}        
\usepackage{amssymb}   
\usepackage{epsfig}
\usepackage{epstopdf}
\usepackage{mdframed}
\usepackage{amsmath}
\usepackage{wasysym}
\usepackage{tabu}
\usepackage[toc]{appendix}
\usepackage{color,soul} 
\usepackage{datetime}
\usepackage{physics}
\usepackage{enumerate}
\usepackage{tablefootnote}
\usepackage{multirow}

\usepackage{footmisc}

\usepackage{hyperref}

\newcommand{\be}{\begin{equation}}
\newcommand{\ee}{\end{equation}}
\newcommand{\bea}{\begin{eqnarray}}
\newcommand{\eea}{\end{eqnarray}}
\newcommand{\bean}{\begin{eqnarray*}}
\newcommand{\eean}{\end{eqnarray*}}
\newcommand{\nn}{\nonumber}

\def\upmarg{4ex}
\def\downmarg{1.8ex}

 \def\indirect{\emph{indirect} }
 \def\Indirect{\emph{Indirect} }
 \def\direct{\emph{direct BPS} }
  \def\Direct{\emph{Direct BPS} }
  
\begin{document}

\begin{titlepage}

\numberwithin{equation}{section}
\begin{flushright}
\small

\normalsize
\end{flushright}
\vspace{0.8 cm}

\begin{center}

\mbox{{\LARGE \textbf{Black Holes Lessons from Multipole Ratios}}}

\medskip

\vspace{1.2 cm} {\large Iosif Bena, Daniel R. Mayerson}\\

\vspace{1cm} {Universit\'e Paris-Saclay, CNRS, CEA,\\ Institut de Physique Th\'eorique, Orme des Merisiers\\ 91191, Gif-sur-Yvette CEDEX, France.}

\vspace{0.5cm}

\vspace{.5cm}
iosif.bena, daniel.mayerson @ ipht.fr

\vspace{2cm}

\textbf{Abstract}
\end{center}

We explain in detail how to calculate the gravitational mass and angular momentum multipoles of the most general non-extremal four-dimensional black hole with four magnetic and four electric charges. We also calculate these multipoles for generic supersymmetric four-dimensional microstate geometries and multi-center solutions. Both for Kerr black holes and BPS black holes many of these multipoles vanish. However, if one embeds these black holes in String Theory and slightly deforms them, one can calculate an infinite set of ratios of vanishing multipoles which remain finite as the deformation is taken away, and whose values are independent of the direction of deformation. For supersymmetric black holes, we can also compute these ratios by taking the scaling limit of multi-center solutions, and for certain black holes the ratios computed using the two methods agree spectacularly. For the Kerr black hole, these ratios pose strong constraints on the parameterization of possible deviations away from the Kerr geometry that should be tested by future gravitational wave interferometers.

\end{titlepage}

\newpage

\setcounter{tocdepth}{2}
\tableofcontents

\newpage
\section{Introduction and Summary}\label{sec:intro}

There are many arguments that black holes can only restore the information that has fallen into them to our universe if there exists a structure at the scale of the horizon from where this information can be imprinted onto Hawking radiation \cite{Mathur:2005zp,Mathur:2009hf, Almheiri:2012rt}. According to these fuzzball/firewall arguments, black hole evaporation is very similar to the burning of a piece of coal, and restores all information. However, since the horizon is a null surface, a horizon-scale structure must have very unusual properties that prevent its collapse into a black hole. 

Such a non-collapsing horizon-replacing structure can be constructed in String Theory, and consists of complicated topologically non-trivial bubbles wrapped by fluxes \cite{Bena:2006kb,Bena:2015dpt,Bena:2017xbt,Heidmann:2017cxt,Bena:2017fvm,Heidmann:2019xrd,Bena:2020yii}. Such bubbles allow for the structure to be horizon-sized without it collapsing under its own gravitational force \cite{Gibbons:2013tqa,deLange:2015gca}.

Besides the presence of a structure that allows information to escape, the region of the black hole horizon is also the place where the the full non-linear glory of general relativity manifests itself. Both the current observations of gravitational waves emitted by black holes mergers \cite{Abbott_2009}, and future satellite-based experiments \cite{Danzmann_1996}, are geared towards exploring how much the physics of this region deviates from what general relativity predicts. In particular, the future observations of gravitational waves emitted during Extreme Mass-Ratio Inspiral (EMRI) events \cite{Audley:2017drz,Barack:2006pq,Babak:2017tow,Ryan:1995wh,Krishnendu:2018nqa} are expected to reveal whether the mass multipoles and angular momentum multipoles are the same as those predicted by Classical General Relativity, or are perhaps modified. We can distinguish two possible causes for such modifications: \\
{\em a.) ~~~horizon-scale structure that allows information to escape \\
b.) ~~~modifications of General Relativity that become important at the horizon scale. }

Microstate geometries generically have multipole moments that differ from those of the black hole whose horizon they are replacing. The first purpose of this paper is to compute these multipoles explicitly for some of the microstate geometries that have been constructed so far, and to compare them to the multipoles of their corresponding black hole. At the moment, this comparison can only be done for supersymmetric black holes, which are very different from those observed using gravitational wave interferometers. However, we believe it can reveal important aspects of how the microstructure at the horizon can modify these multipoles even for non-extremal black holes. 

To see this, recall that from the perspective of low-energy effective theory, the matter that forms the horizon-replacing structure has very unusual properties. In particular, it has to be very stiff to prevent its gravitational collapse.  This is very different from ``normal'' matter, which cannot support itself so close to a null surface. It is certainly possible that a spinning horizon-sized ball of this stiff unusual matter would have different gravitational multipoles than a spinning ball or dust, a spinning shell, or a spinning Kerr black hole.  Understanding this stiff horizon-replacing matter for extremal black holes should therefore illuminate its properties and allow us to understand its physics for non-extremal black holes as well. This is much in the same spirit as the study of the Quark-Gluon Plasma: even if the ${\cal N}=4$ Super-Yang-Mills theory is different from the real world, analyzing the ${\cal N}=4$ plasma reveals highly unusual properties (like the low viscosity-to-entropy ratio \cite{Policastro:2001yc}) that also characterize the real-world Quark-Gluon Plasma.

In addition to modifications of the gravitational multipoles by structure at the scale of the horizon, these multipoles can also be modified in the absence of such structure.\footnote{It is also possible is that this structure somehow succeeds in giving rise to the same physics for macroscopic observers as the black hole \cite{Mathur:2012jk}.} Indeed, although General Relativity is very well tested in the linear regime, it is still possible its non-linear structure will be different at the scale of the black hole horizon, resulting in different gravitational multipoles. For the Kerr black hole, many gravitational multipoles are exactly zero, so detecting even a small deviation for one of these multipoles would be a smoking gun for horizon-scale modifications of GR.

However, when parameterizing the possible deviations of these multipoles away from zero and trying to compute signals that could be visible in gravitational waves emitted from EMRI's, it is important to understand how to do this parameterization correctly. Indeed, given $n$ multipole moments that are identically zero in the Kerr solution and that may give rise to a measurable signal in EMRI gravitational waves, one may be tempted to parameterize them using $n$ independent parameters. The second purpose of this paper is to show that the ratios of these vanishing multipoles is a well-defined property of small deviations from the Kerr solution computed by embedding this black hole in String Theory. Hence, String Theory predicts that the possible \emph{small} departures from zero of these $n$ multipoles are in fact parameterized by a {\em single} parameter, and not by $n$ parameters ! 

\medskip

We will present two methods to compute ratios of vanishing multipole moments. The first method, which we call the \direct method, works for supersymmetric black holes, and consists of first calculating multipole moments and ratios thereof for supersymmetric multi-center microstate geometries. For these geometries, there exists a so-called scaling limit, for which the length of the throat of the microstate geometries becomes longer and longer, so that they resemble the black hole geometry more and more. Unsurprisingly, the microstate geometry multipole moments that are absent for the black hole become smaller and smaller in this limit. However, there exist an infinite number of \emph{ratios} of such vanishing multipole moments which stay finite in the scaling limit, and which we can compute.  Again, these ratios cannot be computed directly in the supersymmetric four-dimensional black hole geometry, where they are zero over zero.

The second method of computing ratios of vanishing multipoles is to slightly deform the black hole by giving it non-trivial charges, angular momentum and temperature, compute the multipole ratios for the deformed black hole, and then take the deformation back to zero. This \indirect method can be applied to compute ratios of vanishing multipoles both for supersymmetric black holes, but also for Schwarzschild, Kerr, and Kerr-Newman black holes. 

One example of such a ratio, which is ill-defined both for supersymmetric black holes and for Kerr black holes, is the product of the angular momentum with the second current multipole moment divided by the product of the mass and the third mass (octopole) moment:
\be \label{eq:introratio}
{S_1 S_2 \over M_3 M_0}\,.
\ee
Using our \indirect method, this becomes a well-defined quantity for both the Kerr and supersymmetric black holes.
Since there are many ways to deform a black hole, one might legitimately worry that this \indirect procedure could give different results depending on how the black hole is deformed. One important result of our investigation is that this does not happen, and hence the ratios of vanishing multipoles are well defined quantities.

For supersymmetric black holes we can also compare the multipole ratios computed using the \direct and the \indirect methods. In principle, these methods explore different regimes: the \indirect method works by considering small departures from the black hole solution and only changes the geometry at the scale of the horizon a small amount --- in other words it captures the physics of (small) modifications of GR at the horizon scale, described in option \emph{(b)} above. In contrast, the \direct method uses bubbling horizonless solutions that differ drastically from the black hole solution at the scale of the horizon (option \emph{(a)} above). Interestingly enough, we find that for certain families of supersymmetric black holes, the two methods give amazingly close results. In our opinion, this is a strong indication that these ratios of vanishing multipoles, computed here for the first time, are well-defined intrinsic properties of black holes. 

\subsection{Summary}

In this paper, we outline a new window into black hole physics by calculating multipole ratios for supersymmetric and non-supersymmetric black holes, as well as for supersymmetric multi-center solutions. For many black holes, the multipoles themselves vanish, but we argue that their \emph{ratios} are well-defined physical properties of these black holes. 

We derive the general formula for the gravitational multipoles of supersymmetric multi-center solutions in Section \ref{sec:bubbles}. These solutions are constructed by specifying eight harmonic functions $V, K^I, L_I, M$ in $\mathbb{R}^3$, of the form $H = h^0 + \sum_{i}h^i/r_i$. When all of the centers are on the $z$-axis, and when we choose a gauge where the constants in these harmonic functions are those of the D2-D2-D2-D6 black hole ($v^0=l_I^0 = 1$ and $k_I^0=m^0=0$), the gravitational multipoles are given by the simple and elegant formulae (\ref{eq:multipolesMmod-simple})-(\ref{eq:multipolesSmod-simple}):
\begin{align}
 M_l &= \frac14 \sum_i \left[ v^i + l_1^i+l_2^i+l_3^i)\right] z_i^l,\\
S_l &= \frac14 \sum_i \left[ - 2 m^i + k_1^i+k_2^i+k_3^i\right] z_i^l.
\end{align}
The formulae for general moduli and for multi-center solutions that are not axisymmetric are derived in Section \ref{sec:multicentermultipoles} and in Appendix \ref{app:generalmultipoles}.

We also derive the multipoles of the most general spinning non-extremal STU black hole, whose charges in String Theory correspond to D6, D4, D2 and D0 branes wrapping the corresponding cycles on the internal manifold. We obtain very simple and compact formulae for the mass and the current (angular momentum) multipoles, given in equations (\ref{eq:truemultipolesRLM}) and (\ref{eq:truemultipolesRLS}).

Many of the gravitational multipoles vanish both for the Kerr black hole and for the four-dimensional supersymmetric black hole. We nevertheless argue that one can associate to these black holes well-defined and finite ratios of these vanishing multipoles; these ratios encode black hole properties that cannot be computed by just examining the solution.

We give two possible ways of calculating these dimensionless multipole ratios. The \indirect method consists of embedding the black holes into String Theory, calculating multipole ratios in the most more general black hole, and taking the Kerr limit or the supersymmetric limit. By contrast, the \direct method allows us to calculate the multipole ratios directly in the scaling limit of microstate geometries corresponding to the (supersymmetric) black hole.

For supersymmetric black holes, we compare these two methods extensively (see Section \ref{NewWindow}). Even if the \indirect method calculates multipole ratios by slightly deforming the black hole horizon, while the \direct method calculates these ratios by completely replacing the black hole horizon with a horizonless structure, we find that for certain black holes the two methods agree surprisingly well. This agreement is correlated to a small entropy parameter $\mathcal{H}$, which quantifies the smallness of the black hole's entropy compared to a black hole with the same electric charges and no magnetic charges. One can also argue that for certain BPS black holes for which the two methods do not agree well, the \direct method is more reliable, as the \indirect method often suffers from large discontinuous jumps when the multipole order is varied in multipole ratios.

Finally, we point out that our \indirect method can also be applied to non-extremal black holes, such as Kerr, to calculate a plethora of new, previously ill-defined multipole ratios for such black holes. Since these numbers are calculated by embedding the Kerr black hole in a supergravity theory that descends from String Theory, their values should be thought of as predictions of String Theory.\footnote{Indeed, one can also imagine embedding the Kerr black in some effective low-energy theory that does not come from String Theory; the multipole ratios computed using that embedding can be different from the ones we compute.}

These numbers put very strong theoretical constrains on the parameterization of gravitational multipoles that differ from those of the Kerr solution and that one might hope to measure using gravitational waves emitted from EMRI's. Indeed, these numbers imply that all the \emph{small} deformations of the gravitational multipoles that vanish in the Kerr solution are controlled by a {\em single} small parameter ! 
\be \label{eq:constraint} M_{2n+1}=-a S_{2n} = n M a (-a^2)^n\, \epsilon,\ee
Furthermore, the deformations of the multipoles that are finite in the Kerr solution are also controlled by the same small parameter. The explicit dependence is given in equation (\ref{bigfat1}).

This result provides a benchmark for determining whether a deviation from the multipoles of the Kerr black hole that might be measured using gravitational waves indicates a \emph{small} or a \emph{large} modification of the geometry at the horizon. Indeed, if the measured deviation from the Kerr multipoles satisfy the constraint (\ref{bigfat1}), obtained using small perturbations in a String Theory embedding of the black hole, the deviation of the metric away from the Kerr solution at the scale of the horizon is also small. On the other hand, if the measured multipole deviations do not obey the constraint (\ref{bigfat1}), this could indicate that the deviations away from the Kerr metric at the horizon scale are \emph{large}, which is compatible with horizon-scale structure --- this could be confirmed using the direct method to calculate multipole ratios in microstate geometries of the Kerr black hole, once they are known.\footnote{A second possibility, if multipole deviations do not obey the constraint (\ref{bigfat1}), is that there are \emph{small} deviations from the Kerr metric at the horizon scale but that the theory describing them \emph{is not} String Theory; we clearly do not prefer this option, although it remains a logical possibility.}

\paragraph{Paper overview}
This is a companion paper to \cite{Bena:2020see} by the same authors. In particular, in \cite{Bena:2020see}  we presented the formulas for the multipoles of the general STU black hole (\ref{eq:truemultipolesRLM})-(\ref{eq:truemultipolesRLS}), and for the multipoles of axisymmetric microstates at special moduli (\ref{eq:multipolesMmod-simple})-(\ref{eq:multipolesSmod-simple}); here, we give the detailed and general derivation and formulas for multipoles at generic moduli and also for non-axisymmetric microstates. In \cite{Bena:2020see}, we also briefly introduced the \direct and \indirect methods for BPS black holes and discussed their matchings for the geometries $A,B,C$; here, we give a detailed and expanded discussion involving many other geometries. Finally, in \cite{Bena:2020see} we briefly presented the \indirect method for the Kerr black hole and mentioned that it constrains deviations from the Kerr multipoles; here, we expand this discussion and its possible theoretical and observational consequences in great detail.

\medskip

In Section \ref{sec:multipoles} we review in detail the general AC(MC) formalism we use to compute mass and current (angular) multipoles moments of four-dimensional solutions. Then, in Section \ref{sec:BHs} we discuss a few non-extremal, rotating black holes in four-dimensional supergravity.  In Section \ref{sec:bubbles} we present the smooth, supersymmetric multi-center black hole microstate geometries whose multipoles we will compute, and give the explicit derivation of the multipole formulae for general axisymmetric multi-center solutions. In our main analysis in Section \ref{NewWindow}, we first introduce the \direct and \indirect methods of calculating dimensionless multipole ratios for both the four-dimensional (static) BPS black hole and more general non-extremal black holes in Section \ref{sec:methods}, before comparing the two methods (for BPS black holes) in Section \ref{sec:comparing}, and computing explicitly new multipole ratios for Kerr in Section \ref{sec:ratiosnonext}. Gravitational wave physicists and astrophysicists interested in parameterizing measurable deviations from the Kerr solution can skip directly to Section \ref{Kerr-deviations}, where we give the explicit constraints these deviations must satisfy. 
We conclude, conjecture, and point to further avenues of research in Section \ref{sec:discussion}.  Appendix \ref{app:generalmultipoles} computes multipole moments for multi-center solutions that are not axisymmetric. Appendix \ref{app:indirect} contains further details of our \indirect method calculations for black holes, while Appendix \ref{app:moreratios} has a more detailed analysis of the \direct method for our microstates; Appendix \ref{app:chargeparams} gives the charge parameters relevant for the most general STU black hole.

\medskip

\textbf{\emph{Note added:}} After our initial paper \cite{Bena:2020see} was released, which already contained the formulas (\ref{eq:multipolesMmod-simple})-(\ref{eq:multipolesSmod-simple}) for the multipoles of axisymmetric microstate geometries with moduli $h=(1,0,0,0,1,1,1,0)$, and as this paper was being prepared, the paper \cite{Bianchi:2020bxa} appeared which contains the multipole formulas for non-axisymmetric four-dimensional microstate geometries at the same special values of the moduli. Here, we derive the general formulae (\ref{eq:generalMmult}) and (\ref{eq:generalSmult}) for the multipoles of non-axisymmetric microstate geometries for arbitrary moduli; these reduce to the values found in \cite{Bianchi:2020bxa}  (given in \ref{eq:generalmultipolesMmod-simple})-(\ref{eq:generalmultipolesSmod-simple}) below) for the special values of the moduli. We are informed that a companion paper to \cite{Bianchi:2020bxa} is also in preparation \cite{Bianchi:2020new}.

\medskip

\section{Gravitational Multipoles in Four Dimensions}\label{sec:multipoles}
For an asymptotically flat four-dimensional metric, one can define multipole moments of the gravitational field in a manifestly coordinate-invariant way. This was developed by Geroch \cite{Geroch:1970cd} and Hansen \cite{Hansen:1974zz} for vacuum solutions, and later generalized to non-vacuum spacetimes including for example a Maxwell field \cite{Sotiriou:2004ud} or a scalar field \cite{Pappas:2014gca}. In the Geroch-Hansen formalism, one extends the spacetime to include the point at infinity; the multipole moments are then introduced as tensors at this point at infinity, generated by a set of potentials. As they are tensors at infinity, the multipole moments are manifestly coordinate-invariant.

Thorne \cite{Thorne:1980ru} introduced an alternative way to define multipole moments in a stationary, asymptotically flat spacetime, which was later proven to be equivalent to the Geroch-Hansen definitions \cite{Gursel:equiv}. In Thorne's method, one writes the metric in a so-called ACMC-$N$ (asymptotically Cartesian and mass centered to order $N$) coordinates; in such coordinates, one  can read off the multipoles from the metric expansion at spatial infinity.

Since most of this paper focuses on four-dimensional stationary, axisymmetric spacetimes with Killing vectors $\partial_t, \partial_\phi$, for which the $(l,m)$ multipoles are only non-zero for $m=0$, in this Section we will only present Thorne's method for such spacetimes. We leave the more involved derivation of the multipole moments of non-axisymmetric microstate geometries to Appendix \ref{app:generalmultipoles}).

Defining a slight generalization of ACMC-$N$ coordinates, we give the metric in ``asymptotically Cartesian'' AC-$N$ coordinates\footnote{We will use the name ``Cartesian coordinates'' loosely, not distinguishing between actual Cartesian coordinates $(x,y,z)$ and their spherical counterparts $(r,\theta,\phi)$, related in the usual way as $(x,y,z)=r(\sin\theta\cos\phi,\sin\theta\sin\phi,\cos\theta)$.} as:
\begingroup
\allowdisplaybreaks
\begin{align}
\nonumber g_{tt} & = -1 + \frac{2M}{r}+ \sum_{l\geq 1}^{N}  \frac{2}{r^{l+1}} \left( \tilde M_l P_l + \sum_{l'<l} c^{(tt)}_{ll'} P_{l'} \right)\\
\label{eq:themultipoleexpansiongtt} & + \frac{2}{r^{N+2}} \left(\tilde M_{N+1} P_{N+1} + \sum_{l'\neq N+1} c^{(tt)}_{(N+1)l'} P_{l'} \right) + \mathcal{O}\left(r^{-(N+3)}\right),\\
\nonumber g_{t\phi} &= -2r\sin^2\theta\left[ \sum_{l\geq 1}^{N} \frac{1}{r^{l+1}} \left( \frac{\tilde S_l}{l} P'_l +\sum_{l'<l} c_{ll'}^{(t\phi)}  P'_{l'}\right) \right.\\
\label{eq:themultipoleexpansiongtphi} & \left.+  \frac{1}{r^{N+2}} \left( \frac{\tilde S_{N+1}}{N+1} P'_{N+1} +\sum_{l'\neq N+1} c_{(N+1)l'}^{(t\phi)}  P'_{l'}\right) + \mathcal{O}\left(r^{-(N+3)}\right)\right]  ,\\
\label{eq:themultipoleexpansiongrr} g_{rr} & = 1 + \sum_{l\geq 0}^N\frac{1}{r^{l+1}} \sum_{l'\leq l} c_{ll'}^{(rr)} P_l + \frac{1}{r^{N+2}} \sum_{l'} c_{(N+1)l'}^{(rr)} P_l + \mathcal{O}\left(r^{-(N+3)}\right),\\
\label{eq:themultipoleexpansiongthetatheta} g_{\theta\theta} & = r^2\left[ 1 + \sum_{l\geq 0}^N\frac{1}{r^{l+1}} \sum_{l'\leq l} c_{ll'}^{(\theta\theta)} P_l + \frac{1}{r^{N+2}} \sum_{l'} c_{(N+1)l'}^{(\theta\theta)} P_l + \mathcal{O}\left(r^{-(N+3)}\right)\right],\\ 
\label{eq:themultipoleexpansiongphiphi}  g_{\phi\phi} & = r^2\sin^2\theta\left[ 1 + \sum_{l\geq 0}^N\frac{1}{r^{l+1}} \sum_{l'\leq l} c_{ll'}^{(\phi\phi)} P_l + \frac{1}{r^{N+2}} \sum_{l'} c_{(N+1)l'}^{(\phi\phi)} P_l + \mathcal{O}\left(r^{-(N+3)}\right)\right],\\
  \label{eq:themultipoleexpansiongrtheta} g_{r\theta} & = (-r\sin\theta) \left[ \sum_{l\geq 0}^N\frac{1}{r^{l+1}} \sum_{l'\leq l} c_{ll'}^{(r\theta)} P'_l + \frac{1}{r^{N+2}} \sum_{l'} c_{(N+1)l'}^{(r\theta)} P_l' + \mathcal{O}\left(r^{-(N+3)}\right) \right],
\end{align}
\endgroup
These coordinates are also mass-centered and thus correspond to Thorne's ACMC-$N$ coordinates if and only if the mass dipole vanishes, $\tilde M_1=0$. This can always be achieved by a simple shift of the origin $r=0$ along the $z$-axis of symmetry.

The argument of the Legendre polynomials $P_l$ (and their derivatives) appearing above is always  $\cos\theta$. The terms that contain $c^{(ij)}_{ll'}$ correspond to non-physical ``harmonics'', and depend on the particular AC(MC)-$N$ coordinates used. Note that, {\em by definition}, for a given $l\leq N$, the only $c_{ll'}$ terms that appear in the expansion have $l'<l$. If one had a coordinate system in which at a level $l_{\rm max}\leq N$ one had a harmonic with a non-zero coefficient $c_{ll'}$ such that $l'>l_{\rm max}$, this coordinate system would not be AC(MC)-$N$ but rather (at most) AC(MC)-($l_{\rm max}-1$). 

For an ACMC-$N$ coordinate system where $\tilde M_1=0$, the true gravitational multipoles $M_l, S_l$ are simply given by $M_l = \tilde M_l, S_l=\tilde S_l$. In a more general AC-$N$ coordinate system where $\tilde M_1\neq 0$, it is easy to see that a simple shift of the origin (with distance $z_0=-\tilde M_1/\tilde M_0$) will relate $M_l,S_l$ to $\tilde M_l,\tilde S_l$ as follows:
\begin{align}
 \label{eq:truemultipoles} M_l &= \sum_{k=0}^l \left(\begin{array}{c} l\\ k\end{array}\right)  \tilde M_k \left(-\frac{\tilde M_1}{\tilde M_0}\right)^{l-k}, & S_l & = \sum_{k=0}^l \left(\begin{array}{c} l\\ k\end{array}\right)  \tilde S_k \left(-\frac{\tilde M_1}{\tilde M_0}\right)^{l-k}
\end{align}
Note that the gravitational multipoles $M_l, S_l$ are truly coordinate-independent; $M_l$ are the ``mass multipoles'' while $S_l$ are the ``current multipoles'' of the metric. The most familiar ones are the mass $M = M_0$ and angular momentum $J = S_1$. For coordinates that are AC(MC)-$N$, the highest-level multipoles we can read off in the expansion of the metric are $M_{N+1},S_{N+1}$.  Note that in order to ascertain that a coordinate system is ACMC-$N$, one needs to check that \emph{all} the metric components admit an expansion of the form (\ref{eq:themultipoleexpansiongtt})-(\ref{eq:themultipoleexpansiongrtheta}), and not just the $g_{tt}, g_{t\phi}$ components which we use to compute $M_l$ and $S_l$.

For a modern review of multipoles in general relativity and their application to astrophysical observables, see \cite{Cardoso:2016ryw} (especially Section 5.1 and the Appendix).

\subsection{Kerr: a simple but cautionary example}
The well-known Kerr metric provides a simple but interesting example of how to apply (both incorrectly and correctly) the  Thorne procedure to extract gravitational multipoles. In standard Boyer-Lindquist coordinates, the metric is given by:
\begin{align} \nn ds^2 &= -\frac{(\Delta-a^2\sin^2\theta)}{\Sigma}dt^2  - 2a \sin^2\theta \frac{(r^2 +a^2-\Delta)}{\Sigma}dt\ d\phi \,  + \frac{\left((r^2+a^2)^2-\Delta a^2\sin^2\theta\right)}{\Sigma}d\phi^2\\
& +\frac{\Sigma}{\Delta}dr^2 + \Sigma d\theta^2,
\label{eq:Kerrds2}
\end{align}
with:
\be \Sigma \equiv r^2 + a^2\cos^2\theta, \qquad \Delta \equiv r^2 -  2M r + a^2.\ee
It is illustrative to consider certain components of this metric at infinity:
\begin{align}
\label{eq:KerrgttBL} g_{tt} &= -1 + \frac{2M}{r} + \frac{1}{r^3}\left( \left[-\frac23 a^2 M\right]  P_2(\cos\theta) - \frac13 a^2 M\right) + \mathcal{O}\left(r^{-4}\right),\\
 g_{t\phi} & = (-2r \sin^2\theta)\left[ \frac{a M}{r^2} + \mathcal{O}\left(r^{-4}\right)\right],\\
\label{eq:KerrgrrBL}  g_{rr} &= 1 + \frac{2M}{r} + \frac{1}{r^2}\left( \left[\frac23 a^2\right] P_2(\cos\theta) + \left[ 4M^2-\frac23a^2\right] \right) + \mathcal{O}\left(r^{-3}\right).
\end{align}
If we had only looked at $g_{tt}$ and $g_{t\phi}$ and had compared them to (\ref{eq:themultipoleexpansiongtt})-(\ref{eq:themultipoleexpansiongtphi}), we might have (mistakenly) concluded that these coordinates are (at least) ACMC-2 and therefore we can read off the quadrupole moment from (\ref{eq:KerrgttBL}) as $M_2 = -2/3 a^2 M$. However, the expansion (\ref{eq:KerrgrrBL}) of $g_{rr}$, and in particular the presence of the $P_2/r^2$ term, reveals that these coordinates are at most ACMC-0, and thus the quadrupole $M_2$ cannot be read off in Boyer-Lindquist coordinates.

A better coordinate system is given by the transformation from the asymptotically prolate spheroidal Boyer-Lindquist coordinates $(r, \theta)$ to the asymptotically spherical coordinates $r_S, \theta_S$:
\be \label{eq:prolatetospher} r_S \sin\theta_S = \sqrt{r^2 + a^2}\sin\theta, \qquad r_S \cos\theta_S  = r\cos\theta.\ee
The Kerr metric (\ref{eq:Kerrds2}) in the asymptotically spherical coordinates $(t, r_S, \theta_S, \phi)$, can be easily checked to be ACMC-$N$ for arbitrary $N$, and the multipoles of Kerr can be read off as:
\be \label{eq:Kerrmultipoles} M_{2n} = M(-a^2)^n, \qquad S_{2n+1} = Ma(-a^2)^n,\ee
with the other multipoles vanishing.

\section{Non-Supersymmetric Black Holes in String Theory}\label{sec:BHs}
In this section we compute the gravitational multipoles of several four-dimensional non-extremal black holes in String Theory and Supergravity. We start with the Kerr-Newman black hole, to give a point of reference to compare to other geometries. Then, we consider the four-charge STU black hole (which can be considered to be an extension of Kerr-Newman), and the Rasheed-Larsen D0/D6 black hole. We then discuss the most general black hole in 4D STU supergravity \cite{Chow:2014cca}, which is a generalization of both the four-charge STU and the Rasheed-Larsen black hole.

\subsection{Warm-up: Kerr-Newman}\label{sec:kerrnewman}
The Kerr-Newman black hole is obtained from the Kerr metric (\ref{eq:Kerrds2}) by the substitution $\Delta\rightarrow \Delta + Q^2$. The metric is a solution to Einstein-Maxwell theory with a gauge field corresponding to an electric charge $Q$. Then, the same coordinate transformation (\ref{eq:prolatetospher}) is required to read off the multipoles. As is well-known, the Kerr-Newman solution to Einstein-Maxwell gravity has the \emph{same} multipoles as the Kerr solution \cite{Sotiriou:2004ud}. Hence the Kerr-Newman multipoles are also given by (\ref{eq:Kerrmultipoles}). Curiously, the coupling of the electric charge, $Q$, to gravity conspires in such a way that \emph{all} the gravitational multipoles are insensitive to $Q$.

Note that we can rewrite the Kerr(-Newman) multipoles in terms of the physical charges $M,J$:
\be \label{eq:KNmultipolesMJ} M_{2n}  = M\left(-\frac{J^2}{M^2}\right)^n, \qquad S_{2n+1} = J \left(-\frac{J^2}{M^2}\right)^n .\ee

\subsection{Four-charge STU black hole}\label{sec:STUBH}
The STU black hole can be written as a solution of String Theory compactified on $T^6$ \cite{Cvetic:1995kv}. This rotating black hole has four independent charges, which can correspond for example to 3 sets of D2 branes wrapping orthogonal $T^2$'s inside $T^6$, and a set of D6 branes wrapping the whole $T^6$. In the notation of \cite{Cvetic:2011dn}, the four-dimensional part of the metric is:
\begin{align}
\label{eq:STUBH} ds^2 & = -\Delta^{-1/2} G (dt + A)^2 + \Delta^{1/2}\left(\frac{dr^2}{X} + d\theta^2 + \frac{X}{G}\sin^2\theta d\phi^2\right),\\
 X &= r^2 - 2 m r + a^2,\\
 G &= r^2 - 2 m r + a^2\cos^2\theta,\\
 A &= \frac{2m a \sin^2\theta}{G} \left[ (\Pi_c - \Pi_s)r + 2 m \Pi_s\right]d\phi,\\
 \Delta &= \prod_{I=0}^3 (r + 2 m \sinh^2 \delta_I) + 2 a^2\cos^2\theta \left[ r^2 + m r \sum_{I=0}^3\sinh^2\delta_I + 4m^2 (\Pi_c-\Pi_s)\Pi_s \right.\\
 & \left. -2m^2 \sum_{I<J<K} \sinh^2\delta_I \sinh^2\delta_J\sinh^2\delta_K\right] + a^4\cos^4\theta,\\
 \Pi_c &= \prod_{I=0}^3 \cosh\delta_I, \qquad \Pi_s = \prod_{I=0}^3 \sinh\delta_I.
\end{align}
The parameters of the solution are $m,a, \delta_I$ (for $I=0,1,2,3$). The $U(1)$ charges of the solution are given by (with $G_4=1$):
\be Q_I = \frac14 m \sinh 2\delta_I.\ee
This black hole is a solution of the STU model and is supported by three scalars and four $U(1)$ gauge fields.\footnote{The pseudoscalars of the STU model are set to zero, but they imply a non-trivial constraint on the solutions; this constraint is automatically satisfied if we take $Q_0$ to be magnetic and $Q_i$ ($i=1,2,3$) to be electric (see for example \cite{Virmani:2012kw,Baggio:2012db}).}

The coordinates in (\ref{eq:STUBH}) are asymptotically spheroidal, and the coordinate transformation (\ref{eq:prolatetospher}) takes them to the coordinates $(t,r_S,\theta_S,\phi)$ which are ACMC-$N$ to arbitrary order $N$. This allows us to calculate all the gravitational multipoles:\footnote{Note that in practice we have only computed the multipoles up to order $l= 6$, which is enough to reveal their general structure. The computation of higher $M_l$ or $S_l$ is tedious but straightforward, and so far we could not find a general proof of \eqref{eq:STUmultipoles}, but it would be quite shocking if this formula stopped working for some higher $l$. Similar statements are true for the other multipoles calculated for the black holes of Section \ref{sec:BHs}.} 
\be \label{eq:STUmultipoles} M_{2n} = \left( \frac14 m\sum_{I=0}^N \cosh 2\delta_I\right) (-a^2)^n ,\qquad S_{2n+1} = a m (\Pi_c - \Pi_s) (-a^2)^n.\ee
The uncharged Kerr black hole (\ref{eq:Kerrds2}) is obtained from (\ref{eq:STUBH}) when $\delta_I=0$; in this limit the multipoles in  (\ref{eq:STUmultipoles}) reduce to the Kerr ones (\ref{eq:Kerrmultipoles}).

For general charges, the multipoles (\ref{eq:STUmultipoles}) can be rewritten in a form that resembles (\ref{eq:KNmultipolesMJ}),
\be \label{eq:STUmultMJ} M_{2n}  = M\left(-a^2\right)^n, \qquad S_{2n+1} = J \left(-a^2\right)^n ,\ee
except that now $a$ is a black hole parameter that is generally different from $J/M$. It is interesting to note that we only obtain $a=\pm J/M$ when the charge parameters satisfy:
\be \sinh \delta_0 = - \sinh \delta_1 \sinh \delta_2 \sinh \delta_3 \pm \cosh\delta_1\cosh\delta_2\sinh\delta_3 \pm \sinh(\delta_1+\delta_2) \cosh\delta_3,\ee
where assumed w.l.o.g. $\delta_3\geq 0$, and one can choose each of the $\pm$ signs separately.

The STU black hole (\ref{eq:STUBH}) admits a BPS limit where we take $m\rightarrow 0$ and $\delta_I\rightarrow \infty$ keeping the charges $Q_I$ fixed. The only regular geometry in this limit is the BPS black hole with zero angular momentum, $a\rightarrow 0$.

\subsection{Rasheed-Larsen black hole}\label{sec:RLBH}
The Rasheed-Larsen black hole \cite{Rasheed:1995zv, Larsen:1999pp} is a four-dimensional black hole with metric:
\begin{align}
\label{eq:RLBH} ds^2 &= -\frac{H_3}{\sqrt{H_1H_2}}(dt + B)^2 + \sqrt{H_1 H_2}( \frac{dr^2}{\Delta} + d\theta^2 + \frac{\Delta}{H_3}\sin^2\theta 
d\phi^2),\\
B &= \sqrt{p q}    \frac{\left(r \left(4 m^2+p q\right)-m (p-2 m) (q-2 m)\right)}{2 H_3 m (p+q)} a\sin^2\theta d\phi,\\
H_1 &= r^2+a^2 \cos^2\theta +\frac{p (p-2 m) (q-2 m)}{2 (p+q)}+r (p-2 m)-\frac{p \sqrt{\left(p^2-4 m\right) \left(q^2-4 m^2\right)}}{2 m (p+q)} a\cos\theta,\\
H_2 &= r^2+a^2 \cos^2\theta +\frac{q (p-2 m) (q-2 m)}{2 (p+q)}+r (q-2 m)+\frac{q \sqrt{\left(p^2-4 m\right) \left(q^2-4 m^2\right)}}{2 m (p+q)} a\cos\theta,\\
H_3 &= r^2+a^2 \cos^2\theta-2mr,\\
\Delta &= r^2 + a^2 -2mr
\end{align}
This metric is sourced by a gauge field and a dilaton $e^{-2\Phi_4} = \sqrt{H_2/H_1}$ \cite{Larsen:1999pp}. There are four parameters in this solution,  $m,a,p$, and $q$, which determine the mass and angular momentum, as well as the asymptotic electric charge $Q$ and magnetic charge $P$ associated to the gauge field \cite{Larsen:1999pp}:
\be Q^2 = \frac{q (q^2 -4m^2)}{4(p+q)},\qquad P^2 = \frac{p(p^2-4m^2)}{4(p+q)}.\ee
These can be interpreted as D0 and D6 charges when the solution is uplifted to Type IIA String Theory. Note that the parameters of this solution have a lower bound, $q,p\geq 2 m$, which is reached if and only if one of the charges vanishes.

The coordinates in (\ref{eq:RLBH}) are similar to the prolate spheroidal coordinates used in the Kerr metric (\ref{eq:Kerrds2}); the coordinate transformation (\ref{eq:prolatetospher}) gives us coordinates $(t,r_S,\theta_S,\phi)$. These are AC-$N$ coordinates (to arbitrary $N$) that are not ACMC-$N$,    since $\tilde M_1\neq 0$ in these coordinates.  We find:
\begin{align} \label{eq:RLmultipolesM} \tilde M_{2n} &= \left[\frac{p+q}{4}\right] (-a^2)^n, & \tilde M_{2n+1} &= \left[\frac{a}{8m}\frac{p-q}{p+q}\sqrt{(p^2-4m^2)(q^2-4m^2)}\right] (-a^2)^n  ,\\
  \label{eq:RLmultipolesS} \tilde S_{2n} &= 0, & \tilde S_{2n+1} &= \left[ \frac{a}{4m}\frac{\sqrt{p q}(p q + 4m^2)}{p+q} \right]  (-a^2)^n ,
\end{align}
Note that the dipole $\tilde M_1$ (and all higher-order odd coefficients $\tilde M_{2n+1}$) vanishes if and only if $p=q$, (equal electric and magnetic charges), or when one of the charges vanishes ($p=2m$ or $q=2m$). 
 The true gravitational multipoles are then given by (\ref{eq:truemultipoles}).\footnote{The formulae we will derive below, (\ref{eq:truemultipolesRLM})-(\ref{eq:truemultipolesRLS}), are also applicable for the Rasheed-Larsen black hole.}
Note that the even $\tilde M_{2n}$ in (\ref{eq:RLmultipolesM}) and the odd $\tilde S_{2n+1}$ in (\ref{eq:RLmultipolesS}) have the same form as those of the STU black hole (\ref{eq:STUmultMJ}). The coefficient, $a$, is also generically not equal to $a=\pm J/M$ unless:
\be \frac{p q(p q + 4m^2)^2}{m^2(p+q)^4} = 1.\ee
This equation is satisfied for example when both charges vanish ($p=q=2m$) and the Rasheed-Larsen black hole becomes the Kerr black hole. 

The Rasheed-Larsen black hole admits an under-rotating extremal limit, where we take $a\rightarrow 0, m\rightarrow 0$ while keeping $a/m$ fixed. In this limit, it is clear that only $\tilde M_0,\tilde M_1$ and $\tilde S_1$ remain non-zero in (\ref{eq:RLmultipolesM})-(\ref{eq:RLmultipolesS}).

The under-rotating extremal limit of the Rasheed-Larsen black hole is U-dual to a special family of the almost-BPS \cite{Goldstein:2008fq} black holes constructed in  \cite{Bena:2009ev}. When compactified to four dimensions the most general almost-BPS black hole in \cite{Bena:2009ev} is characterized its by angular momentum and five charges, which in String Theory correspond to $\overline{\text{D6}}$, D2, D2, D2 and D0 branes.  It is easy to see that the multipole structure of the general almost-BPS extremal black hole in \cite{Bena:2009ev} is precisely the same as that of the under-rotating extremal limit of the Rasheed-Larsen black hole, and thus only $\tilde M_0, \tilde M_1$ and $\tilde S_1$ are non-zero.

\subsection{Most general STU black hole}\label{app:generalBH}
Finally, we also compute the multipole moments of the most general rotating STU black hole described in \cite{Chow:2014cca} (see especially Section 5.2 therein), which reduces to all the previous black hole solutions in special limits. The metric is given by:
\begin{align}
 \label{eq:app:generalBH} ds^2 & =  -\frac{R-U}{W}(dt+\omega)^2  + W\left( \frac{dr^2}{R} + \frac{du^2}{U} + \frac{R U}{a^2(R-U)}d\phi^2\right),\\
 R &= r^2 -  2mr +a^2 - n^2, \qquad U = a^2 - (u-n)^2,\\
 W^2 &= (R-U)^2 + (2Nu+L)^2 + 2 (R-U)(2Mr+V),\\
 \omega &= \frac{2N(u-n)R + U (L+ 2N n)}{a(R-U)}d\phi,\\
 L &= 2(-n \nu_1 + m\nu_2)r + 4(m^2+n^2)\mathcal{D}, \qquad V = 2(n\mu_1 -m\mu_2)u + 2(m^2 + n^2)C.
 \end{align}
 The solution depends on 11 parameters: the mass, NUT-charge, and rotation parameters, $m,n$ and $a$, as well as four electric/magnetic charge parameters, $\delta_I$/$\gamma_I$, for $I=0,\cdots,3$. The actual mass and NUT charge of the solution are given by:
 \be \label{eq:appgeneral:MN} M = m\mu_1 + n \mu_2, \qquad N = m\nu_1 + n\nu_2,\ee
 and $\mu_{1,2},\nu_{1,2},C,\mathcal{D}$ are complicated parameters, whose dependence on the electric and magnetic parameters $\delta_I,\gamma_I$ we give explicitly in  Appendix \ref{app:chargeparams}.
 
As we are only considering asymptotically flat metrics, we must set the NUT charge to zero, $N=0$, implying that the NUT parameter is set to:
\be \label{eq:fixn} n = -m\frac{\nu_1}{\nu_2}.\ee
We can then obtain asymptotically spheroidal coordinates by taking:
\be u = n + a \cos\theta.\ee
To extract the multipoles, we must pass to asymptotically spherical coordinates by further using the coordinate transformation (\ref{eq:prolatetospher}). Then, the coordinates $(t,r_S,\theta_S,\phi)$ are again AC-$N$ to all orders, but not ACMC-$N$, just as for the Rasheed-Larsen black hole of Section \ref{sec:RLBH}. Using the quantities and definitions:\footnote{Note that we are using conventions where $J=ma$ in the Kerr limit when all charges vanish. This is related to the black hole conventions of \cite{Bena:2020see} by flipping the sign of all current multipoles $S_l$.}
\begin{align}\label{eq:generalBHMDJ}
 M &= M_0= m\left(\mu_1 -\frac{\nu_1}{\nu_2} \mu_2\right), & J &= S_1 = ma(\frac{\nu_1^2}{\nu_2}+\nu_2),\\
 \nonumber D &\equiv \frac{\tilde M_1}{a} = m\left(\mu_2 + \frac{\nu_1}{\nu_2}\mu_1\right),\\
\label{eq:Zdef} Z &\equiv D - i M, & \bar{Z} &= D + i M,
\end{align}
we can rewrite the multipole moments of this black hole as:
 \begin{align}
 \label{eq:truemultipolesRLM} M_{l} &= -\frac{i}{2} \left(-\frac{a}{M}\right)^l  Z \bar{Z}\left(Z^{l-1} - \bar{Z}^{l-1}\right),\\
 \label{eq:truemultipolesRLS} S_{l} &= \frac{i}{2} \left(-\frac{a}{M}\right)^{l-1} \frac{J}{M}   \left(Z^{l} - \bar{Z}^{l}\right)
\end{align}
Note that the multipoles $M_l,S_l$ of the Rasheed-Larsen black hole are also given by (\ref{eq:truemultipolesRLM})-(\ref{eq:truemultipolesRLS}) for $M=M_0,J=S_1,D=\tilde M_1/a$ taken from (\ref{eq:RLmultipolesM})-(\ref{eq:RLmultipolesS}).

This general black hole thus generalizes the Rasheed-Larsen black hole discussed in Section \ref{sec:RLBH}, which corresponds to $\delta_{1,2,3}=\gamma_{1,2,3}=N=0$ in (\ref{eq:app:generalBH}). It is also a generalization of the four-electric-charge STU black hole discussed in Section \ref{sec:STUBH}, to which it reduces when when $\gamma_I=N=0$.

\section{Supersymmetric Bubbling Multi-Center Geometries}\label{sec:bubbles}
In this section, we discuss the gravitational multipoles of bubbling microstate geometries that have the same mass and charge as black holes but have no angular momentum  \cite{Bena:2004de, Berglund:2005vb, Bena:2007kg}. These smooth geometries as usually built as solutions to five-dimensional supergravity, and, depending on certain parameters can have $\mathbb{R}^{4,1}$ or  $\mathbb{R}^{3,1} \times S^1$ asymptotics, and thus represent microstate geometries of five or four-dimensional black holes. When the asymptotics is $\mathbb{R}^{3,1} \times S^1$ these solutions can be reduced to certain families of the four-dimensional multi-center solutions constructed by Denef and Bates \cite{Denef:2000nb,Denef:2002ru,Bates:2003vx}.\footnote{See Appendix  A of \cite{Bena:2015dpt} for the relation between the ``Denef four-dimensional conventions" and the five-dimensional conventions we use here.} 

We introduce these geometries in Section \ref{sec:4Dmulticenter} and we derive analytic formulae for general multi-center solutions with all centers on the $z$-axis in Section \ref{sec:multicentermultipoles}. Finally, we apply our multipoles formulae to  ``scaling'' geometries in Section \ref{sec:scalinggeom} (including ``pincer'' geometries in Section \ref{sec:pincers}), and we compute dimensionless ratios of multipoles that stay finite in the scaling limit and thus characterize the resulting black hole.

\subsection{Four-dimensional multi-center geometries} \label{sec:4Dmulticenter}

We want to compute the multipoles of a class of supersymmetric bubbling multi-center solutions whose four-dimensional metric is:
\be \label{eq:ds2multicenter} ds^2 = - (\mathcal{Q}(H))^{-1/2}(dt +  \omega)^2 + (\mathcal{Q}(H))^{1/2}\left( dr^2 + r^2d\theta^2 + r^2\sin^2\theta d\phi^2\right).\ee
The solution is completely determined by 8 harmonic functions $H = (V, K^I, L_I, M)$ ($I=1,2,3$) on the flat $\mathbb{R}^3$ basis defined by $(r,\theta,\phi)$ \cite{Gauntlett:2004qy,Gauntlett:2004wh, Bena:2005ni}. These harmonic functions are determined by the locations of their poles, $\vec{r}_i$ ($i=1,\cdots, N$), which are commonly known as ``centers''. The coefficients $h_i$ together are the charges associated to the center $i$, collectively denoted in the charge vector $\Gamma^i$:
\be \Gamma^i = \left(v^i, k_1^i, k_2^i, k_3^i, l_1^i, l_2^i, l_3^i, m^i\right),\ee
and the moduli at infinity are collectively grouped in $h$ as:
\be h = \left( v^0, k_1^0, k_2^0, k_3^0, l_1^0, l_2^0, l_3^0, m^0 \right).\ee
The harmonic functions are then collectively given by:
\be H = h + \sum_{i=1}^N \frac{\Gamma^i}{r_i},\ee
where $r_i\equiv |\vec{r}-\vec{r}_i|$ is the distance in $\mathbb{R}^3$ to the $i$'th center. We further need to specify the completely symmetric constant tensor $C_{IJK}$ that characterizes the five-dimensional supergravity theory in which we work; for the STU model $C_{IJK} = |\epsilon_{IJK}|$. The warp factors and KK rotation parameters of the five-dimensional solution are:
\be Z_I = L_I + \frac12 C_{IJK} \frac{K^J K^K}{V}, \quad \mu = M + \frac{1}{2V} K^I L_I + \frac{1}{6V^2}C_{IJK}K^I K^J K^K, \ee
and the warp factor of the four-dimensional solution, $\mathcal{Q}(H)$, is given by the simple expression:
\be \label{eq:quarticinvdef} \mathcal{Q}(H) = Z_1 Z_2 Z_3 V - \mu^2 V^2.\ee

 In Appendix \ref{app:generalmultipoles} we will derive the multipole moments for the most general multicenter solution, but in this Section will only consider axisymmetric bubbling geometries, where all centers are on the $z$-axis (at positions $z_i$) and $\partial_\phi$ is a Killing vector of the full solution. For these solutions the four-dimensional rotation parameter, $\omega$, is:
\be \omega = \sum_{i<j} \omega_{ij} d\phi,\ee
with each pair of centers $(i,j)$ contributing:\footnote{We assume $z_i>z_j$ for $i<j$ in (\ref{eq:omegaij}). If the centers are in the reverse order, this is related to our expressions by flipping the sign of the $\phi$ coordinate, $\phi\rightarrow -\phi$. Note that this would also flip the sign of all resulting current multipoles, $S_l\rightarrow -S_l$.}
\begin{align}
\nn \omega_{ij} &=  \frac{\langle \Gamma^i,\Gamma^j\rangle}{|z_i-z_j|}\left( \frac{z_i -r\cos\theta}{\sqrt{r^2 + z_i^2 - 2 r z_i \cos\theta}} - (z_i\leftrightarrow z_j) \right.\\
\label{eq:omegaij} & \left.+ \frac{r^2 + z_iz_j -r(z_i+z_j)\cos\theta}{\sqrt{r^2 + z_i^2 - 2 r z_i \cos\theta}\sqrt{r^2 + z_j^2 - 2 r z_j \cos\theta}} -1  \right) ,\end{align}
where we have used the symplectic product of two charge vectors:
\be \langle \Gamma^i, \Gamma^j \rangle \equiv m^i v^j - \frac12 k_I^i l_I^j - (i\leftrightarrow j).\ee

Finally, to avoid Dirac-Misner strings at any center, $i$, the positions of the centers and their charges must satisfy 
the so-called bubble equations:
\be \label{eq:bubbleeqs} \sum_{j\neq i} \frac{\langle\Gamma^i, \Gamma^j\rangle}{ |\vec{r}_i - \vec{r}_j|} = \langle h, \Gamma^i\rangle\, .\ee
From a five-dimensional perspective, for a certain choice of the parameters of the harmonic functions \cite{Bena:2004de, Berglund:2005vb, Bena:2007kg}, the solutions described by the system above are smooth and have no horizon. They are therefore microstate geometries, in which the horizon of the black hole is replaced by a smooth cap. This is the only know example of structure at the scale of the horizon that can be constructed in a theory where gravity is taken into account. The reason why this structure does not collapse into the black horizon is the presence of topologically-nontrivial cycles wrapped by fluxes, and it was argued by Gibbons and Warner that this mechanism is the only one that can prevent the collapse of horizon-sized-structure into a black hole \cite{Gibbons:2013tqa,deLange:2015gca}. 

When compactified to four dimensions, this topologically-nontrivial structure corresponds to multi-center configurations of D6 branes with Abelian worldvolume flux. Each of these D6 branes preserves 16 supercharges, but together they preserve only the 4 supercharges of the black hole.  Some of these D6 branes have positive charge and mass and some have negative charge and negative mass (like orientifolds).  From a four-dimensional perspective, it is this latter feature that prevents the collapse into the black hole. 

It is important to note that our calculations are independent of the smoothness of the multi-center bubbling solutions in five dimensions, as long as the bubble equations, \eqref{eq:bubbleeqs} are satisfied. Hence, the formulae which we derive apply also to the most general multicenter solution in four dimensions \cite{Bates:2003vx}, including for example the descendants of black rings in Taub-NUT \cite{Bena:2005ni,Elvang:2005sa,Gaiotto:2005xt}.

\subsection{Gravitational multipoles} \label{sec:multicentermultipoles}
Every term in the harmonic functions with a pole on the $z$-axis at the position $z_i$ has the multipole expansion:
\be \frac{1}{r_i} = \frac{1}{\sqrt{r^2 + z_i^2-2r z_i \cos\theta}} = \sum_{l=0}^{\infty} z_i^l \frac{P_l(\cos\theta)}{r^{l+1}},\ee
and the harmonic functions can be expanded as:
\be \label{eq:harmfuncexpand} H = h + \sum_i \frac{h^i}{r_i} = h + \sum_i h^i \sum_{l=0}^\infty  z_i^l \frac{P_l(\cos\theta)}{r^{l+1}}.\ee
When we multiply two such harmonic functions together, at $O(r^{-(l+1)})$ we get:
\be \left(H_A H_B\right)_{\mathcal{O}(r^{-l-1})} = (h_A^0 \sum_i h_B^i z_i^l+h_B^0 \sum_i h_A^i z_i^l) \frac{P_l(\cos\theta)}{r^{l+1}} + \frac{\textrm{(lower harmonics)}}{r^{l+1}},\ee
where the lower harmonics are proportional to $P_{l'}(\cos\theta) P_{l''}(\cos\theta)$ with $(l'+1)+(l''+1)=l+1$ or $l'+l''+1=l$ so that in particular $l'+l''<l$. This implies that these do \emph{not} contribute to the $l$-th multipole of the product $H_A H_B$. To extract the $l$-th multipole from a generic function $f(H_A)$ of harmonic functions $H_A$, the formula above generalizes to:
\be \left. f(H_A)\right|_{\mathcal{O}(r^{-l-1})} = \frac{P_l(\cos\theta)}{r^{l+1}}\sum_B \partial_{h_B^0}\left[f(H_A)_\infty\right] \sum_i h_B^i z_i^l + \frac{\textrm{(lower harmonics)}}{r^{l+1}},\ee
where we introduced the notation $f(H_A)_\infty:=\lim_{r\rightarrow\infty}f(H_A)$ to denote the functional evaluated when the radius is taken to infinity; this can be thought as a function of the moduli $h_A^0$.

\paragraph{Mass multipoles}
For a general multi-center solution with centers on a line, the reasoning above allows us to find all the mass multipole by using the expansion (\ref{eq:themultipoleexpansiongtt}):\footnote{Note that we must choose the position of the centers relative to the coordinate origin such that $M_1=0$ so that these are indeed the true gravitational multipoles $M_l$.}
\be M_l = -\frac12 \sum_A \partial_{h^0_A}\left[ \mathcal{Q}^{-1/2}_\infty \right] \sum_i h_A^i z_i^l,\ee
where $\mathcal{Q}_\infty \equiv \mathcal{Q}(h^0)$ is the quartic invariant (\ref{eq:quarticinvdef}) evaluated on the moduli $h^0$ (instead of on the harmonic functions $H$ as in (\ref{eq:quarticinvdef})); note that $\mathcal{Q}_\infty=\lim_{r\rightarrow\infty} \mathcal{Q}(H)$. Explicitly, we have (note that we are using the 5D notation for the moduli):
\begin{align}
 4\mathcal{Q}_\infty &= -\sum_I (k_I^0)^2 (l_I^0)^2 + 4v^0 l_1^0l_2^0l_3^0 - 4(v^0)^2(m^0)^2 - 4v^0m^0\sum_I k_I^0l_I^0-8m^0 k_1^0k_2^0k_3^0\\
\nn &+ 2\left[ (k_1^0l_1^0)(k_2^0l_2^0) + (k_1^0l_1^0)(k_3^0l_3^0) + (k_2^0l_2^0)(k_3^0l_3^0)\right].
\end{align}
Note that the actual value of the moduli must be such that $\mathcal{Q}_\infty=1$, to ensure that the metric is indeed asymptotically flat.

We can rewrite the mass multipole function a bit more explicitly as:\footnote{Note that in this equation we define $(z_i^l)_{l=0}\equiv 1$; this is important to avoid order of limits problems when a center is at the origin ($z_i=0$), or when we take the scaling limit, $z_i\rightarrow 0$.}
\begin{align}
 \nn M_l &= -\frac12 \left( \partial_{v^0} \left[ \mathcal{Q}^{-1/2}_\infty \right] \sum_i v^i z_i^l + \sum_I \partial_{k^0_I} \left[ \mathcal{Q}^{-1/2}_\infty \right] \sum_i k_I^i z_i^l \right.\\
 \label{eq:multipolesMmc}& \left. + \sum_I \partial_{l^0_I} \left[ \mathcal{Q}^{-1/2}_\infty \right] \sum_i l_I^i z_i^l +  \partial_{m^0} \left[ \mathcal{Q}^{-1/2}_\infty \right] \sum_i m^i z_i^l \right).
 \end{align}

\paragraph{Current multipoles}
These will be completely determined by $\omega_\phi$. First, we consider the first line of (\ref{eq:omegaij}), and in particular the term of the form:
\begin{align}
\nonumber \left(\frac{z_i -r\cos\theta}{\sqrt{r^2 + z_i^2 - 2 r z_i \cos\theta}}\right)_{\mathcal{O}(r^{-l})}& = z_i \frac{z_i^{l-1} P_{l-1}(\cos\theta)}{r^l} - \cos\theta \frac{z_i^l P_l(\cos\theta)}{r^l}  + \cdots \\
\label{eq:omegaijrewrite}&= \frac{1}{r^l} z_i^l \left(\frac{\sin^2\theta}{l}\right)P_l'(\cos\theta) + \cdots,
\end{align}
where the $\cdots$ stand for lower harmonics, and we have used the recurrence relation:
\be \frac{x^2-1}{n} P_n'(x) = x P_n(x) - P_{n-1}(x).\ee
It is not hard to see after a bit of thought that the second line of (\ref{eq:omegaij}) (which goes as $1/(r_ir_j)$) only contains lower harmonics at a given order of $l$, so it will not contribute to the relevant multipole.

Comparing to (\ref{eq:themultipoleexpansiongtphi}), we can therefore conclude that the current multipoles are given by:
\be \label{eq:multipolesSmc1} S_l = \frac12\sum_{i<j} \frac{\langle \Gamma^i,\Gamma^j\rangle}{|z_i-z_j|} (z_i^l-z_j^l), \ee
where we used the fact that $\mathcal{Q}_\infty = 1$. Note in particular that $\mathcal{Q}$ does not contribute at all to the current multipoles (it only gives lower harmonics in the expansion of $g_{t\phi}$). 
We can rewrite this expression using the bubble equations (\ref{eq:bubbleeqs}) as:
\begin{align}\nn S_l& = \frac14 \sum_{i,j} \frac{\langle \Gamma^i,\Gamma^j\rangle}{|z_i-z_j|} (z_i^l-z_j^l)\\
 \nn &= \frac14 \sum_i \left(\sum_j \frac{\langle \Gamma^i,\Gamma^j\rangle}{|z_i-z_j|}\right)z_i^l - \frac14 \sum_j \left(\sum_i \frac{\langle \Gamma^i,\Gamma^j\rangle}{|z_i-z_j|}\right)z_j^l\\
 \label{eq:multipolesSmc2} &= \frac12 \sum_i \langle h, \Gamma^i\rangle\,  z_i^l.
\end{align}
where for convenience we always use the convention that the $i=j$ term vanishes in the double sums.

From the above considerations about $\mathcal{Q}$ and $\omega_\phi$, we can also easily conclude that the purely spatial metric components in (\ref{eq:ds2multicenter}) have an expansion that satisfies the AC-$N$ condition (\ref{eq:themultipoleexpansiongrr})-(\ref{eq:themultipoleexpansiongrtheta}) for arbitrary $N$. If additionally we choose the position of the origin $z=0$ such that $M_1=0$, then we can conclude that the coordinates in which the metric (\ref{eq:ds2multicenter}) is written are ACMC-$\infty$~! 

\subsection{Particular scaling geometries} \label{sec:scalinggeom}

We are now in the measure to introduce several explicit multi-center geometries whose gravitational multipoles we will compute. In Section \ref{sec:pincers}, we introduce so-called ``pincer'' geometries (both $\mathbb{Z}_2$-symmetric and asymmetric), and in Section \ref{sec:4cent} we introduce four different four-center scaling solution. We will label the pincers by a pair of integers, $(n_1,n_2)$, where the pincer's number of centers is $N=\left[(2n_1+2n_2)+3\right]$; the four-center geometries will be labelled $A,B,C,D$.

\subsubsection{Pincer geometries: \texorpdfstring{$(n_1,n_2)$}{(n1,n2)}} \label{sec:pincers}
As a concrete example, we will consider the ``pincer'' geometries, of which the original 5-center (asymmetric) pincer was given in \cite{Bena:2006kb}, later generalized to a 7-center $\mathbb{Z}_2$-symmetric pincer in \cite{Bena:2006kb,Bena:2012zi} and generalized to symmetric $N=4n+3$ center pincers for $n=1,\cdots,6$ ($N=7,\cdots,27$) in \cite{Bena:2015dpt}. We will also consider an (asymmetric) 9-center pincer, which has not appeared before in the literature.

Our pincer geometries have $N=\left[(2n_1+2n_2)+3\right]$ centers on the $z$-axis; we will consider the symmetric pincers with $n_1=n_2=1,\cdots,6$ (hence with $7,11,15,19,23$ and $27$ centers) and the asymmetric pincers with $n_2=n_1+1=1,2$ (with  $5$ and $9$ centers). 

\begingroup
\allowdisplaybreaks
The $v^i, k_I^i$ charges of the centers are given by:
\begin{align}
  v^i &= (n_1\times\{20,-20\}, -12,25,-12,\{-20,20\}\times n_2),\\
 k_I^i &= \tilde{k}_I^i -v^i\tilde{k}^{\text{(tot)}}_I,
 \label{eq:kikitilde}
\end{align}
with:
\begin{align}
 \tilde{k}_1^i &= (n_1\times\{1375, -1325\}, \frac{5}{2} 12, \frac{5}{2}12, \frac52 12, \{-1325, 1375\}\times n_2),\\
 \tilde{k}_2^i &= (n_1\times\{-\frac{1835}{2}+980 \hat{k}, \frac{1965}{2}-980\hat{k}\}, 12\hat{k}, 25\hat{k}, 12\hat{k},\\
 \nonumber &\quad\{\frac{1965}{2}-980\hat{k}, -\frac{1835}{2}+980\hat{k}\}\times n_2),\\
 \tilde{k}_3^i &= (n_1\times\{-\frac{8260}{3}, \frac{8380}{3}\}, \frac13 12, \frac13 25, \frac13 12, \{\frac{8380}{3}, -\frac{8260}{3}\}\times n_2),
 \end{align}
 and finally:
 \begin{align}
 \label{eq:kitildetot} \tilde{k}^{\text{(tot)}}_I &= \sum_i \tilde{k}^i_I.
\end{align}
\endgroup
The gauge of the $k$ charges has been chosen such that $\sum_i k_I^i = 0$ for all $I$ (note that $\sum_i v^i=1$). The $l_I^i, m^i$ charges are given by:
\be \label{eq:lmcharges} l_I^i = -\frac12 C_{IJK} \frac{k^i_J k^i_K}{v^i}, \qquad m^i = \frac{1}{12} C_{IJK}\frac{k^i_I k^i_J k^i_K}{(v^i)^2}, \qquad \forall i \textrm{ (no sum)},\ee
which are the conditions needed to ensure the uplift to a five-dimensionala multi-center bubbling solution with a Gibbons-Hawking base space is smooth (modulo orbifolds).
We will always take the origin $z=0$ to be such that the mass dipole moment of the solution vanishes, $M_1=0$. For symmetric pincers with $n_1=n_2$, this implies the $(2n_1+1)$-th center is at the origin and that the solution is $\mathbb{Z}_2$ symmetric under $z\leftrightarrow -z$. Since we would like the metric to have flat asymptotics, we will choose the constants in the harmonic functions to be:\footnote{Since we have flat $\mathbb{R}^{3,1}\times S^1$ asymptotics, these constants are different from those in \cite{Bena:2015dpt,Bena:2006kb,Bena:2012zi}, which have have flat $\mathbb{R}^{4,1}$ asymptotics; in particular, we also have non-zero $v^0$ and $k^0_1$.}:
\be\label{eq:pincermoduli}
(v^0, k^0_1,k^0_2,k^0_3, l^0_1,l^0_2,l^0_3,m^0) = (1,-2m^0,0,0,1,1,1,m^0 ),
\ee
where $m^0$ is fixed (as a function of $\hat{k}$) to the the value necessary to ensure that the sum of the bubble equations (\ref{eq:bubbleeqs}) vanishes:
\be \label{eq:m0cond} \sum_i \langle h, \Gamma^i \rangle = 0, \quad \rightarrow \quad m^0 = \frac{\sum_i m^i}{1+\sum_i l_1^i}.\ee
Note that $\lim_{r\rightarrow\infty} \mathcal{Q}(H) = \mathcal{Q}_\infty = 1$ is automatically satisfied by our choice of moduli, ensuring that the metric is asymptotically flat when compactified to four dimensions.

We can also perform a further gauge transformation (see e.g. (94) of \cite{Bena:2007kg}):
\be \label{eq:gaugetransf} K^1\rightarrow K^1 + c\, V, \qquad L_I\rightarrow L_I - c\, C_{IJ1}  K^J, \qquad M\rightarrow M -\frac{c}2 L_1, \ee
with gauge parameter:
\be \label{eq:gaugeparam} c = 2m^0,\ee
and then the moduli become:
\be\label{eq:pincermodulinom0}
(v^0, k^0_1,k^0_2,k^0_3, l^0_1,l^0_2,l^0_3,m^0) = (1,0,0,0,1,1,1,0).
\ee
Note however that in this gauge, (still) $\sum_i m^i \neq 0$; this is because the equation to the right of the arrow in (\ref{eq:m0cond}) is not gauge-invariant. This gauge for which the moduli are (\ref{eq:pincermodulinom0}) is a natural gauge in four dimensions for solutions with D2,D2,D2 and D6 charges. 

The pincer geometries depend on the value of the parameter $\hat{k}$ and admit a ``scaling'' limit for $\hat{k}\rightarrow k^*$ where:
\be k^* \approx 3.1797.\ee
Note that this value this is approximately independent of $n_1,n_2$. In the scaling limit the distances between the centers vanish, while the solutions develop a longer and longer black hole-like throats. This is achieved by a very small change of the constant  $k^*$ which gives rise to solutions in which the  inter-center distances $r_{ij}$ scales as $r_{ij}\rightarrow \epsilon\, r_{ij}$. The \emph{scaling limit} is when all centers coincide, $\epsilon\rightarrow 0$.

\subsubsection{Four geometries with four centers: \texorpdfstring{$A-D$}{A-D}} \label{sec:4cent}
We will introduce four four-center geometries, which we will label by the letters $A,B,C,D$.

\paragraph{Geometry $A$}
We consider the four-center scaling solution constructed in \cite{Heidmann:2017cxt}. The $v^i, \tilde k_I^i$ charges (with the $k_I^i$ charges determined by (\ref{eq:kikitilde}) and (\ref{eq:kitildetot})) of the centers are given by:
\begin{align}
 v^i &= \left( 1, 1, 12, -13 \right),\\
 \tilde k^i_1 &= \left( -\frac{2087}{10000}, -\frac{678089}{1250}, \frac{55636379}{10000} + \hat{k}, \frac{3445309}{2000}\right),\\
 \tilde k^i_2 &= \left( -\frac{491}{2500}, \frac{4712993}{1250}, \frac{30306499}{5000}, \frac{32175101}{5000} \right),\\
 \tilde k^i_3 &= \left( \frac{1}{10000}, -\frac{49939}{10000}, -\frac{311181}{5000}, \frac{133657}{2000} \right).
\end{align}
The $l_I^i, m^i$ charges are again given by\footnote{Note that the $\hat{k}=0$ solution given explicitly in eq. (46) of \cite{Heidmann:2017cxt} has a typo in the sign of the $m$-charge of center 1.} (\ref{eq:lmcharges}), and we use the same asymptotically flat moduli (\ref{eq:pincermoduli}) with $m_0$ determined by (\ref{eq:m0cond}). The scaling solution is at:
\be \hat{k} \approx -0.804597.\ee
We again take the origin of our coordinate system, $z=0$, to be such that the mass dipole moment of the solution vanishes, $M_1=0$.

\paragraph{Geometry $B$}
The charges for this solution\footnote{\label{ftnthankP}We thank Pierre Heidmann for sharing the charges of this unpublished scaling geometry with us.} are given by  (again with (\ref{eq:kikitilde}), (\ref{eq:kitildetot}), (\ref{eq:lmcharges}), (\ref{eq:pincermoduli}), and (\ref{eq:m0cond})):
\begin{align}
 v^i &= \left(1.000, -156.96, 159.0, -2.04 \right),\\
  \tilde k^i_1 &= \left(0.4951 + \hat{k}, -217.1, 166.6, -6.899\right),\\
  \tilde k^i_2 &= \left(0.9053, -474.0, 461.6, -6.905\right),\\
  \tilde k^i_3 &= \left(1.226, -68.79, 50.96, -0.6686\right).
\end{align}
The scaling solution is at:
\be \hat{k} \approx 0.5354.\ee

\paragraph{Geometry $C$}
The charges for this solution\footref{ftnthankP} are given by (again with (\ref{eq:kikitilde}), (\ref{eq:kitildetot}), (\ref{eq:lmcharges}), (\ref{eq:pincermoduli}), and (\ref{eq:m0cond})):
\begin{align}
 v^i &= \left(1.000, -1.896, 2.000, -0.104 \right),\\
  \tilde k^i_1 &= \left(0.7796 + \hat{k}, -20.99, 15.88, -7.329\right),\\
  \tilde k^i_2 &= \left(0.4543, 2.452, -9.061, 0.1448\right),\\
  \tilde k^i_3 &= \left(-0.09249, -5.241, 3.364, -0.2651\right).
\end{align}
The scaling solution is at:
\be \hat{k} \approx -1.6122.\ee

\paragraph{Geometry $D$}
The charges of this solution were given in  \cite{Bena:2017fvm} (eq. (4.3)). This solution consists of three (smooth) Gibbons-Hawking centers and one supertube at the second center. The $v^i, \tilde k^i_I$ charges for this solution are given by (again with (\ref{eq:kikitilde}), (\ref{eq:kitildetot}), (\ref{eq:pincermoduli}), and (\ref{eq:m0cond})):
\begin{align}
 v^i &= \left(1, 0, 1, -1 \right),\\
  \tilde k^i_1 &= \left(-184 + \hat{k}, -60, 27, 361\right),\\
  \tilde k^i_2 &= \left(-145, 0, 10909, 5308\right),\\
  \tilde k^i_3 &= \left( 1, 0, -68, 67\right).
\end{align}
For the Gibbons-Hawking centers $1,3,4$, the $l^i_I,m^i$ charges are given by (\ref{eq:lmcharges}). For the supertube center $2$ (with $v^2=0$), these charges are given by:
\be \left(l^2_1, l^2_2, l^2_3; m^2\right) = \left(0, -1300, 1229796; 13322790\right). \ee
The scaling solution is at:
\be \hat{k} \approx -0.000034117.\ee

This geometry is not smooth in five dimensions, because the Gibbons-Hawking charge of the second center is zero. However, if one dualizes this solution to the duality frame in which the three D2 charges correspond to D1 and D5 charges and to momentum along their common direction, the second center will have a dipole charge corresponding to a KK monopole whose special direction is along the D1-D5 common direction. In this duality frame the solution is smooth.

\subsection{Multipoles}\label{sec:pincermultipoles}
For moduli given by (\ref{eq:pincermoduli}), the multipole formulae (\ref{eq:multipolesMmc}) and (\ref{eq:multipolesSmc2}) simplify to:\footnote{Note that (\ref{eq:multipolesMmod})-(\ref{eq:multipolesSmod}) are explicitly asymmetric in the three different species of charge $I$ because of the way in which the moduli (\ref{eq:pincermoduli}) are chosen. However, we checked that the results of Section \ref{NewWindow} (and Appendix  \ref{app:moreratios}) stay qualitatively the same if we change (\ref{eq:pincermoduli}) to have $k^0_I=-2m^0$ for $I=2,3$ (and the other $k^0_{J\neq I}=0$) instead.}
\begin{align}
\label{eq:multipolesMmod} M_l &= \frac14 \sum_i \left[ v^i - 2m^0(k_2^i + k_3^i) + (l_1^i+l_2^i+l_3^i)\right] z_i^l,\\
\label{eq:multipolesSmod} S_l &= \frac12 \sum_i \left[ m^0(v^i + l_1^i) - m^i + \frac12(k_1^i+k_2^i+k_3^i)\right] z_i^l.
\end{align}
If we further perform the gauge transformation (\ref{eq:gaugetransf})-(\ref{eq:gaugeparam}), such that $m^0\rightarrow 0$ and the moduli of the solution have the canonical values corresponding to a D2-D2-D2-D6 black hole in four dimensions (\ref{eq:pincermodulinom0}), these expressions simplify even further and take the very illustrative form:
\begin{align}
\label{eq:multipolesMmod-simple} M_l &= \frac14 \sum_i \left[ v^i + l_1^i+l_2^i+l_3^i)\right] z_i^l,\\
\label{eq:multipolesSmod-simple} S_l &= \frac14 \sum_i \left[ - 2 m^i + k_1^i+k_2^i+k_3^i\right] z_i^l.
\end{align}
For any scaling geometry, in the scaling limit where $z_i = \epsilon z_i$ with $\epsilon\rightarrow 0$, all multipoles except the mass, $M_0$, vanish. This includes the four-dimensional angular momentum, $S_1=J$. Since in the scaling limit the throat becomes infinite and the solution becomes identical to the black hole, this is consistent with the fact that for supersymmetric black holes in four dimensions (which are non-rotating), all the multipoles except $M_0$ vanish. 

Note that for the $\mathbb{Z}_2$-symmetric pincers with $n_1=n_2$, these multipole formulae imply immediately that all odd multipole moments vanish:
\be \left[M_{2n+1}\right]_{n_1=n_2} = \left[S_{2n+1}\right]_{n_1=n_2} = 0.\ee

\section{A New Window into Black Hole Physics}
\label{NewWindow}

The (static) BPS black hole does not have any non-vanishing multipole moments (except the mass $M_0$). This means that any ratio of multipole moments is also in principle ill-defined. Nevertheless, we will introduce two ways in which we can associate a well-defined, finite multipole ratio to the static, BPS black hole. These two ways will be the \direct and \indirect methods. The \direct method involves taking the scaling limit of supersymmetric microstate geometries, while the \indirect method involves considering the extremal limit of a family of non-extremal black holes.

Our \indirect method can also easily be extended to calculate multipole ratios for non-extremal black holes, such as Kerr. This allows us also to associate to these non-extremal black holes multipole ratios that were previously unknown and that characterize their physics.

We introduce these two methods in the next subsection, after which we compare and discuss them for BPS black holes. Finally, we also discuss the ratios of vanishing multipoles for non-extremal black holes.

\subsection{Two methods of computing ratios of vanishing multipoles}\label{sec:methods}

\paragraph{\Direct method for BPS microstates}
The \direct method involves taking supersymmetric, scaling multi-center solutions such as those introduced in Section \ref{sec:scalinggeom}. In the scaling limit $\epsilon\rightarrow 0$, the centers coincide, and the metric becomes that of the corresponding (static, BPS) black hole, and all multipoles vanish except $M_0$. However, certain ratios of multipoles are independent of the scaling parameter $\epsilon$ and thus remain well-defined and finite in the scaling limit $\epsilon\rightarrow 0$. Hence, the microstate geometries allow us to obtain information about this black hole that cannot be obtained from the black hole solution (where these ratios are undefined as they are zero over zero).

For example, we can consider dimensionless ratios of mass multipoles:
 \be
 \label{eq:massmultipoleratiodef}\mathcal{M}_{\{a_1,a_2,\cdots\}/\{b_1,b_2,\cdots\}} := \frac{M_{a_1} M_{a_2}\cdots}{M_{b_1} M_{b_2}\cdots}  .
 \ee
 The dimension of $M_l$ (or $S_l$) is $[\textrm{length}]^{l+1}$, so to construct dimensionless ratios we need to demand that $\sum_i (a_i+1) = \sum_i (b_i+1)$. Note that a ratio calculated in the scaling microstate geometries will not be zero or infinite in the scaling limit in which the centers coincide ($\epsilon\rightarrow 0$) if and only if $\sum_i a_i = \sum_i b_i$.
 A simple example of such a dimensionless ratio is the square of the quadrupole moment $M_2$ divided by the product of the hexadecapole $M_4$ moment and the mass $M_0$:
 \be
  \label{eq:massmultipoleratioex}\mathcal{M}_{2,2/4,0} := \frac{M_{2} M_{2}}{M_{4} M_{0}} .
 \ee
 
 We can similarly define $\mathcal{S}_{\{a_1,a_2,\cdots\}/\{b_1,b_2,\cdots\}}$  as dimensionless ratios of current multipoles\footnote{For convenience, if one of the $a_i,b_i$ are zero, we use  $M_0$ instead of $S_0$.} $S_l$. Furthermore, we can consider mixed ratios, involving mass and current multipoles, such as
    \be
  \label{eq:massmultipoleratioex2} \frac{S_{6} M_{0}}{S_{2} M_{2} M_2} .
 \ee
 All these ratios remain finite as the throat of the microstate geometry is taken to infinity, and hence these ratios constitute non-trivial characteristics of the black hole that cannot be inferred directly from the black hole solution. 
 
 We call this method of associating multipole ratios to the static, BPS black hole the \direct method, as it only involves considering supersymmetric geometries.

\paragraph{\Indirect method for BPS black holes}

We can consider a family of non-extremal, rotating STU black holes (those of Section \ref{app:generalBH}) with its electric and magnetic charges held fixed. Every (rotating) black hole in this family has non-zero multipoles, which vanish in the non-rotating ($a\rightarrow 0$) and extremal BPS ($m\rightarrow 0$) limits\footnote{For more details on how these limits are taken, see Section 8.1 of \cite{Chow:2014cca}.}. However, many ratios of these vanishing multipoles remain finite in this limit, and one can argue that these finite ratios characterize the corresponding BPS black hole, even if they cannot be computed directly in the BPS black hole solution.

These finite ratios will be a function of the black hole parameters $\mu_1,\mu_2,\nu_1,\nu_2$, as can be seen from (\ref{eq:generalBHMDJ}). More computational details of this method are given in Appendix  \ref{app:indirect-susy}, including the explicit values of these parameters for the black holes we consider. From (\ref{eq:generalBHMDJ}) and the multipole formulae (\ref{eq:truemultipolesRLM})-(\ref{eq:truemultipolesRLS}), it is clear that the same ratios that are well-defined and finite for this method, are those that are also well-defined and finite in the \direct method.

We call this method the \indirect method of calculating multipole ratios for the BPS black hole as it involves departing from extremality, and thus deforming the black hole in question.

\paragraph{\Indirect method for non-extremal black holes}
The \indirect method can also be applied to calculate ratios of vanishing multipoles for any black hole in which some of the multipoles vanish, such as the Kerr or Kerr-Newman black hole, or the other black holes discussed in Section \ref{sec:STUBH} and \ref{sec:RLBH}. We simply deform the black hole in question to a more general STU black hole with four electric and four magnetic charges (of the kind presented in Section \ref{app:generalBH}), calculate the multipole ratios, and take the limit of these multipole ratios as we go back to the original black hole.

Any multipole calculated for a general black hole in Section \ref{app:generalBH} is a function of the four parameters $M,J,D,a$ as can be seen from (\ref{eq:truemultipolesRLM})-(\ref{eq:truemultipolesRLS}) with (\ref{eq:Zdef}) (with $M,J,D$ given in (\ref{eq:generalBHMDJ})). Thus, any multipole ratio $\mathcal{M}$ (which may involve both mass and current multipoles) of such a black hole is a function of these four parameters:
\be \mathcal{M} = \mathcal{M}(M,J,D,a).\ee
From (\ref{eq:truemultipolesRLM})-(\ref{eq:truemultipolesRLS}), (\ref{eq:Zdef}), and (\ref{eq:generalBHMDJ})), it is clear that \emph{generically all multipoles are non-vanishing} (except $M_1$), which means any multipole ratio $\mathcal{M}$ is well-defined. Then, one can simply take the limit of $\mathcal{M}$ to the (special) black hole in question.

For example, for Kerr or Kerr-Newman black holes, the appropriate limit is:
\be \label{eq:generalratioKerr} \mathcal{M}_{\text{Kerr}} = \lim_{J\rightarrow M a}\lim_{D\rightarrow 0} \mathcal{M}(M,J,D,a).\ee
Beyond Kerr, the above procedure can also easily be extended to the four-charge black hole of Section \ref{sec:STUBH}. The multipoles of this black hole differ from those of the Kerr(-Newman) black hole because  $J\neq M a$, so the appropriate limit is only:
\be \label{eq:generalratioSTU4charge} \mathcal{M}_{\text{4-charge}} = \lim_{D\rightarrow 0} \mathcal{M}(M,J,D,a).\ee

This limits (\ref{eq:generalratioKerr}) or (\ref{eq:generalratioSTU4charge}) are always well-defined (although the results may be infinite), since they are insensitive to which direction in charge-space we take it. Thus, they associate well-defined multipole ratios with the black hole in question. Note that these limits can also be used for computing ratios of non-vanishing multipoles which can also be computed directly in the black hole geometry.

We can interpret the \indirect method as placing the (special) black hole, such as the BPS or the Kerr black hole, in a larger phase space of black holes given by the most general STU black holes of Section \ref{app:generalBH}. Then, the statement that \emph{generically} all multipoles are non-vanishing means that the sub-space of black holes with (some) vanishing multipoles is of codimension larger than zero in this large phase space. Our \indirect method then simply corresponds to defining the multipole ratios over the entire (large) phase space by continuity.

\subsection{Comparing methods for BPS black holes}\label{sec:comparing}
As we can see from the multipole moments of multi-center geometries (\ref{eq:multipolesMmod})-(\ref{eq:multipolesSmod}), multipole ratios calculated by the \direct method will generically be finite and well-defined, unless an accidental symmetry of the microstates causes one or more multipole moments to vanish (even before taking the scaling limit). This is precisely what happens for the $\mathbb{Z}_2$ symmetric pincers $(n,n)$ introduced in Section \ref{sec:pincers}; from (\ref{eq:multipolesMmod})-(\ref{eq:multipolesSmod}) we can immediately see that for these geometries, all odd multipoles vanish, $M_{2n+1}=S_{2n+1} = 0$. Thus, here we will only consider the other six geometries --- the two asymmetric pincers $(1,0)$ and $(2,1)$ from Section \ref{sec:pincers} and the four four-center geometries $A,B,C,D$ of Section \ref{sec:4cent}. For these six geometries, we will calculate multipole ratios and compare them to those of the corresponding static, BPS black hole as calculated with the  \indirect methods. Note that we discuss multipole ratios of the symmetric pincers (calculated using the \direct method) further in Appendix \ref{app:moreratios}.

We give the explicit numerical values of several ratios of vanishing multipole of relatively low multipole order, calculated via the \direct and \indirect methods, both in Table \ref{tab:simplemultipolevals-aspinc} for the two asymmetric pincer geometries $(1,0)$ and $(2,1)$, and in Table \ref{tab:simplemultipolevals-4c} for the four-center geometries $A,B,C,D$.

\begin{figure}[p]\centering
\begin{subfigure}{0.70\textwidth}\centering
 \includegraphics[width=\textwidth]{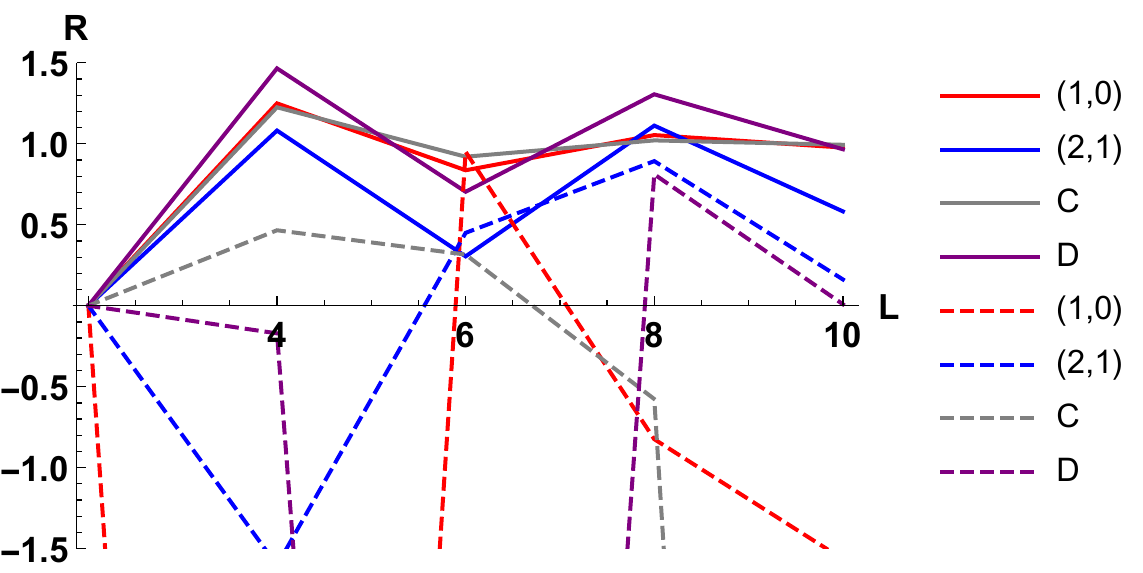}
 \caption{$R_2^{(M)}(L,\delta=1/2)$ as a function of $L$ for geometries $(1,0),(2,1),C$ and $D$.}
 \label{fig:R2Mdirectvsindirect-fournomatch}
\end{subfigure}\\
\begin{subfigure}{0.48\textwidth}\centering
  \includegraphics[width=\textwidth]{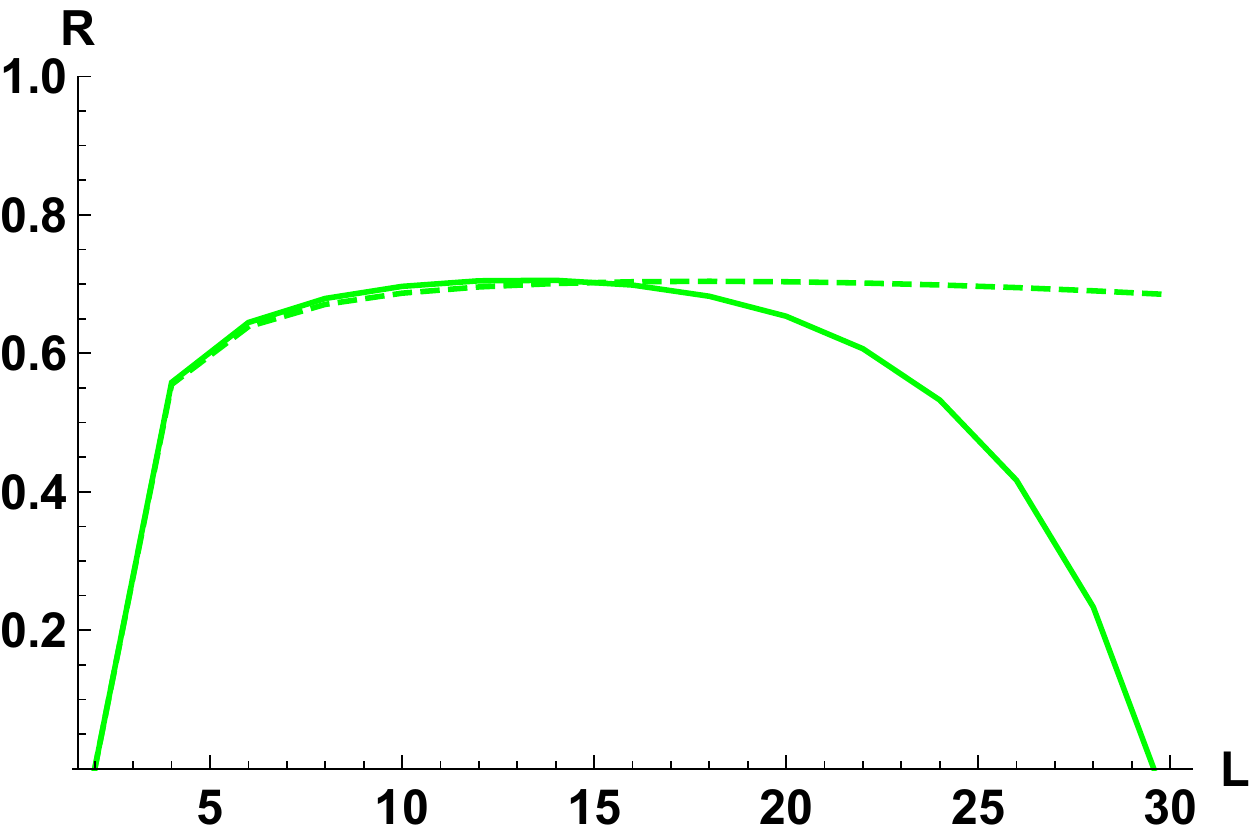}
\caption{$R_2^{(M)}(L,\delta=1/2)$ as a function of $L$ for geometry $A$}
  \label{fig:R2Mdirectvsindirect-matchA}
\end{subfigure}\hspace*{0.01\textwidth}
\begin{subfigure}{0.48\textwidth}\centering
  \includegraphics[width=\textwidth]{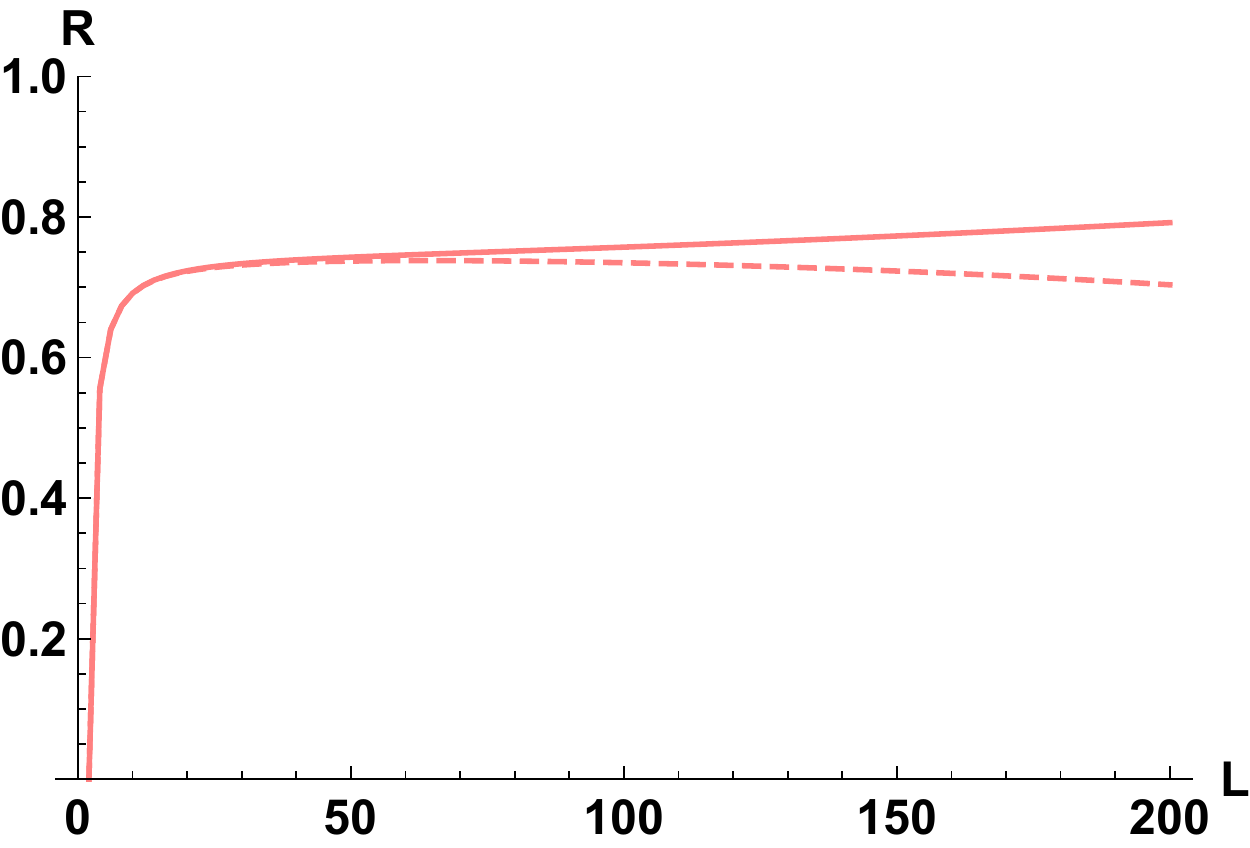}
\caption{$R_2^{(M)}(L,\delta=1/2)$ as a function of $L$ for geometry $B$}
  \label{fig:R2Mdirectvsindirect-matchB}
\end{subfigure}
 \caption{The ratio of vanishing multipoles $R_2^{(M)}(L,\delta=1/2)$ plotted as a function of $L$ for the six geometries we are considering. \emph{\textbf{Solid}} lines correspond to the values calculated by the \direct method, \emph{\textbf{dashed}} lines to the values calculated using the \indirect method.\newline Note the scale difference of the $L$ axis between the three figures.}
 \label{fig:R2Mdirectvsindirect}
\end{figure}

One thing that is immediately clear in Tables \ref{tab:simplemultipolevals-aspinc} and \ref{tab:simplemultipolevals-4c} when comparing the \direct and \indirect methods, is that the values the two methods give for the multipole ratios match \emph{extremely well} for geometries $A$ and $B$, and rather poorly for the others. We can also see this in more detail graphically. Defining the function:
\be  \label{eq:R2def} R_2^{(M)}(L,\delta) \equiv \mathcal{M}_{\{L(1+\delta),L(1-\delta)\}/\{L,L\}} = \frac{M_{L(1+\delta)} M_{L(1-\delta)}}{(M_L)^2},\ee
we give the graph of $R_2^{(M)}(L,\delta)$ at $\delta=1/2$ for even $L$ for the six geometries in fig. \ref{fig:R2Mdirectvsindirect}. All geometries (calculated in either \direct or \indirect methods) have $R_2^{(M)}(L=2,\delta=1/2)=0$ since $M_1=0$, but otherwise we can immediately see in fig. \ref{fig:R2Mdirectvsindirect-fournomatch} that the \direct and \indirect methods give wildly different results for each of the four geometries $(1,0),(2,1),C,D$. On the other hand, we can see in fig. \ref{fig:R2Mdirectvsindirect-matchA} that for geometry $A$, the two methods match very nicely until $L\sim 14$, where they start to give different values for $R_2^{(M)}(L,\delta)$. Geometry $B$ does even better; in fig. \ref{fig:R2Mdirectvsindirect-matchB} we see that the two methods match until $L\sim 40$.

To parameterize how much the ratios of vanishing multipoles differ when calculated using our two methods, we can  define a so-called \emph{entropy parameter} $\mathcal{H}$ as \cite{Heidmann:2017cxt}:
\be \label{eq:entropypar} \mathcal{H} = \frac{\mathcal{Q}(Q_I,P_I)}{Q_1 Q_2 Q_3 Q_4},
\ee
where $\mathcal{Q}$ is the quartic invariant (\ref{eq:quarticinvdef}), evaluated on the charges $Q_I,P_I$ (instead of on the harmonic functions as in (\ref{eq:quarticinvdef})); explicitly, we have \cite{Chow:2014cca}:
\be \mathcal{Q}(Q_I,P_I) = \frac14 \left( 4(\prod_I Q_I + \prod_I P_I) + 2\sum_{I<J} Q_IQ_J P_I P_J - \sum_I Q_I^2P_I^2\right).\ee
The four-dimensional BPS black hole has entropy $S = \pi \sqrt{\mathcal{Q}(Q_I,P_I)}$; thus the dimensionless parameter $\mathcal{H}$ quantifies how small is the entropy of our black holes compared to the entropy of a purely electric black hole with the same charges. If one uplifts our four-dimensional black holes to five-dimensional BMPV black holes, $\mathcal{H}$ parameterizes how close to the cosmic censorship bound these BMPV black holes are. Indeed, $S_{BMPV}=2\pi \sqrt{Q_1Q_2Q_3-J^2}$ and\footnote{For our black holes, note that in four dimensions our magnetic charges satisfy $P_2=P_3=0$ and $P_1=-P_4$, and moreover $Q_4=1$ (since for all geometries we consider, $\sum_i v^i=1$), so that:
\be \mathcal{Q}(Q_I,P_I) = Q_1 Q_2 Q_3 Q_4 -\frac14(Q_1+Q_4)^2(P_1)^2.\ee
} $\mathcal{H} = (Q_1 Q_2 Q_3 - J^2)/(Q_1 Q_2 Q_3)$.

\begin{figure}[p]\centering
 \includegraphics[width=0.7\textwidth]{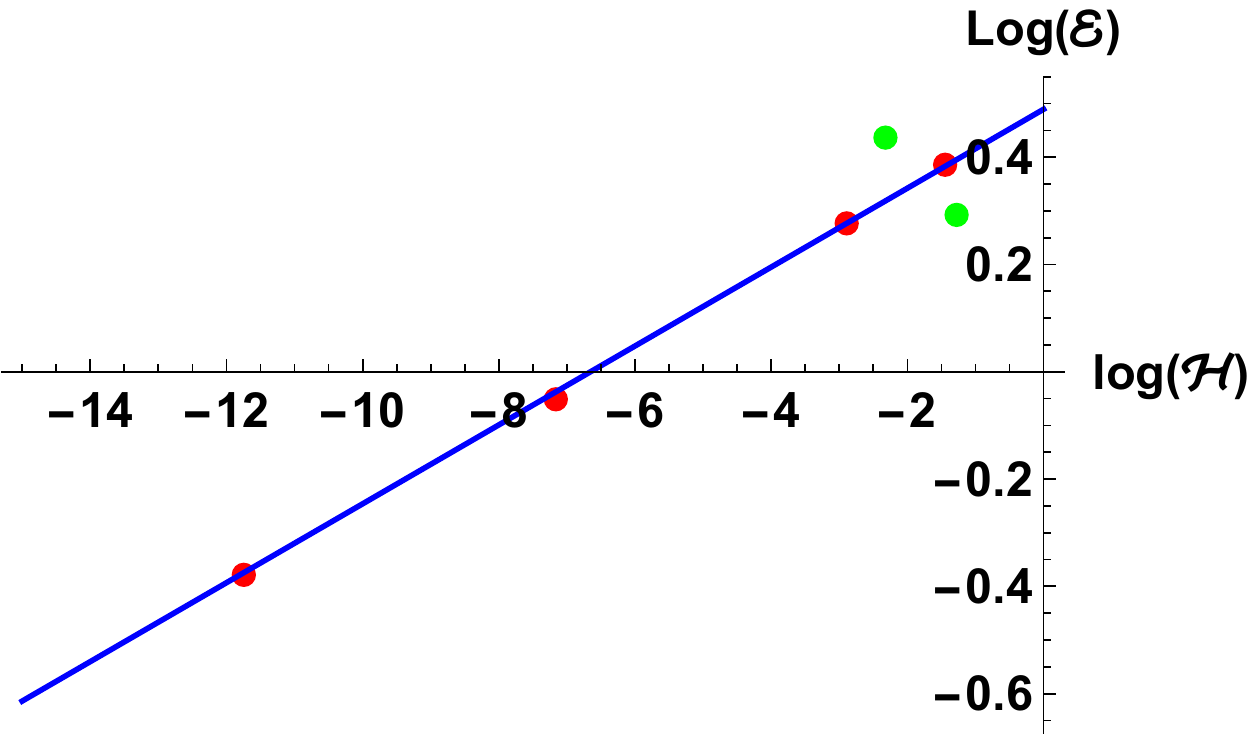}
 \caption{A log-log plot of $\mathcal{E}_{\text{(ave)}}$ vs. $\mathcal{H}$. The values are those in table \ref{tab:entropyvserror}. The red dots are the geometries $A,B,C,D$ and the green dots are the pincers $(1,0),(2,1)$. The blue line is the best-fit linear regression for these six points, given by $  y= 0.49+0.07 x$ with statistical parameters $R^2 =0.950$ and $p=9.5\times 10^{-4}$. (The fit without the two pincer datapoints gives approximately the same line with $R^2 =0.9997$ and $p=1.6\times 10^{-4}$.)  }
 \label{fig:loglogHE}
\end{figure}

The value of the entropy parameter for the six geometries in question can be found in table \ref{tab:entropyvserror}. We see that geometries $A$ and $B$ have an extremely small  $\mathcal{H}$, and hence it is possible that there is a correlation between a small $\mathcal{H}$ and a good matching between \direct and \indirect methods of calculating multipole ratios. To make this correlation quantitative we can define a ``mismatch parameter'' $\mathcal{E}$, for a given ratio $\mathcal{M}$:
\be \label{eq:errordef} \mathcal{E}^{(\mathcal{M})} \equiv \left| \frac{ \mathcal{M}^{\text{(dir)}} - \mathcal{M}^{\text{(ind)}}}{\mathcal{M}^{\text{(ind)}}}\right|,\ee
where $\mathcal{M}^{\text{(dir)}}$ is the ratio calculated using the \direct method and $\mathcal{M}^{\text{(ind)}}$ using the \indirect method. Thus, $\mathcal{E}^{(\mathcal{M})}$ gives the relative difference in value between the two methods for a given multipole ratio. In table \ref{tab:entropyvserror}, we give the value for $\mathcal{E}_{\text{(ave)}}$, which is the average of (\ref{eq:errordef}) over the 12 multipole ratios considered in tables \ref{tab:simplemultipolevals-aspinc},  \ref{tab:simplemultipolevals-4c}. The correlation between these two quantities can be seen very clearly in a log-log plot, see fig. \ref{fig:loglogHE}.

From the above considerations, it is not a giant leap to conjecture that this is a general phenomenon: the multipole ratios calculated using the \direct and \indirect methods will agree more when $\mathcal{H}$ is small, and hence the entropy coming from the electric charges is mostly eaten by the electric-magnetic interactions (or five-dimensional angular momentum). We have checked that there is  no correlation between the relative error given in table \ref{tab:entropyvserror} and any other parameter, including the black hole entropy itself, or the maximum scale separation between centers in the corresponding microstate geometry\footnote{We can define a dimensionless parameter $\mathcal{S}$ that quantifies the scale separation in a microstate as follows:
\be \mathcal{S} = (n-1) \frac{\min_{(i,j)} | z_i-z_j|}{\max_{(i,j)} | z_i-z_j|}, \ee
where both the minimum and maximum distances are taken over pairs of centers, and $n$ is the number of centers of the microstate geometry. Thus, this quantity is independent of the scaling parameter $\epsilon$ (for scaling solutions). We have $\mathcal{S}\leq 1$ and  $\mathcal{S}=1$ when all the centers are spaced equally (and thus there is precisely no scale separation). We can calculate  $\mathcal{S}_{(A,B,C,D)} = (0.64, 0.088, 0.022, 0.97)$ and see immediately that there is no correlation between $\mathcal{S}$ and $\sum_i \mathcal{E}^{(\mathcal{M}_i)}$.}.

Having discussed when the \direct and \indirect methods give the same answers for multipole ratios, we can also consider the regime in which their answers differ. For such ratios, one can ask which of the two methods gives the most trustworthy result?

\begin{figure}[p]\centering
\begin{subfigure}{0.48\textwidth}\centering
 \includegraphics[width=\textwidth]{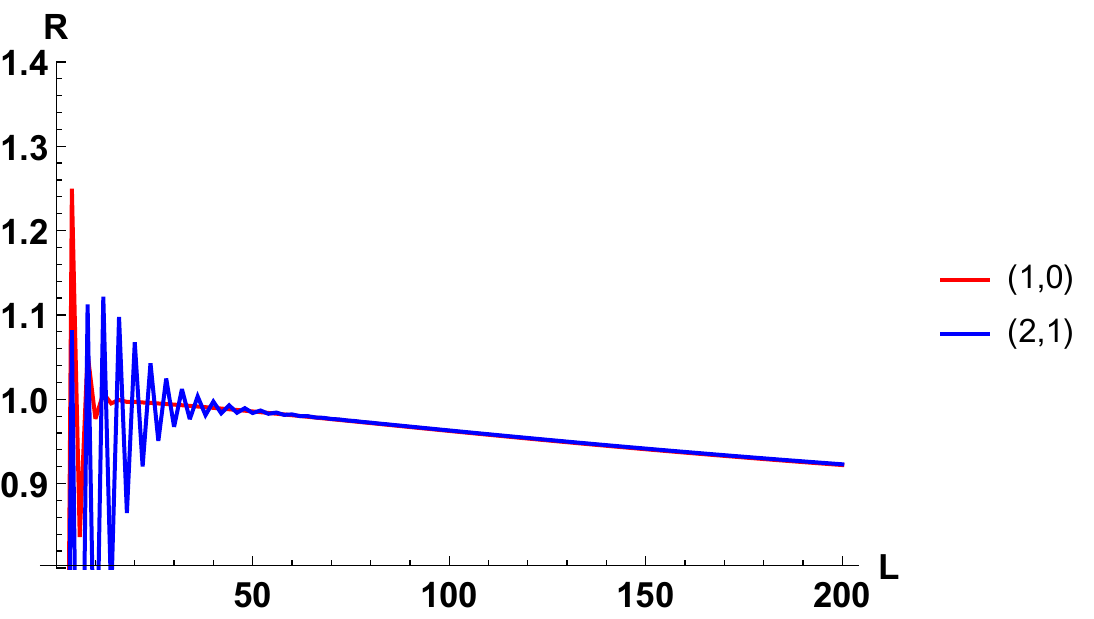}
 \caption{\direct for $(1,0)$ and $(2,1)$}
\end{subfigure}\hspace*{0.01\textwidth}
\begin{subfigure}{0.48\textwidth}\centering
  \includegraphics[width=\textwidth]{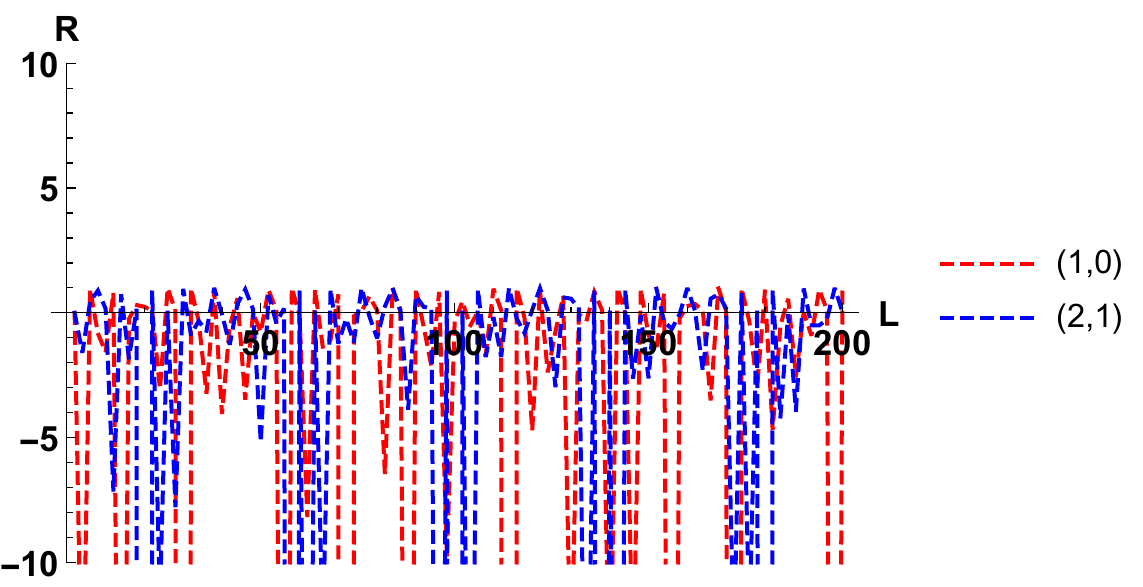}
 \caption{\indirect for $(1,0)$ and $(2,1)$}
\end{subfigure}\\
\begin{subfigure}{0.48\textwidth}\centering
 \includegraphics[width=\textwidth]{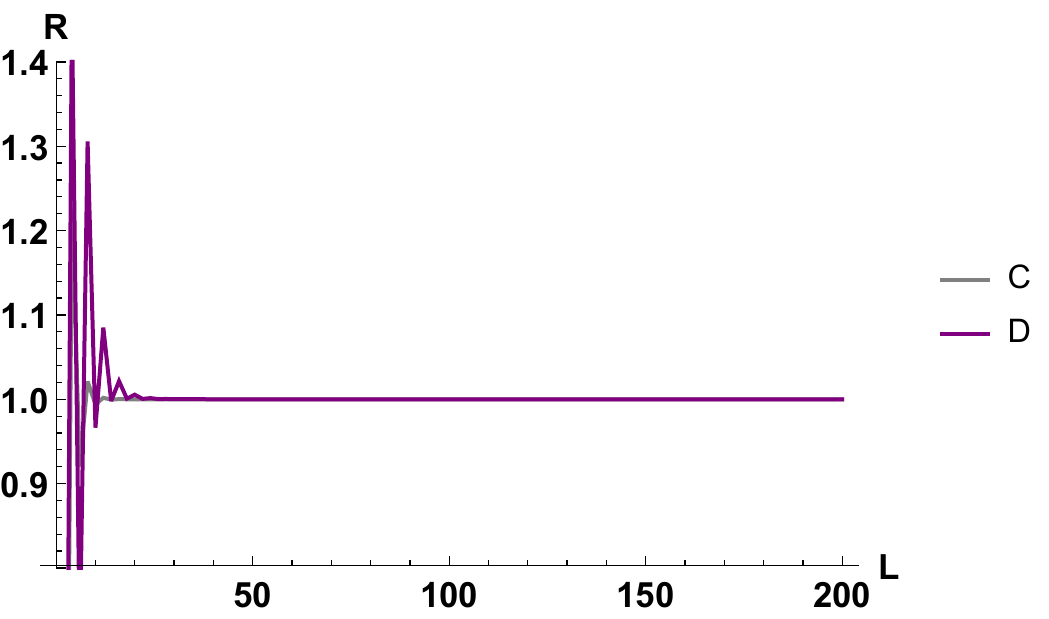}
 \caption{\direct for $C$ and $D$}
\end{subfigure}\hspace*{0.01\textwidth}
\begin{subfigure}{0.48\textwidth}\centering
  \includegraphics[width=\textwidth]{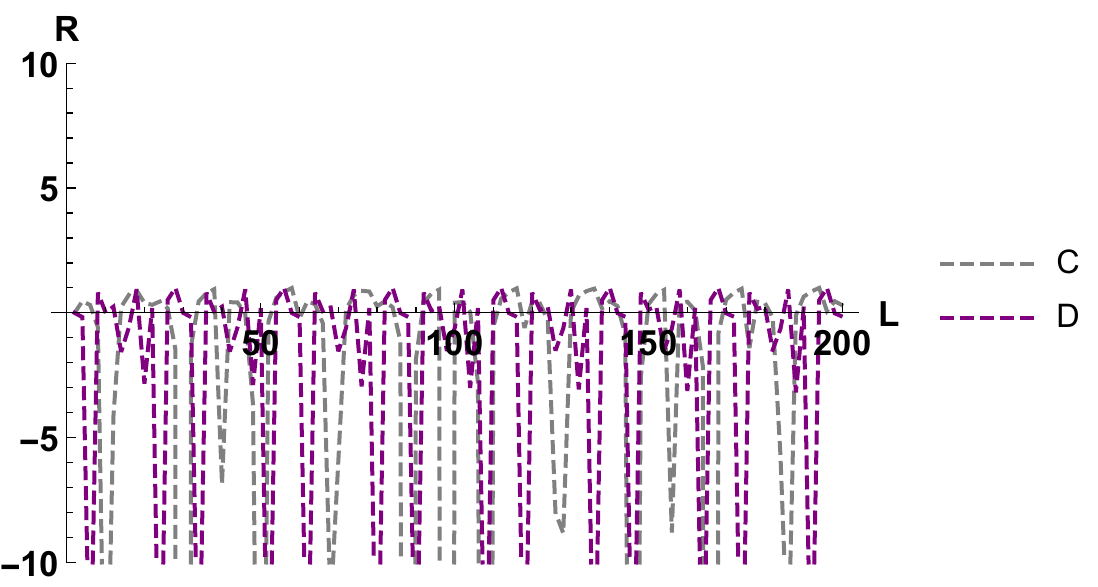}
 \caption{\indirect for $C$ and $D$}
\end{subfigure}\\
\begin{subfigure}{0.48\textwidth}\centering
 \includegraphics[width=\textwidth]{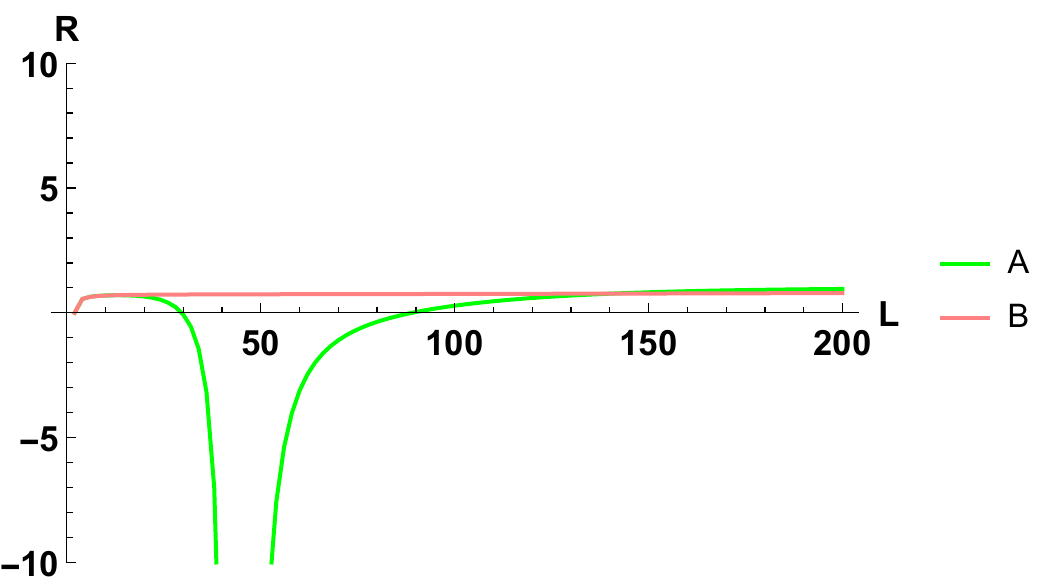}
 \caption{\direct for $A$ and $B$}
\end{subfigure}\hspace*{0.01\textwidth}
\begin{subfigure}{0.48\textwidth}\centering
  \includegraphics[width=\textwidth]{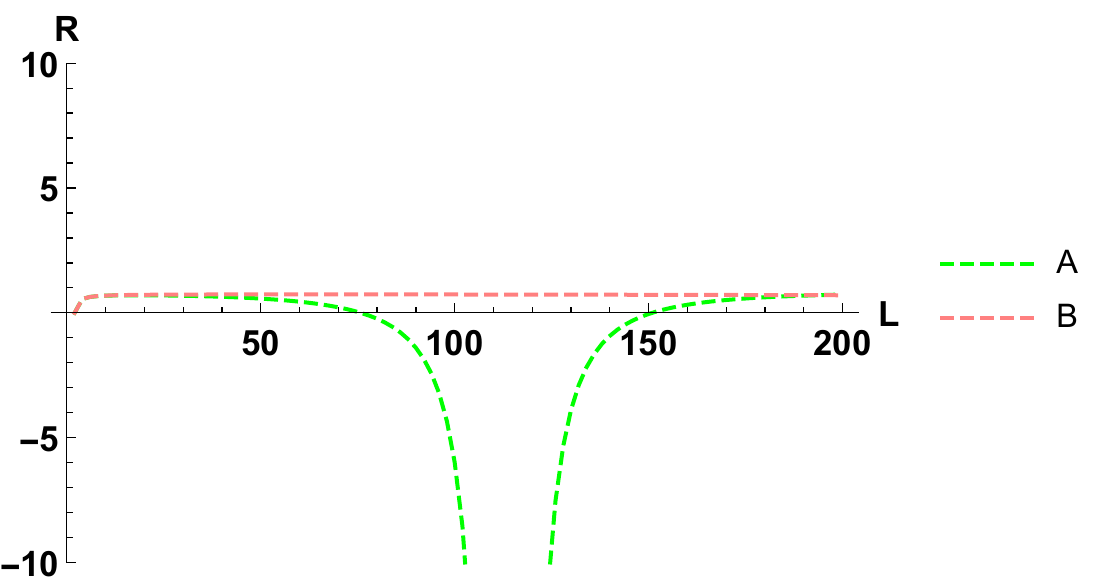}
 \caption{\indirect for $A$ and $B$}
\end{subfigure}
 \caption{$R_2^{(M)}(L,\delta=1/2)$ as a function of $L$ for the six geometries we are considering. We see that the \indirect method produces highly oscillatory results  while the \direct method produces smooth graphs.}
 \label{fig:dirvsindirect-osc}
\end{figure}

Consider fig. \ref{fig:dirvsindirect-osc}, where we plot $R_2^{(M)}(L,\delta=1/2)$ for even $L$ between $2$ and $200$ for our six geometries, using both the \direct and the \indirect methods. For the $A,B$ geometries, with small $\mathcal{H}$ and thus a good matching between the two methods (at low multipole order), we see that both the \direct and \indirect methods produce smooth, continuous graphs. However, for the other four geometries, where the two methods disagree even at low multipole order, we see that the \direct method (also) produces smooth, continuous graphs whereas the multipole ratios calculated using the \indirect method are highly oscillating and their graphs are discontinuous.

For this reason, we conjecture that for supersymmetric black holes, the \direct method of calculating multipole ratios is more reliable and gives more sensible results than the \indirect method. We discuss and analyze more multipole ratios calculated using the \direct method, including numerous figures, in Appendix \ref{app:moreratios}.

This being said, the wild oscillations of multipole ratios computed using the \indirect method for supersymmetric black hole does not seem to be present in the multipole ratios computed using this method for the Kerr black hole (and for other non-extremal black holes). It is therefore possible that the \direct method works better than the \indirect one close to the BPS bound, but the two methods give closer results for non-supersymmetric black holes. To settle this issue one would need to compute multipole moments of microstate geometries for Kerr black hole. Such microstate geometries were constructed for the NHEK solution \cite{Heidmann:2018mtx}, but these do not yet have flat asymptotics.

\begin{table}[t]\centering \small
\begin{tabular}{|c||c||c|c|c|c||c|c|c|c|}
\hline
 & & \multicolumn{4}{c||}{\direct} &  \multicolumn{4}{c|}{\indirect} \\ 
\emph{Row} & \emph{Ratio} & $A$ & $B$ & $C$ & $D$&  $A$ & $B$ & $C$ & $D$ \\ \hline\hline
%
  \rule{0pt}{\upmarg}\emph{1}& $\dfrac{S_1 S_1}{M_2 M_0}$ & -28.579 & -134.89  & 1.0629 & 695.95 & -22.615 & -135.58  & -6.1987 & -5272.8\\[\downmarg]  \hline
   \rule{0pt}{\upmarg}\emph{2}& $\dfrac{S_1 S_3}{M_2 M_2}$ & 87.089 & 404.67  & 3.2652 & 712.00 & 67.774 & 406.74  & 16.083 & 7621.0 \\[\downmarg] \hline
   \rule{0pt}{\upmarg}\emph{3}& $\dfrac{M_2 S_2}{M_3 S_1}$ &  1.0048 & 1.0000  & 1.1358 & 1.8496 & 1 & 1 & 1 & 1\\[\downmarg]  \hline
  \rule{0pt}{\upmarg}\emph{4}& $\dfrac{M_2 S_2}{M_0 S_4}$ & -0.49061 & -0.49998  & 0.26742 & 0.80569 & -0.50078 & -0.50002  & -0.62713 & -2.2455 \\[\downmarg] \hline
 \rule{0pt}{\upmarg}\emph{5}& $\dfrac{M_2 S_3}{M_4 S_1}$ & 1.0100 & 1.0000  & 1.0564 & 0.79636 & 1 & 1 & 1 & 1    \\[\downmarg]  \hline
    \rule{0pt}{\upmarg}\emph{6}& $\dfrac{M_3 S_2}{M_2 S_3}$ & 1.3247 & 1.3333  & 0.70317 & 0.47370 & 1.3337 & 1.3333  & 1.3854 & 1.6919 \\[\downmarg] \hline
  \rule{0pt}{\upmarg}\emph{7}& $\dfrac{S_3 S_2}{S_1 S_4}$ & 1.4950 & 1.5000  & 0.82155 & 0.82427 & 1.5008 & 1.5000  & 1.6271 & 3.2455 \\[\downmarg]  \hline   
 \rule{0pt}{\upmarg}\emph{8}& $\dfrac{M_3 M_3}{M_6 M_0}$ & -0.79111 & -0.79995  & 0.18349 & 0.10829 & -0.80187 & -0.80004  & -1.1458 & 6.8622  \\[\downmarg] \hline 
  \rule{0pt}{\upmarg}\emph{9}& $\dfrac{M_3 M_3}{M_4 M_2}$ & 1.3317 & 1.3333  & 0.65398 & 0.20396 & 1.3337 & 1.3333  & 1.3854 & 1.6919  \\[\downmarg] \hline
 \rule{0pt}{\upmarg}\emph{10}& $\dfrac{M_2 S_4}{M_4 S_2}$ & 0.67559 & 0.66668  & 1.2858 & 0.96614 & 0.66632 & 0.66666  & 0.61458 & 0.30812\\[\downmarg] \hline
 \rule{0pt}{\upmarg}\emph{11}& $\dfrac{M_4 S_4}{M_2 S_6}$ & 1.9622 & 1.9999  & 0.80104 & 0.72370 & 2.0031 & 2.0001  & 2.6819 & -0.80288 \\[\downmarg]  \hline
  \rule{0pt}{\upmarg}\emph{12}& $\dfrac{M_5 S_3}{M_3 S_5}$ & 1.1833 & 1.2000  & 1.0787 & 1.8572 & 1.2006 & 1.2000  & 1.3188 & -1.8062  \\[\downmarg] \hline 
\end{tabular}
\caption{Several mixed mass-current (dimensionless) multipole ratios for the four-center solutions of Sec. \ref{sec:4cent}, computed using the \direct and the \indirect methods.}
\label{tab:simplemultipolevals-4c}
\end{table}

\clearpage

\begin{table}[t]\centering \small
\begin{tabular}{|c||c||c|c||c|c|}
\hline
 & & \multicolumn{2}{c||}{\direct} &  \multicolumn{2}{c|}{\indirect} \\ 
\emph{Row} & \emph{Ratio} & $(1,0)$ & $(2,1)$ & $(1,0)$ & $(2,1)$ \\ \hline\hline
%
  \rule{0pt}{\upmarg}\emph{1}& $\dfrac{S_1 S_1}{M_2 M_0}$ & 3228.9 & 484.36 & -4701.2 & -35506.\\[\downmarg]  \hline
   \rule{0pt}{\upmarg}\emph{2}& $\dfrac{S_1 S_3}{M_2 M_2}$ & 6189.9 & 1319.4 & 1250.0 & 29854. \\[\downmarg] \hline
   \rule{0pt}{\upmarg}\emph{3}& $\dfrac{M_2 S_2}{M_3 S_1}$ &  0.97500 & 7.4722 & 1 & 1\\[\downmarg]  \hline
  \rule{0pt}{\upmarg}\emph{4}& $\dfrac{M_2 S_2}{M_0 S_4}$\tablefootnote{For the six $\mathbb{Z}_2$-symmetric solutions in Sec. \ref{sec:pincers}; the resulting ratios lie between 0.11967 and 0.42966.} & 0.34472 & 0.32794 & 1.3622 & 6.2822 \\[\downmarg] \hline
 \rule{0pt}{\upmarg}\emph{5}& $\dfrac{M_2 S_3}{M_4 S_1}$ & 0.96772 & 0.89312 & 1 & 1   \\[\downmarg]  \hline
    \rule{0pt}{\upmarg}\emph{6}& $\dfrac{M_3 S_2}{M_2 S_3}$ & 0.49901 & 0.98095 & 4.7610 & 2.1893  \\[\downmarg] \hline
  \rule{0pt}{\upmarg}\emph{7}& $\dfrac{S_3 S_2}{S_1 S_4}$ & 0.66084 & 0.89329 & -0.36219 & -5.2822 \\[\downmarg]  \hline
 \rule{0pt}{\upmarg}\emph{8}& $\dfrac{M_3 M_3}{M_6 M_0}$ & 0.20005 & 0.035533 & 1.0592 & 1.6235 \\[\downmarg] \hline 
  \rule{0pt}{\upmarg}\emph{9}& $\dfrac{M_3 M_3}{M_4 M_2}$ & 0.49528 & 0.11725 & 4.7610 & 2.1893  \\[\downmarg] \hline
 \rule{0pt}{\upmarg}\emph{10}& $\dfrac{M_2 S_4}{M_4 S_2}$\tablefootnote{This ratio can also be computed for the six $\mathbb{Z}_2$-symmetric solutions in Sec. \ref{sec:pincers}; the resulting ratios all lie between 0.87117 and 0.96834.} & 1.4644 & 0.99981 & -2.7610 & -0.18931\\[\downmarg] \hline
 \rule{0pt}{\upmarg}\emph{11}& $\dfrac{M_4 S_4}{M_2 S_6}$\tablefootnote{For the six $\mathbb{Z}_2$-symmetric solutions in Sec. \ref{sec:pincers}; the resulting ratios lie between 1.0244 and 1.1620.} & 0.77199 & 0.96938 & 0.42334 & 0.13732 \\[\downmarg]  \hline
  \rule{0pt}{\upmarg}\emph{12}& $\dfrac{M_5 S_3}{M_3 S_5}$ & 1.2445 & 1.0620 & 0.16331 & 0.11804  \\[\downmarg] \hline
\end{tabular}
\caption{Several mixed mass-current (dimensionless) multipole ratios for the asymmetric pincers of Sec. \ref{sec:pincers},computed using the \direct and the \indirect methods.}
\label{tab:simplemultipolevals-aspinc}
\end{table}

\clearpage

\clearpage

\begin{table}[p]\centering
\begin{tabular}{|c||c|c|}
\hline
Geometry & $\mathcal{H}$ & $\mathcal{E}_{\text{(ave)}}$\\ \hline \hline
$(1,0)$ & 0.28 & 2.79\\ \hline
$(2,1)$ & 0.098 & 15.7\\ \hline
$A$ & $7.7\times 10^{-4}$ & 0.0451 \\ \hline
$B$ & $7.9\times 10^{-6}$ & 0.000888\\ \hline
$C$ & 0.055 & 2.31\\ \hline
$D$ & 0.24  & 8.56 \\ \hline
\end{tabular}
\caption{The entropy parameter $\mathcal{H}$ (see (\ref{eq:entropypar})) and the average error between the \direct and the \indirect methods (see (\ref{eq:errordef})) computed for the six geometries we consider in this paper.}
\label{tab:entropyvserror}
\end{table}

\clearpage
\subsection{Ratios for the Kerr black hole}\label{sec:ratiosnonext}
As we explained above, our \indirect method gives a way to associate well-defined values of multipole ratios to any non-extremal black hole. We will focus on the Kerr black hole, for which the \indirect method gives the multipole ratios (\ref{eq:generalratioKerr}).

Using (\ref{eq:generalratioKerr}), we can find new expressions for many multipole ratios that are otherwise ill-defined in the Kerr geometry. Some examples are:
\begin{align}
\frac{ M_{l+1} M_{l+2} }{ M_{l} M_{l+3} } &= \frac{ S_{l} S_{l+1} }{ S_{l-1} S_{l+2} } =  \frac{ M_{l+2} S_{l} }{ M_{l} S_{l+2} } =  1 - \frac{4}{ 3 + (-1)^l (2l+1) },\\
\frac{ S_{l+1} S_{l+2} }{M_l M_{l+3} } &=  -\frac{1+ (-1)^l (2 l+3)}{3 + (-1)^l (2 l+1)},\label{eq:Kerrratiofamily1}\\
\frac{ M_{l+1} M_{l+2} }{S_l S_{l+3} } &= -\frac{1-(-1)^l(2l+1)}{3-(-1)^l(2l+3)}.
\end{align}
All of these ratios are naively zero over zero for the Kerr solution since $S_{2l}=M_{2l+1} = 0$ for all $l$, but can be calculated using our \indirect method and are well-defined  for all integer $l\geq 0$ (except $l=0$ for the ratio involving $S_{l-1}$).
One can also use the \indirect method to calculate the following ratios:
\begin{align} 
\label{eq:vanishingdev} \frac{\left[M_{2n} - M(-a^2)^n\right]}{S_{2n}} &= \frac{\left[S_{2n+1} - J(-a^2)^n\right]}{M_{2n+1}}    = 0,\\
\label{eq:M2dev} \frac{\left[M_{2n} - M(-a^2)^n\right]M_{2n}}{\left(S_{2n}\right)^2} &= -\frac12 + \frac{3}{4n} ,\\
\label{eq:S3dev} \frac{\left[S_{2n+1} - J(-a^2)^n\right]S_{2n+1}}{\left(M_{2n+1}\right)^2} &= -\frac{2n+1}{4n}.
\end{align}
The ratios (\ref{eq:vanishingdev})-(\ref{eq:S3dev}) are constructed in such a way that the Kerr limit of these ratios using the \indirect method is unique and unambiguous. We discuss these subtle ``subtracted'' ratios in more detail in appendix \ref{app:indirect-subtracted}.
We do not believe there is an alternative way to define the ratios (\ref{eq:vanishingdev})-(\ref{eq:S3dev}) in such a way that they remain unambiguous and well-defined in the Kerr limit. 

Finally, we note that there are multipole ratios that are independent of the type of black hole we consider, which follows immediately from (\ref{eq:truemultipolesRLM})-(\ref{eq:truemultipolesRLS}). One such ratio is:
\be \label{eq:Kerrratiofamilyl}  \frac{ M_2 S_l}{ M_{l+1} S_1} = 1,\ee
which is valid for all integer $l>0$ and is independent of all four parameters $M,D,J,a$ of the general black hole. Note that $l=2,3$ corresponds to rows 3 and 5 of tables \ref{tab:simplemultipolevals-4c} and \ref{tab:simplemultipolevals-aspinc}.

\subsection{Constraining deviations of Kerr}
\label{Kerr-deviations}

The ratios of vanishing multipole moments computed for the Kerr black hole in the previous section can be used to place very strong constraints on the parameterization of deviations of the gravitational multipoles from the Kerr geometry that one expects to constrain using gravitational-wave measurements.  

The most general parameterization of a small  perturbation of the multipoles away from the Kerr values is:
\be \label{eq:Kerrdev} M_l = (M_l)_{\rm Kerr} + m_l^{(1)}\, \epsilon, \qquad S_l = (S_l)_{\rm Kerr} + s_l^{(1)}\,\epsilon, \ee
where for simplicity we have introduced a small dimensionless parameter $\epsilon\ll 1$ to organize the expansion around the Kerr values.

It is then easy to see that, for even $l=2n$, (\ref{eq:Kerrratiofamily1}) gives:
\be \frac{ M_2 S_{2n}}{ M_{2n+1} S_1} = -a\frac{s_{2n}^{(1)}}{m_{2n+1}^{(1)}} + \mathcal{O}(\epsilon).\ee
Thus, the reasonable demand that the perturbation of Kerr has a continuous limit to Kerr as the perturbation is turned off ($\epsilon\rightarrow 0$) means that this ratio must be equal to $1$ at $\mathcal{O}(\epsilon^0)$, which unambiguously fixes all of the perturbed odd mass multipoles in terms of the perturbed even current multipoles:
\be\label{eq:Kerrmconstraint} m_{2n+1}^{(1)} = -a s_{2n}^{(1)}.\ee
In the same way, equation (\ref{eq:Kerrratiofamilyl}) for even $l=2n$ gives:
\be \frac{ M_{2n+2} S_{2n} }{ M_{2n} S_{2n+2} } = -a^2 \frac{s_{2n}^{(1)}}{s_{2n+2}^{(1)}} + \mathcal{O}(\epsilon).\ee
Hence, demanding that this be equal to (\ref{eq:Kerrratiofamilyl}) gives:
\be\label{eq:Kerrsconstraint} s_{2n+2}^{(1)} = -\frac{n+1}{n}a^2 s_{2n}^{(1)}.\ee

For the multipoles $M_{2n}, S_{2n+1}$, using (\ref{eq:vanishingdev}), it is easy to see that the $\mathcal{O}(\epsilon)$ term in (\ref{eq:Kerrdev}) must vanish:
\be \label{eq:m2n1} m_{2n}^{(1)}= s_{2n+1}^{(1)} = 0.\ee
The first non-zero perturbations of $M_{2n}, S_{2n+1}$ must then come at $\mathcal{O}(\epsilon^2)$ and are fixed by (\ref{eq:M2dev}) and (\ref{eq:S3dev}) (using expansion coefficients $m_{2n}^{(2)}, s_{2n+1}^{(2)}$ defined analogously to (\ref{eq:Kerrdev})):
\begin{align}
 \frac{\left[M_{2n} - M(-a^2)^n\right]M_{2n}}{\left(S_{2n}\right)^2} &= M(-a^2)^n \frac{m_{2n}^{(2)}}{\left(s_{2n}^{(1)}\right)^2} + \mathcal{O}(\epsilon),\\
 \frac{\left[S_{2n+1} - J(-a^2)^n\right]S_{2n+1}}{\left(M_{2n+1}\right)^2} &=  Ma(-a^2)^n \frac{s_{2n+1}^{(2)}}{\left(m_{2n+1}^{(1)}\right)^2} + \mathcal{O}(\epsilon),
\end{align}
so that we have:
\begin{align}
 \label{eq:m2n2dev} m_{2n}^{(2)} &= \frac{\left(s_{2n}^{(1)}\right)^2}{M(-a^2)^n} \left(-\frac12 + \frac{3}{4n}\right),\\
 \label{eq:s2n2dev} s_{2n+1}^{(2)} &= \frac{ \left(m_{2n+1}^{(1)}\right)^2}{Ma(-a^2)^n} \left(-\frac{2n+1}{4n}\right).
\end{align}

Hence, the dimensionless ratios we compute constrain small deviations away from the Kerr values of all multipole moments to a single, one-parameter family of allowed deviations. Explicitly, combining (\ref{eq:Kerrmconstraint}), (\ref{eq:Kerrsconstraint}), (\ref{eq:m2n1}), and (\ref{eq:m2n2dev})-(\ref{eq:s2n2dev}), we can parametrize this family in terms of the dimensionless parameter $\epsilon$, determined by $S_2\equiv M a^2 \epsilon$ (or equivalently $s_2^{(1)}\equiv Ma^2$ in (\ref{eq:Kerrdev})):
\begin{align}
S_{2n}&= -n M (-a^2)^n \epsilon, \nonumber \\
M_{2n+1} &= n Ma(-a^2)^n  \epsilon,\label{bigfat1} \\
M_{2n} - \left(M_{2n}\right)_{\rm Kerr} &= -n^2M (-a^2)^n  \left(\frac{2n-3}{4n}\right)\epsilon^2, \nonumber \\
S_{2n+1} - \left(S_{2n+1}\right)_{\rm Kerr} &= - n^2(-a^2)^n Ma  \left(\frac{2n+1}{4n}\right) \epsilon^2. \nonumber
\end{align}
%
%
%
These constraints have three important consequences:

First, these constraints come from embedding the Kerr black hole in a supergravity theory which is the low-energy effective action of String Theory. It is well possible that other theoretical models that modify the multipoles of the Kerr solution predict small multipole deviations that do not satisfy the constraints (\ref{bigfat1}). This indicates that these models are incompatible with String Theory.

Second, the lowest gravitational multipoles of rotating black holes are expected to be constrained by future observations of EMRI gravitational waves \cite{Audley:2017drz,Barack:2006pq,Babak:2017tow,Ryan:1995wh}, and possibly even sooner \cite{Krishnendu:2018nqa}. If these multipoles differ from those of the Kerr solution in a manner consistent with our String-Theory-derived constraints, this would be a big predictive success for String Theory, and would indicate that whatever effect gives rise to these modification comes from String Theory.

If, however, gravitational wave measurements find multipoles that differ from those of the Kerr solution, but do not satisfy the constraints (\ref{bigfat1}), there are two possible explanations. One would be that the Kerr solution is still modified very slightly, but the modifications are incompatible with String Theory. The second (and more likely in our opinion) would be that the modifications of the Kerr solution are very large, and hence away from the linear regime in which we worked to compute Kerr multipole ratios using the \indirect method. 

Thus, within String Theory, the multipole ratios (\ref{bigfat1}) constitute a benchmark for distinguishing between \emph{strong} modifications of the Kerr geometry and \emph{weak} modifications. Indeed, all weak modifications will give rise to multipole moments that obey (\ref{bigfat1}), while strong modifications, like those brought about by structure at the scale of the horizon, will give rise to modifications that are significantly different from those of  (\ref{bigfat1}). The discrepancy of the \direct and \indirect results for supersymmetric black holes confirms the fact that a structure at the scale of the horizon can give rise to multipole ratios that differ drastically from those computed using slight perturbations of the horizon. 

Third, when two black hole merge the resulting configuration relaxes into a Kerr black hole, and at late times this relaxation can be described perturbatively using the quasinormal modes  of the Kerr geometry. This final ring-down should be characterized by the final relaxation of the multipole moments $M_{2n}$ and $S_{2n+1}$, and we believe it should be possible to use our result (\ref{bigfat1}) to argue that during the relaxation these decaying moments are not independent but rather related with each other in a universal way. Such universality of the final relaxation should be visible in gravitational-wave observations of black hole mergers in the (late) ring-down phase.

\section{Discussion}\label{sec:discussion}

The most important goal of our research into gravitational multipoles is to understand whether the horizon-scale structure which is necessary to preserve quantum unitarity gives rise to gravitational multipoles that are different from those of the classical black hole solution. For the supersymmetric black holes we have considered in this paper, this structure is given by scaling bubbling microstate geometries. In the scaling limit, in which the charges and angular momenta of the microstate geometries match those of the black hole, the gravitational multipoles become identical to those of the classical black hole solution.

However, we have found that the ratios of the vanishing multipoles computed by a small (linear) perturbation of the classical black hole solution (the \indirect method) can differ a fair bit from the ratios of vanishing multipoles computed 
in scaling microstate geometries (the \direct method). This indicates that \emph{large} changes in the geometry at the scale of the horizon modify the ratios of vanishing multipoles.

The key question is how this story will extend to non-extremal black holes. One possibility is that for non-extremal black holes, the multipole moments of the microstate geometries will be different from those of the classical black hole solution. After all, a spinning ball of dust or a spinning lump of liquid have different multipole moments than the Kerr black hole, so it is possible that the nontrivial topology and fluxes that will make up the non-extremal microstate geometry will also spin differently from the Kerr black hole.  Unfortunately, there are no examples yet of explicit microstate geometries that have the same mass, charges and angular momentum as non-extremal four-dimensional black holes, so one cannot offer at this point any support for this possibility.

The other possibility is that, even away from extremality, the multipoles of the microstate geometries will still be the same as those of the black hole. However, much like for supersymmetric black holes, we do not expect the ratios of vanishing multipoles to be the same. Hence, if any other effect will cause a small measurable deformations of the multipoles away from the Kerr values, our calculation gives a benchmark for ascertaining whether the deviations from the Kerr geometry at the scale of the horizon are small or large. 

\paragraph{Comparing the two methods for BPS black holes}
We have seen that there is an impressive correlation between the agreement of the  \direct and \indirect methods for computing ratios of vanishing multipole and the smallness of the entropy parameter, $\mathcal{H}$, defined in (\ref{eq:entropypar}); see table \ref{tab:entropyvserror} and fig. \ref{fig:loglogHE}. We believe it is important to obtain a deeper understanding of this correlation, which is simply an empirical observation in our results at the moment.

One possibility, which we mentioned above, is that the \indirect method gives the ratios of vanishing multipoles when the black hole is only slightly perturbed around the BPS solution, while the \direct method gives these ratios when the horizon is completely replaced by a bubbling microstructure. Hence, the disagreement from the result of the \indirect method can be considered as a benchmark for parameterizing how much the structure at the scale of the horizon (that is needed to allow information to escape) differs from the black hole solution.

Another interesting possibility is that, for a particular black hole, the \indirect method gives the \emph{average} value of the multipole ratios calculated using the \direct method for all the possible microstates of this black hole. The spread or \emph{variance} of the multipole ratios for all these microstates could be determined by $\mathcal{H}$ -- in other words, a smaller $\mathcal{H}$ would imply that the microstate multipole ratios calculated using the \direct method will generally lie closer to the would-be ``average'' value calculated for the black hole using the \indirect method.

We can wonder also why the matching of the \direct and \indirect methods is correlated to the value of the entropy parameter $\mathcal{H}$, and not to the actual black hole entropy, $S$. Note that $\mathcal{H}$ can be intuitively thought of as a ``relative'' entropy, which measures the entropy of the black hole relative to a black hole that has only the electric charges; moreover, our mismatch parameter $\mathcal{E}$ defined in (\ref{eq:errordef}) is also a \emph{relative} difference parameter. It would be interesting to see if there may exist a possible connection between the \emph{absolute} difference between multipole ratios and the (absolute) entropy of the black hole, $S$. We leave this investigation to further study.

\paragraph{Multipole ratios in observations}

As we saw in Section \ref{Kerr-deviations}, the \indirect method can be used to calculate ratios of vanishing multipoles for the Kerr black hole, and constrain the space of non-Kerr multipoles coming from small deviations away from the Kerr solution at the scale of the horizons. 

By contrast, the ratios of vanishing multipoles computed via the \direct method using microstate geometries do not generically agree with the ratios computed using the \indirect method. This disagreement comes from the fact that microstate geometries drastically modify the physics at the scale of the horizon, and hence cannot be thought of as small perturbations of the black hole solution. 

This indicates that, within String Theory, the multipole ratio constraints derived in Section  \ref{Kerr-deviations} form a benchmark for determining whether the physics at the scale of the horizon is weakly or strongly modified from the physics of the Kerr solution. Hence, a possible gravitational wave measurement of non-Kerr multipoles that obey the constraints \eqref{bigfat1} would be a spectacular confirmation of this \emph{small} modification prediction of String Theory, while a possible measurement that does not obey the constraints \eqref{bigfat1} would be indicative of either a large deviation at the scale of the horizon within String Theory or of a small deviation at the scale of the horizon within another theory. 

Furthermore, the ratios we compute could be related to the (very) late-time relaxation properties in the formation process of black holes.\footnote{We thank the members of the Saclay String Theory Journal Club, and especially Nick Warner, for an elaborate discussion on this idea.} When a black hole (or a particular microstate) is formed, multipole radiation will carry away much of the information of the initial collapsing state. It is possible that the multipole ratios we consider indicate a \emph{universal} late-time relaxation behavior for the (ratios) of such multipole radiation after formation. This could possibly be probed in current and upcoming gravitational wave observations of black hole mergers by LIGO and eLISA \cite{Audley:2017drz,Barack:2006pq,Babak:2017tow,Ryan:1995wh,Krishnendu:2018nqa}; most likely the multipole ratios would correspond to effects in the (late) ring-down phase after the actual merger occurs.


\paragraph{Other directions}
The possibility of observing multipole ratios in the late-time relaxation behavior of the formation of black holes might also be interpreted as taking the limit where the energy of the gravitational radiation after formation becomes soft. Thus, there may be an interesting link between the multipole ratios we compute and the soft charges of a black hole \cite{Bondi:1962px,Sachs:1962wk,Barnich:2011mi}. However, these soft charges are defined as an expansion at null infinity (${ 
\cal I}^+$) while the multipoles (and thus their ratios) are defined at spatial infinity $i^+$. It would be interesting to see if the two expansions could be related and hence if the multipoles could be related to the soft charges of the black hole.

It would also be interesting to compute the multipole ratios using the \indirect method by embedding the Kerr black hole into modified theories of gravity that do \emph{not} descend from String Theory \cite{Papadopoulos:2018nvd}. Based on the consistency of String Theory, one expects that the ratios we compute by embedding a black hole in supergravity should be the same as those computed by embedding the black hole in a theory which is deformed with higher-derivative corrections calculated in String Theory\footnote{This was done for BPS black holes in \cite{Castro:2008ne}.}. However, if one deforms the theory arbitrarily, it is possible that the deformed multipole ratios will end up different than those we have computed using the \indirect method. We believe it is important to understand whether this happens and how. 

It is interesting to remark that, while microstate geometries have been used so far to describe holographically pure CFT states that belong to the black hole ensemble, our work shows that they have another unexpected utility: they can be used in the \direct method to calculate an infinite number of dimensionless multipole ratios that characterize the black hole and the cannot be computed in any other way.

There are two obvious places where one can generalize this \direct method. The first is the computation of multipole moments in almost-BPS multi-center solutions \cite{Goldstein:2008fq,Bena:2009ev,Bena:2009en}, and generalizations thereof \cite{Bossard:2011kz}, and the comparison of these multipole moments and ratios to those computed for the corresponding almost-BPS black hole using the \indirect method. General almost-BPS black holes in four dimensions \cite{Bena:2009ev} have a richer structure compared to BPS black holes, because they can also have a nontrivial angular momentum. It would be interesting to understand whether, for these black holes, there is also a discrepancy between the ratios of vanishing multipoles computed using the two methods, and whether this discrepancy is correlated to a specific feature of these black holes. 

The second place where one can generalize the \direct method is the computation of multipole moments in bubbling microstate geometries corresponding to the extremal Kerr solution in five-dimensions. Such bubbling geometries have only been constructed so far for the five-dimensional NHEK geometry \cite{Heidmann:2018mtx} using the supergravity ansatz of \cite{DallAgata:2010srl, Bena:2012wc}. To compute the asymptotic expansion of the full, asymptotically flat (extremal) Kerr solution, one would have to match the perturbed NHEK UV region of the microstate geometries constructed in \cite{Heidmann:2018mtx} to the NHEK IR region of the Kerr black hole, and see how the asymmetries brought about by the microstructure at the bottom of the NHEK throat modify the Kerr gravitational multipole moments. This procedure can be technically quite involved, but has the advantage of allowing us to compute ratios of vanishing multipoles for the Kerr black hole using the \direct method, and produce the first explicit predictions of the fuzzball proposal to real-life black holes.

\section*{Acknowledgments}
We would like to thank Pierre Heidmann for sharing the unpublished scaling geometries $B$ and $C$. We would also like to thank Massimo Bianchi, Vitor Cardoso, Bogdan Ganchev, Yixuan Li, Andrea Puhm, Bert Vercnocke, and Nick Warner for useful discussions and comments on the manuscript. The  work  of  IB is  supported  by  the  ANR  grant  Black-dS-String  ANR-16-CE31-0004-01, by the  John Templeton Foundation grant 61149, and the ERC Grants 787320-QBH Structure and 772408-Stringlandscape. DRM is supported by the ERC Starting Grant 679278 Emergent-BH.

\appendix
\section{Non-Axisymmetric Microstate Multipoles}\label{app:generalmultipoles}

We derived the general formulae (\ref{eq:multipolesMmc}) and (\ref{eq:multipolesSmc2}) for the gravitational multipoles of axisymmetric microstate geometries with arbitrary four-dimensional moduli in Section \ref{sec:multicentermultipoles}. It is a simple matter to expand this derivation to include microstate geometries that are not axisymmetric, using the ACMC formalism of \cite{Thorne:1980ru}.\footnote{While in (\ref{eq:themultipoleexpansiongtt})-(\ref{eq:themultipoleexpansiongrtheta}) we generalized ACMC-$N$ coordinates to AC-$N$ coordinates to allow for $\tilde M_1\neq 0$, we will simply work in ACMC-$N$ coordinates in this appendix for simplicity, for which $M_{1m}=0$. Further, note that as discussed in Section \ref{sec:multicentermultipoles}, since the purely spatial components of the metric are also determined by $\mathcal{Q}$ and $\omega$, it is trivial to conclude that the microstate geometry metric (\ref{eq:ds2multicenter}) is ACMC-$\infty$, even when it is not axisymmetric.}

\paragraph{Spherical harmonics}
Following \cite{Thorne:1980ru}, we will introduce the (scalar) spherical harmonics as:
\be Y_{lm}(\theta,\phi) = (-1)^m \sqrt{\frac{2l+1}{4\pi}\frac{(l-m)!}{(l+m)!}} e^{im\phi} P_l^m(\cos\theta),\ee
where $P_l^m(\cos\theta)$ are the associated Legendre polynomials. (Note the factor of $(-1)^m$, which differs from other normalizations.) These are orthonormal as:
\be \int Y_{lm}(\theta,\phi)Y_{l'm'}^*(\theta,\phi) \,d\Omega_2 = \delta_{ll'}\delta_{mm'}.\ee
We will also need the vector spherical harmonics:
\be \vec{Y}^B_{lm} = i \sum_{m'=-l}^l \sum_{m''=-1}^1 \left(1\, m'', l\, m'; l\, m\right) \vec{\xi}_{(m'')} Y_{lm'}(\theta,\phi),\ee
where we have used the Clebsch-Gordon coefficient $\left(1\, m'', l\, m'; l\, m\right)$, and the collection of vectors (components given in Cartesian coordinates $(x,y,z)$):
\be \vec{\xi}_{(-1)} = \frac{1}{\sqrt{2}}(1,-i,0), \qquad \vec{\xi}_{(0)} = (0,0,1) , \qquad \vec{\xi}_{(1)} = \frac{1}{\sqrt{2}}(-1,-i,0).\ee
Other vector spherical harmonics are:
\be \vec{Y}^E_{lm} = -\hat{r}\times \vec{Y}^B_{lm}, \qquad \vec{Y}^R_{lm} = \hat{r} Y_{lm}(\theta,\phi),\ee
where $\hat{r}$ is the unit radial vector. All of these vector spherical harmonics are normalized such that (for $J=E,B,R$):
\be \int \vec{Y}^J_{lm}\cdot \vec{Y}^{J'*}_{l'm'} \,d\Omega_2 = \delta_{JJ'}\delta_{ll'}\delta_{mm'}.\ee

\paragraph{Mass multipoles}
The generalization of the ACMC-$N$ expansion of $g_{tt}$ in (\ref{eq:themultipoleexpansiongtt}) to non-axisymmetric spacetimes is \cite{Thorne:1980ru}:\footnote{Note that, for the expansion (\ref{eq:gengtt}) to define an ACMC-$N$ coordinate system, we need $M_{1m}=0$, which can be achieved by a judicious choice of the origin.}
\begin{align}
\label{eq:gengtt} g_{tt} & = -1 + \frac{2M}{r}+ \sum_{l\geq 1}^{N} \sqrt{\frac{4\pi}{2l+1}} \frac{2 }{r^{l+1}} \left(  \sum_{m=-l}^m M_{lm} Y_{lm} + \sum_{l'<l}\sum_{m'=-l'}^{l'} c^{(tt)}_{ll'm'} Y_{l'm'} \right)\\
\nonumber & + \sqrt{\frac{4\pi}{2N+3}} \frac{2}{r^{N+2}} \left( \sum_{m=-N-1}^{N+1}\left(M_{N+1,m} Y_{N+1,m}\right) + \sum_{l'\neq l}\sum_{m'=-l'}^{l'} c^{(tt)}_{(N+1)l'm'} Y_{l'm'} \right) + \mathcal{O}\left(r^{-(N+3)}\right),
 \end{align}
where the argument of all spherical harmonics is always $Y_{lm}=Y_{lm}(\theta,\phi)$. The coefficients $M_{lm}$, which are defined for $l\geq 0$ and $|m|\leq l$, are then the mass multipoles for a generic, non-axisymmetric spacetime. Note that they have been normalized such that, when the spacetime is axisymmetric, $M_{lm}=0$ for $m\neq 0$ and $M_{l0}=M_l$ as defined in Section \ref{sec:multipoles}. Note that from the reality of the metric, it follows that $M_{l(-m)} = (-1)^m M_{lm}^*$. The coefficients $c^{(tt)}_{ll'm'}$ are arbitrary and do not contribute to the multipole structure.

The generalization of the expansion (\ref{eq:harmfuncexpand}) of a harmonic function to include poles that are not only on the $z$-axis is then:
\be \label{eq:genharmfuncexpand} H = h + \sum_i \frac{h^i}{r_i} = h + \sum_i h^i \sum_{l=0}^\infty \sum_{m=-l}^l \mathcal{R}^{(i)}_{lm} \frac{Y_{lm}(\theta,\phi)}{r^{l+1}},\ee
where the coefficient $\mathcal{R}^{(i)}_{lm}$ is defined for the $i$-th center at position $(r_i,\theta_i,\phi_i)$ as:
\be \mathcal{R}^{(i)}_{lm} \equiv (r_i)^l \sqrt{\frac{4\pi}{2l+1}} Y_{lm}^*(\theta_i,\phi_i) .\ee
Note that when the center is on the $z$-axis, $\theta_i=(0\mod \pi)$ so that $\mathcal{R}_{l0} = (r_i)^l=z_i^l$ and $\mathcal{R}_{lm}=0$ for $m\neq 0$, in agreement with the axisymmetric expansion (\ref{eq:harmfuncexpand}).

From comparing (\ref{eq:genharmfuncexpand}) and (\ref{eq:gengtt}), following the same reasoning as in Section \ref{sec:multicentermultipoles}, we arrive at:
\begin{align}
 \nn M_{lm} &= -\frac12 \left( \partial_{v^0} \left[ \mathcal{Q}^{-1/2}_\infty \right] \sum_i v^i \mathcal{R}^{(i)}_{lm} + \sum_I \partial_{k^0_I} \left[ \mathcal{Q}^{-1/2}_\infty \right] \sum_i k_I^i \mathcal{R}^{(i)}_{lm} \right.\\
 & \label{eq:generalMmult} \left. + \sum_I \partial_{l^0_I} \left[ \mathcal{Q}^{-1/2}_\infty \right] \sum_i l_I^i \mathcal{R}^{(i)}_{lm} +  \partial_{m^0} \left[ \mathcal{Q}^{-1/2}_\infty \right] \sum_i m^i \mathcal{R}^{(i)}_{lm} \right).
 \end{align}
which generalizes (\ref{eq:multipolesMmc}).

\paragraph{Current multipoles}
Now, we need the generalization of the expansion of $g_{t\phi}$ in (\ref{eq:themultipoleexpansiongtphi}) to non-axisymmetric spacetimes; in general, all off-diagonal time-space components of the metric will be turned on. Using Cartesian coordinates $x^i$ (or any other asymptotically orthonormal coordinate frame), the ACMC-$N$ expansion of these time-space components is \cite{Thorne:1980ru}:
\begin{align}
\label{eq:gengti} g_{ti} & = \sum_{l\geq 1}^N \frac{2}{r^{l+1}} \left( -\sqrt{\frac{4\pi (l+1)}{l(2l+1)}} \sum_{m=-l}^l  S_{lm}\left(Y^B_{lm}\right)_i + \tilde C^{(ti)}_l+  \sum_{l'<l}C^{(ti)}_{ll'}\right)\\
\nonumber & + \frac{2}{r^{N+2}} \left( -\sqrt{\frac{4\pi (N+2)}{(N+1)(2N+3)}} \sum_{m=-N+1}^{N+1}  S_{N+1,m}\left(Y^B_{N+1,m}\right)_i + \tilde C^{(ti)}_{N+1}+  \sum_{l'\neq N+1}C^{(ti)}_{N+1,l'}\right)\\
\nonumber &+ \mathcal{O}(r^{-(N+3)}),
\end{align}
where we have defined the shorthands:
\begin{align}
\tilde C_l^{(ti)} &\equiv \sum_{m'=-l}^l\left(\tilde e^{(ti)}_{lm'} \left(Y^E_{lm'}\right)_i + \tilde r^{(ti)}_{lm'} \left(Y^R_{lm'}\right)_i \right),\\
C^{(ti)}_{ll'} &\equiv \sum_{m'=-l'}^{l'}\left( b^{(ti)}_{l,l'm'}\left(Y^B_{l'm'}\right)_i +  e^{(ti)}_{l,l'm'} \left(Y^E_{l'm'}\right)_i +  r^{(ti)}_{l,l'm'} \left(Y^R_{l'm'}\right)_i \right).
 \end{align}
The normalization of the multipoles $S_{lm}$ has again been chosen such that for axisymmetric spacetimes, $S_{lm}=0$ when $m\neq 0$ and $S_{l0}=S_l$ as defined in Section \ref{sec:multipoles}. Again from reality of the metric, it follows that $S_{l(-m)} = (-1)^m S_{lm}^*$. The coefficients $b^{(ti)}_{l,l'm'},e^{(ti)}_{l,l'm'},r^{(ti)}_{l,l'm'}$ and $\tilde e^{(ti)}_{lm'},\tilde r^{(ti)}_{lm'}$ are arbitrary and do not contribute to the multipole structure.\footnote{Note that the $\tilde C_l^{ti}$ piece in (\ref{eq:gengti}) contains $Y^E_{lm},Y^R_{lm}$ vector harmonics that appear at the same harmonic order, $l$, in the multipole expansion as $Y^B_{lm}$. Nevertheless, it is only the piece proportional to $Y^B_{lm}$ that contains physical, coordinate-independent information: the $S_{lm}$ current multipoles. Analyzing the linear gravitational-radiation field of a slow-moving source reveals that the pieces $\sim Y^E_{lm},Y^R_{lm}$ can either be gauged away completely or are proportional to the time derivative of the mass multipoles $M_l$. A further analysis shows that $M_{lm}$ and $S_{lm}$ also completely determine the full non-linear gravitational radiation field for any source. See especially sections 8, 9, and 10 of \cite{Thorne:1980ru}.}

As discussed in Section \ref{sec:multicentermultipoles}, $\mathcal{Q}$ does not contribute to the curent multipoles $S_{lm}$, and so the contribution for microstate geometries to $S_{lm}$ comes entirely from the one-form $\omega$ in (\ref{eq:ds2multicenter}). Moreover, $\omega$ is the sum of contributions $\omega_{ij}$ coming from pairs $(ij)$ of centers, with the contribution of each pair given in (\ref{eq:omegaij}) for axisymmetric spacetimes. We can rewrite (\ref{eq:omegaij}) and (\ref{eq:omegaijrewrite}) in a covariant form as:
\be \vec{\omega}_{ij} = \sum_l \frac{1}{r^{l+1}} \left(-\sqrt{\frac{4\pi (l+1)}{l(2l+1)}}\right) \sum_{m=-l}^l \vec{Y}_{lm}^B\left( \mathcal{R}^{(i)}_{lm} - \mathcal{R}^{(j)}_{lm} \right) + \cdots,\ee
which is then also valid when the pair $(ij)$ is not aligned on the $z$-axis. Comparing with (\ref{eq:gengti}), the generalization of (\ref{eq:multipolesSmc2}) to non-axisymmetric spacetimes will then be:
 \be \label{eq:generalSmult} S_{lm} = \frac12 \sum_i \langle h, \Gamma^i\rangle\,  \mathcal{R}_{lm}^{(i)},\ee
 where again for axisymmetric spacetimes we recover $S_{lm}=0$ for $m\neq 0$ and $S_{l0}=S_l$ as defined in Section \ref{sec:multipoles}.
 
When the asymptotic moduli have the canonical value for a solution with D2, D2, D2 and D6 charges, $h=(1,0,0,0,1,1,1,0)$, the general formulae (\ref{eq:generalMmult}) and (\ref{eq:generalSmult}) simplify to:
\begin{align}
\label{eq:generalmultipolesMmod-simple} M_{lm} &= \frac14 \sum_i \left[ v^i + l_1^i+l_2^i+l_3^i)\right] \mathcal{R}_{lm}^{(i)},\\
\label{eq:generalmultipolesSmod-simple} S_{lm} &= \frac14 \sum_i \left[ - 2 m^i + k_1^i+k_2^i+k_3^i\right] \mathcal{R}_{lm}^{(i)}.
\end{align}
These formulae generalize (\ref{eq:multipolesMmod-simple})-(\ref{eq:multipolesSmod-simple}) to non-axisymmetric spacetimes, and appeared first in \cite{Bianchi:2020bxa}. 

It would be very interesting to use these formulae to compute multipole moments and ratios of vanishing multipoles using multicenter non-axisymmetric scaling solutions, such as those of \cite{Bena:2007qc,deBoer:2008zn,Bianchi:2017bxl}.

\section{More Details on the \Indirect Method}\label{app:indirect}
In section \ref{sec:methods}, we introduced the \indirect method of calculating multipole ratios for any black hole. This allowed us to compute previously undefined multipole ratios for black holes such as Kerr or static BPS black holes. In this appendix, we will discuss some aspects of the \indirect method in more detail and point out some subtleties involved.

First of all, it is useful to be precise about how the \indirect method works. Consider a black hole that has a certain value of the 10 black hole parameters $(m_0,a_0,\delta_{I0},\gamma_{I0})$ (see section \ref{app:generalBH}). Then, for a ratio which involves any two \emph{monomials} $\mathcal{M}_1,\mathcal{M}_2$ in of the multipoles $M_l,S_l$, the indirect method computes this ratio as the following limit:
\be \label{eq:app:Rdef} \mathcal{R} \equiv \lim_{(m,a,\delta_I,\gamma_I)\rightarrow (m_0,a_0,\delta_{I0},\gamma_{I0})} \left(\frac{\mathcal{M}_1(M_l,S_l)}{\mathcal{M}_2(M_l,S_l)}\right)(m,a,\delta_I,\gamma_I),\ee
where all of the $M_l,S_l$ are functions of the 10 parameters through the expressions (\ref{eq:truemultipolesRLM})-(\ref{eq:truemultipolesRLS}) combined with (\ref{eq:generalBHMDJ}) and the expressions for $\mu_i,\nu_i$ in terms of $\delta_I,\gamma_I$ given in appendix \ref{app:chargeparams}. Some of the parameters $(m_0,a_0,\delta_{I0},\gamma_{I0})$ are allowed to vanish; for example the Kerr black hole has   $\delta_{I0}=\gamma_{I0}=0$. The \indirect method is then summarized by (\ref{eq:app:Rdef}) and the non-trivial statement that any ratio defined in such a way is \emph{uniquely and unambiguously defined}, which means it does not matter in which order one takes the (10-dimensional) limit in (\ref{eq:app:Rdef}).

\paragraph{Effective 4D phase space}
It can sometimes be useful to consider the limit in (\ref{eq:app:Rdef}) within an effective four-dimensional phase space parametrized by $(M,J,a,D)$, defined in terms of the fundamental parameters by (\ref{eq:generalBHMDJ}), although sometimes care must be taken when considering limits using these parameters. For example, when the charge parameters are all small, $\delta_I\sim \epsilon$ and $\gamma_I\sim \epsilon$, we have:
\be \label{eq:app:smallcharges} M = m + \mathcal{O}(\epsilon^2), \qquad D\sim \mathcal{O}(\epsilon^4), \qquad J = M a + \mathcal{O}(\epsilon^4) .\ee
For reference, note also that in this regime (see appendix \ref{app:chargeparams}):
\be \label{eq:app:munusmallcharge} \mu_1 = \nu_2= 1 + \frac12\sum_I (\delta_I^2+\gamma_I^2) + \mathcal{O}(\epsilon^4), \qquad \mu_2 = -\nu_1 = \sum_I \delta_I\gamma_I + \mathcal{O}(\epsilon^4).\ee
From (\ref{eq:app:smallcharges}), it is clear that the three limits $(D,J,a)\rightarrow 0$ (which would give the Schwarzschild black hole) are \emph{not} independent. For example, consider the ratio:
\be \label{eq:app:Rex} \mathcal{R}_{\rm ex} = \frac{M_2 M_3}{S_1 S_4}.\ee
If we express $\mathcal{R}_{\rm ex}$ in terms of $(M,J,a,D)$, we get:
\be \mathcal{R}_{\rm ex} = \frac{a^2(D^2 + M^2)^2}{2J^2(D^2-M^2)},\ee
which seems ill-defined (i.e. depends on the order of limits) when $(D,J,a)\rightarrow 0$. However, when we express this ratio in terms of the fundamental parameters (for small $\delta_I,\gamma_I\sim \epsilon$), we get:
\be \mathcal{R}_{\rm ex} = \frac{\left(\mu_1^2+\mu_2^2\right)^2}{2 \left(\mu_1^2 \left(\nu_1^2-\nu_2^2\right)+4 \mu_1 \mu_2 \nu_1 \nu_2+\mu_2^2 \left(\nu_2^2-\nu_1^2\right)\right)} = -\frac12 + \mathcal{O}(\epsilon^4),\ee
which has a smooth, unambiguous limit as $(\delta_I,\gamma_I,a)\rightarrow 0$.

\subsection{``Subtracted'' multipole ratios}\label{app:indirect-subtracted}
As we mentioned above, the \indirect method gives a unique, unambiguous answer for any ratio (\ref{eq:app:Rdef}) of multipole \emph{monomials}, which in particular is independent of the order of limits in (\ref{eq:app:Rdef}). However, one can construct more complicated ``subtracted'' ratios, in which the numerator or denominator (or both) is no longer a simple monomial. Our interest in these ratios is more mathematical than physical, since the constraints they place on the deviations for the Kerr multipoles are subleading.  

These subtracted ratios have a  more subtle behavior. Take for example the ratio:
\be \mathcal{R}_{\rm sub,ex} \equiv \frac{M_0 M_2 +S_1^2}{S_2} = \frac{m (\mu_1 \nu_2-\mu_2 \nu_1) \left(-\mu_1^2-\mu_2^2+\nu_1^2+\nu_2^2\right)}{2 \nu_2 (\mu_1 \nu_1+\mu_2 \nu_2)}.\ee
For simplicity, start with $\delta_{1,2,3}=\gamma_{1,2,3}=0$. Now, consider the Kerr limit $(\delta_0,\gamma_0)\rightarrow 0$. When we scale the parameters as $(\delta_0,\gamma_0) = (\delta, \gamma) \epsilon$ with $\epsilon\rightarrow 0$, this limit becomes:
\be \mathcal{R}_{\rm sub,ex} = \lim_{\epsilon\rightarrow 0} \frac{M_0 M_2 +S_1^2}{S_2} = \frac{m}{4}\frac{\gamma ^4-6 \gamma ^2 \delta ^2+\delta ^4}{\gamma ^3 \delta -\gamma  \delta ^3}.\ee
We see that this limit explicitly depends on the direction from which we approach the Kerr point $(\delta_0,\gamma_0)=(0,0)$ and so this is not a unique or unambiguous limit.

It is clear one must take great care in defining subtracted ratios to avoid such ambiguities. It is instructive to discuss (\ref{eq:vanishingdev}) for $n=1$ in more detail:\footnote{The analysis for general $n$ and for the subtracted ratios (\ref{eq:M2dev}) and (\ref{eq:S3dev}) proceeds analogously.}
\be \label{eq:app:RKerr} \mathcal{R}^{\text{(Kerr)}} \equiv \lim_{(\delta_I,\gamma_I)\rightarrow 0} \frac{M_{2} - M(-a^2)}{S_{2}}  .\ee
Here, $M_{2},S_{2},M$ are all functions of the \emph{same} parameters $(m,a,\delta_I,\gamma_I)$. This means the limit can be evaluated as:
\be \mathcal{R}^{\text{(Kerr)}} =  \lim_{(\delta_I,\gamma_I)\rightarrow 0}  -\frac12  \frac{\mu_1\nu_1 + \mu_2\nu_2}{\nu_1^2+\nu_2^2} = 0,\ee
where the final equality follows easily from the small-charge limit (\ref{eq:app:munusmallcharge}). This limit is clearly well-defined and unambiguous, since it does not depend on the limit direction taken.

We could have considered an alternative version of (\ref{eq:app:RKerr}):
\be \label{eq:app:RKerralt} \mathcal{R}_{\rm alt}^{\text{(Kerr)}} \equiv \lim_{(m,a,\delta_I,\gamma_I)\rightarrow (m_0,a_0,0,0)} \frac{M_{2} - \hat{M}(-a^2)}{S_{2}},\ee
where the quantities appearing in this limit have the following functional dependence on the limit variables $(m,a,\delta_I,\gamma_I)$ and the Kerr black hole parameters $(m_0,a_0)$:
\be (M_2,S_2)= (M_{2},S_2)(m,a,\delta_I,\gamma_I), \qquad  \hat{M}=\hat{M}(m_0,\delta_I,\gamma_I). \ee
Taking the simplified case $\delta_{1,2,3}=\gamma_{1,2,3}=0$, we can parametrize this limit as:
\be m=m_0(1+\tilde m\,\epsilon^4), \qquad (\delta_0,\gamma_0)=(\delta,\gamma)\epsilon,\ee
and then taking $\epsilon\rightarrow 0$ results in:
 \be \mathcal{R}_{\rm alt}^{\text{(Kerr)}} = \lim_{\epsilon\rightarrow 0}\frac{M_{2} - \hat{M}(-a^2)}{S_{2}} = \frac{\tilde m}{\gamma^3\delta - \gamma \delta^3} ,\ee
 which depends on the particular direction in which the limit is taken. So, although both $\mathcal{R}^{\text{(Kerr)}}$ in (\ref{eq:app:RKerr}) and $\mathcal{R}^{\text{(Kerr)}}_{\rm alt}$ in (\ref{eq:app:RKerralt}) naievely appear to compute the same ratio, it is clear that the ratio $\mathcal{R}^{\text{(Kerr)}}$ is unambiguous and well-defined while $\mathcal{R}^{\text{(Kerr)}}_{\rm alt}$ is not.
 
 Although we have no proof, we believe there is no alternative way to define the subtracted ratios (\ref{eq:vanishingdev})-(\ref{eq:S3dev}) in such a way that they remain well-defined and unambiguous in the Kerr limit \emph{and} result in a different value for the limit. However, if such an alternative subtracted ratio did exist, it would imply that the $\mathcal{O}(\epsilon^2)$ terms in the deformed Kerr multipole formulas (\ref{bigfat1}) are not unique.

\subsection{Ratios for supersymmetric black holes}\label{app:indirect-susy}
Here, we explain in a more detail the \indirect method for the black holes corresponding to the supersymmetric microstate geometries we constructed in Section \ref{sec:scalinggeom}.

For a given multi-center microstate geometry, we first calculate the four electric ($Q_I$) and four magnetic ($P_I$) charges in four dimensions. In the gauge where the moduli are given by (\ref{eq:pincermodulinom0}) (in particular, where $m_0=0$), the charges are given by:
\begin{align}
 Q_{1,2,3}^{(ms)} & =  \sum_i l_{1,2,3}^i, & Q_0^{(ms)} &= \sum_i v^i,\\
 P_{1,2,3}^{(ms)} &=  \sum_i k_{1,2,3}^i, & P_0^{(ms)} &= -2 \sum_i m^i.
 \end{align}
Note that from our gauge choice it follows that $\sum_{I=0}^3 P_I =0$ for all of the microstate geometries discussed in Section \ref{sec:scalinggeom};\footnote{The microstate geometries in Section \ref{sec:scalinggeom} are introduced in a gauge where $\sum_i k^i_I=0$ (for $I=1,2,3$); performing the gauge transformation (\ref{eq:gaugetransf})-(\ref{eq:gaugeparam}) then gives $P_1=-P_4$ and $P_2=P_3=0$.} the four-dimensional mass $M=M_0$ (as one can read off from (\ref{eq:multipolesMmod-simple})) is given by:
 \be M = M_0 = \frac1{4} \sum_{I=1}^4 Q_I ,\ee
 as appropriate for a BPS solution (and in agreement with \cite{Chow:2014cca}).
 
 Now, we take a family of non-extremal, rotating black holes of the most general kind as given in Section \ref{app:generalBH}, with non-extremality parameter $m$, rotation parameter $a$, and 8 charge parameters $\delta_I,\gamma_I$; we set the NUT charge $N=0$ which fixes the parameter $n$ by (\ref{eq:fixn}). We let the parameters $m,a$ vary in our family of black holes, and the parameters $\delta_I,\gamma_I$ are then determined such that the charges $Q_I,P_I$ remain fixed to the microstate values. This means that for each value $m$, we solve:
 \be \label{eq:solvingcharges} Q_I^{(ms)} = Q_I(m,\delta_I,\gamma_I), \qquad P_I^{(ms)} = P_I(m,\delta_I,\gamma_I),\ee
 for $\delta_I,\gamma_I$. This determines a family of (non-extremal, rotating) black holes with the same charges $Q_I^{(ms)},P_I^{(ms)}$ corresponding to the BPS microstate under consideration. For this family of black holes, we can consider e.g. the ratio (\ref{eq:massmultipoleratioex}):
 \be \label{eq:therationonext} 
\frac{M_2 M_2}{M_4 M_0} = - \frac{D^2 + M^2}{3D^2-M^2} = \frac{\left(\mu_1^2+\mu_2^2\right) \left(\nu_1^2+\nu_2^2\right)}{\mu_1^2 \left(3 \nu_1^2-\nu_2^2\right)+8 \mu_1 \mu_2 \nu_1 \nu_2-\mu_2^2 \left(\nu_1^2-3 \nu_2^2\right)}.
\ee
Note that (\ref{eq:therationonext}) does not contain any explicit factors of $m,a$, but does depend on $m$ indirectly through the expressions for the charges $Q_I,P_I$, which depend on $m$ as well as on $\mu_i,\nu_i$.

Finally, we take the limit of (\ref{eq:therationonext}) when going to the non-rotating, BPS extremal limit of the black holes in this family, thus retrieving the value of (\ref{eq:therationonext}) for the BPS black hole that the microstate geometry is a microstate of. The non-rotating limit $a\rightarrow 0$ of (\ref{eq:therationonext}) is clearly trivial. Then, we can take the (BPS) extremal limit $m\rightarrow 0$ of (\ref{eq:therationonext}) by solving (\ref{eq:solvingcharges}) for decreasing values of $m$ and plugging the resulting values of $\delta_I,\gamma_I$ into (\ref{eq:therationonext}). In practice, we decrease the non-extremality parameter to $m\sim 10^{-10}$, for which the ratio (\ref{eq:therationonext}) has clearly converged to its $m\rightarrow 0$ value, at least to a much larger precision than we display in tables \ref{tab:simplemultipolevals-aspinc} and \ref{tab:simplemultipolevals-4c}. For the six black holes we consider in this paper, we give the values of $\mu_i,\nu_i$ (relative to $\mu_1$, since we note that the relevant multipole ratios only depend on these relative values) in table \ref{tab:munus}.

\begin{table}[ht]\centering
 \begin{tabular}{|c||c|c|c|}
 \hline
  BH & $\mu_2/\mu_1$ & $\nu_1/\mu_1$ & $\nu_2/\mu_1$\\ \hline \hline
  $(1,0)$ & 0.409831 & 15.34 & 72.4949\\ \hline
   $(2,1)$ & 0.536201 & 53.5961 & 206.984\\ \hline
  $A$ & 35.8367 & 0.0033209 & 170.488\\ \hline
    $B$ & -219.409 & -1.47008 & 2554.83 \\ \hline
        $C$ & -2.68896 & -0.228787 & 7.13903 \\ \hline
              $D$ & 1.05995 & 8.79139 & 105.449\\ \hline
 \end{tabular}
\caption{The ratios of $\mu_i,\nu_i$ parameters for the six BPS black holes considered in this paper; these values are those obtained in the extremal limit $m\rightarrow 0$, thus are those of the static, BPS black hole.}
\label{tab:munus}
\end{table}

\newpage
\section{Multipole Ratios in Microstate Geometries}\label{app:moreratios}

In this appendix, we analyze multipole ratios calculated by the \direct method introduced in Section \ref{sec:methods}. In Section \ref{NewWindow}, we have already considered various ratios for the asymmetric pincers $(1,0)$ and $(2,1)$ of Section \ref{sec:pincers}, as well as the four-center geometries $A,B,C,D$ of Section \ref{sec:4cent}. Here, we will additionally consider multipole ratios for the symmetric pincers $(n,n)$ (for $n=1,\cdots, 6$) introduced in Section \ref{sec:pincers}, whenever the ratio at hand is well-defined for them (recall that $S_{2n+1}=M_{2n+1}=0$ for these geometries). Including the symmetric pincers in our analysis can give us some insight into how the multipole ratios change when the number of centers $N$ is varied.

\begin{figure}[p]\centering
 \includegraphics[width=0.8\textwidth]{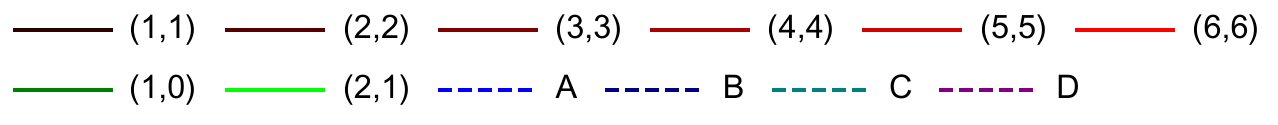}
 \caption{Legend for Figures \ref{fig:bubblemassmultipoles-R2}, \ref{fig:bubblemassmultipoles-R3}, \ref{fig:bubblemassmultipoles-Ri0}, \ref{fig:bubblecurmultipoles-R2}, \ref{fig:bubblecurmultipoles-R3}, and \ref{fig:bubblecurmultipoles-Ri0}. The label $(n_1,n_2)$ refers to the pincers of Section \ref{sec:pincers}; $A,B,C,D$ are the four-center solutions of Section \ref{sec:4cent}.}
 \label{fig:legend}
\end{figure}

To ease our analysis, we will first consider ratios of only mass multipoles $M_l$, and then (only) current multipoles $S_l$. Note that the legend of figs. \ref{fig:bubblemassmultipoles-R2}, \ref{fig:bubblemassmultipoles-R3}, \ref{fig:bubblemassmultipoles-Ri0}, \ref{fig:bubblecurmultipoles-R2}, \ref{fig:bubblecurmultipoles-R3}, and \ref{fig:bubblecurmultipoles-Ri0} is given in fig. \ref{fig:legend}.

\subsection{Ratios of mass multipoles}\label{sec:subsecmass}



\begin{figure}[hp]\centering
\begin{subfigure}{0.48\textwidth}\centering
 \includegraphics[width=\textwidth]{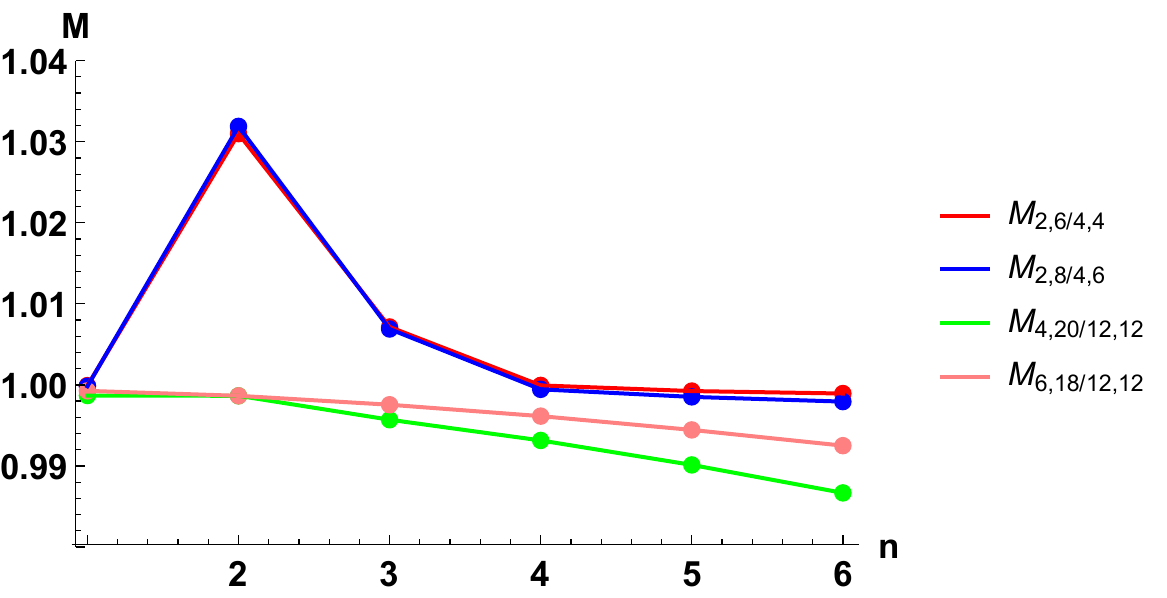}
 \caption{Mass multipole ratios without $M_0$}
 \label{fig:bubblemassmultipolesnoM0}
\end{subfigure}\hspace*{0.01\textwidth}
\begin{subfigure}{0.48\textwidth}\centering
  \includegraphics[width=\textwidth]{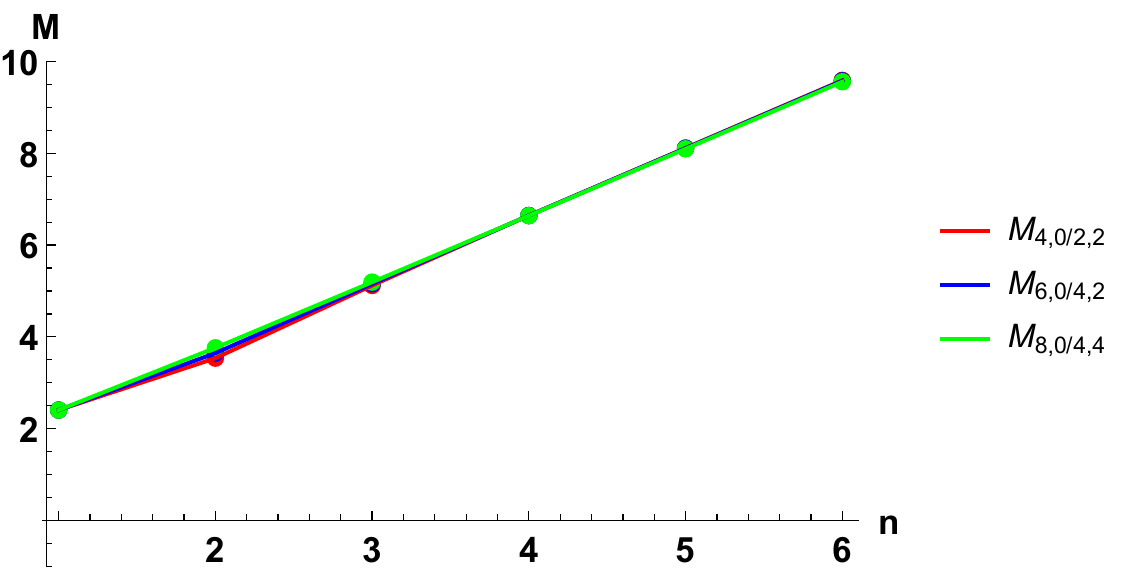}
\caption{Mass multipole ratios with $M_0$}
  \label{fig:bubblemassmultipoleswithM0}
\end{subfigure}
 \caption{Mass multipole ratios computed from symmetric pincers $(n,n)$ with $n=1,\cdots, 6$ (section \ref{sec:pincers}).}
 \label{fig:bubblemassmultipoles}
\end{figure}

\paragraph{Ratios of mass multipoles that do not involve $M_0$}
These ratios are of the form  $\mathcal{M}_{\{a_1,a_2,\cdots\}/\{b_1,b_2,\cdots\}} $ with $a_i,b_i > 0$ for all even $a_i,b_i$.  When  $a_i,b_i$ are of the same order ($a_i/b_j\sim \mathcal{O}(1)$ for all $a_i,b_j$), we note that the \direct method gives values that are always very close to one. This trend is also fairly stable for a small or large number of centers; see Figure \ref{fig:bubblemassmultipolesnoM0}.

There is a qualitatively different behavior of multipole ratios when there is a large difference in the multipole order, obtained for example when  $a_{i_0}/b_j\ll 1$ for a given $a_{i_0}$ and all $b_j$. These ratios tend to be much smaller than one.

\begin{figure}[p]\centering
 \begin{subfigure}{0.48\textwidth}\centering
 \includegraphics[width=\textwidth]{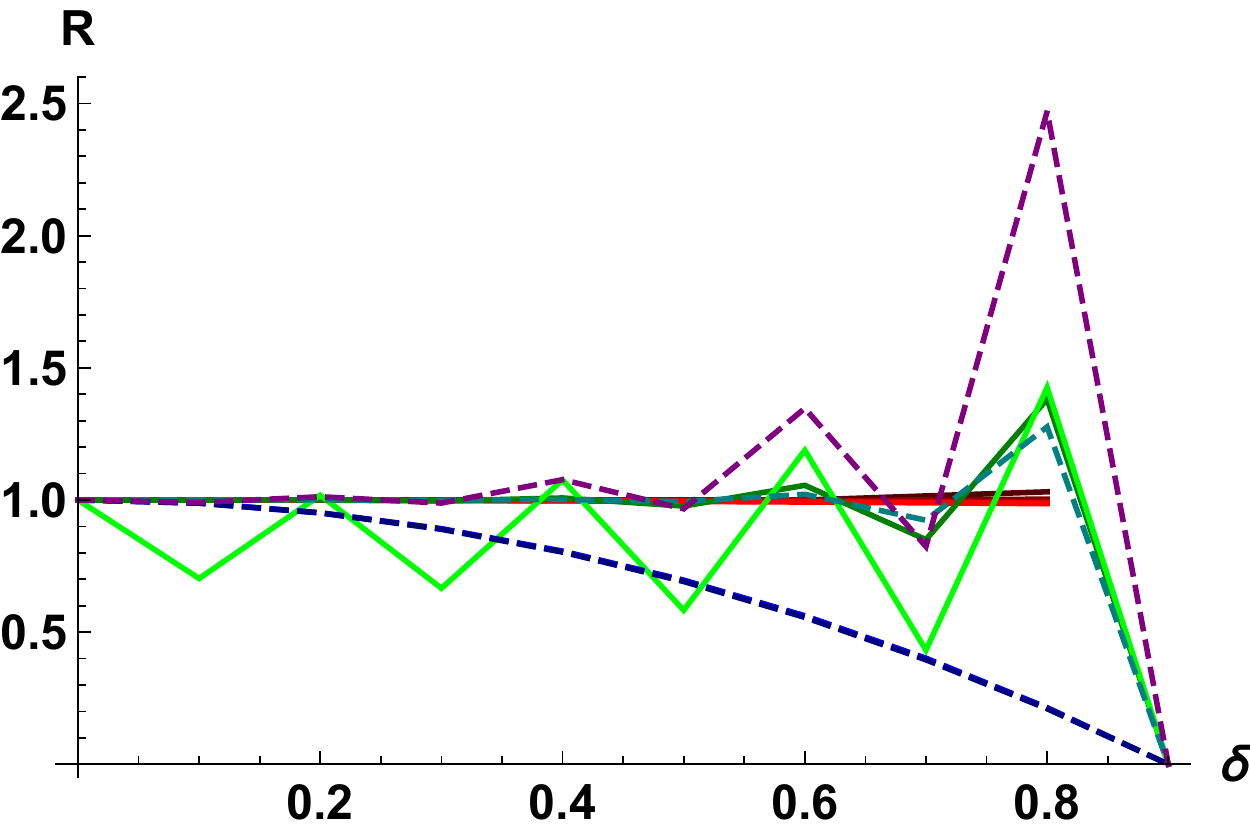}
 \caption{$R^{(M)}_2(L=10,\delta)$ vs. $\delta$}
\end{subfigure}\hspace*{0.01\textwidth}
\begin{subfigure}{0.48\textwidth}\centering
 \includegraphics[width=\textwidth]{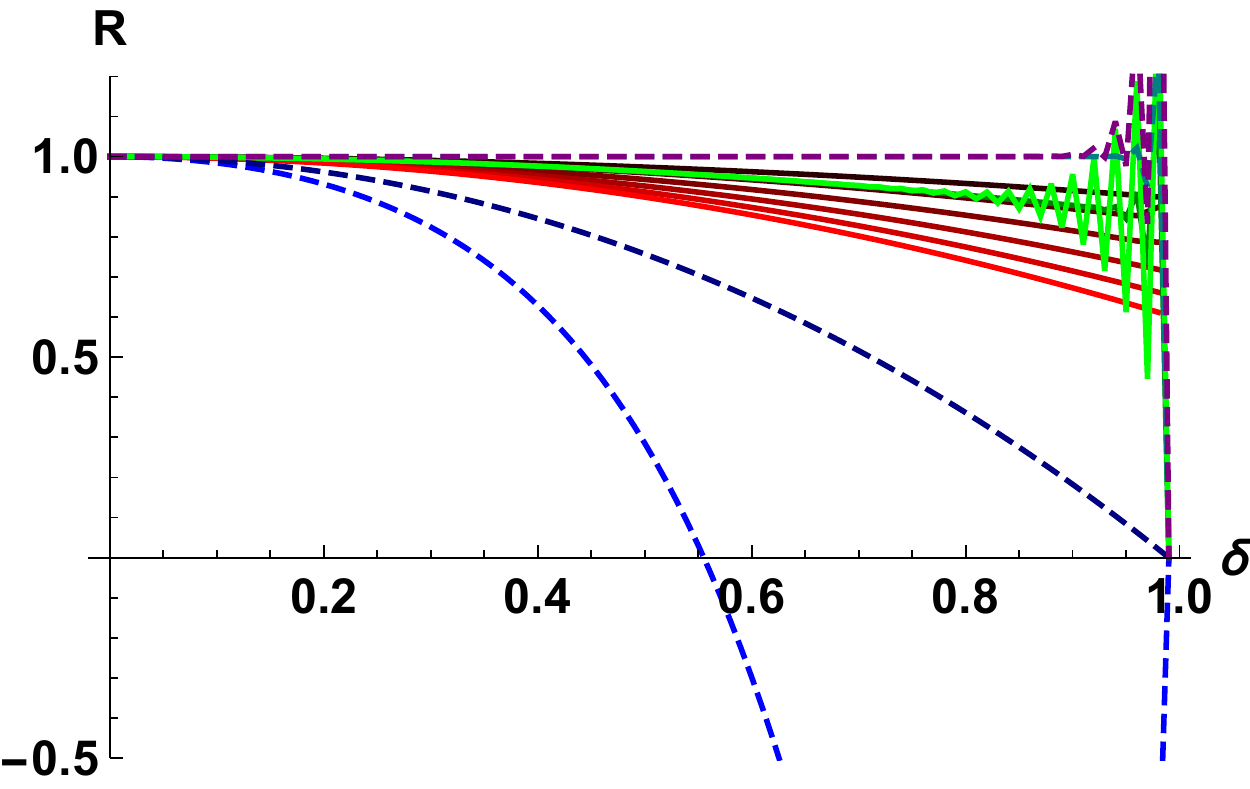}
 \caption{$R^{(M)}_2(L=100,\delta)$ vs. $\delta$}
\end{subfigure}\\
\begin{subfigure}{0.48\textwidth}\centering
 \includegraphics[width=\textwidth]{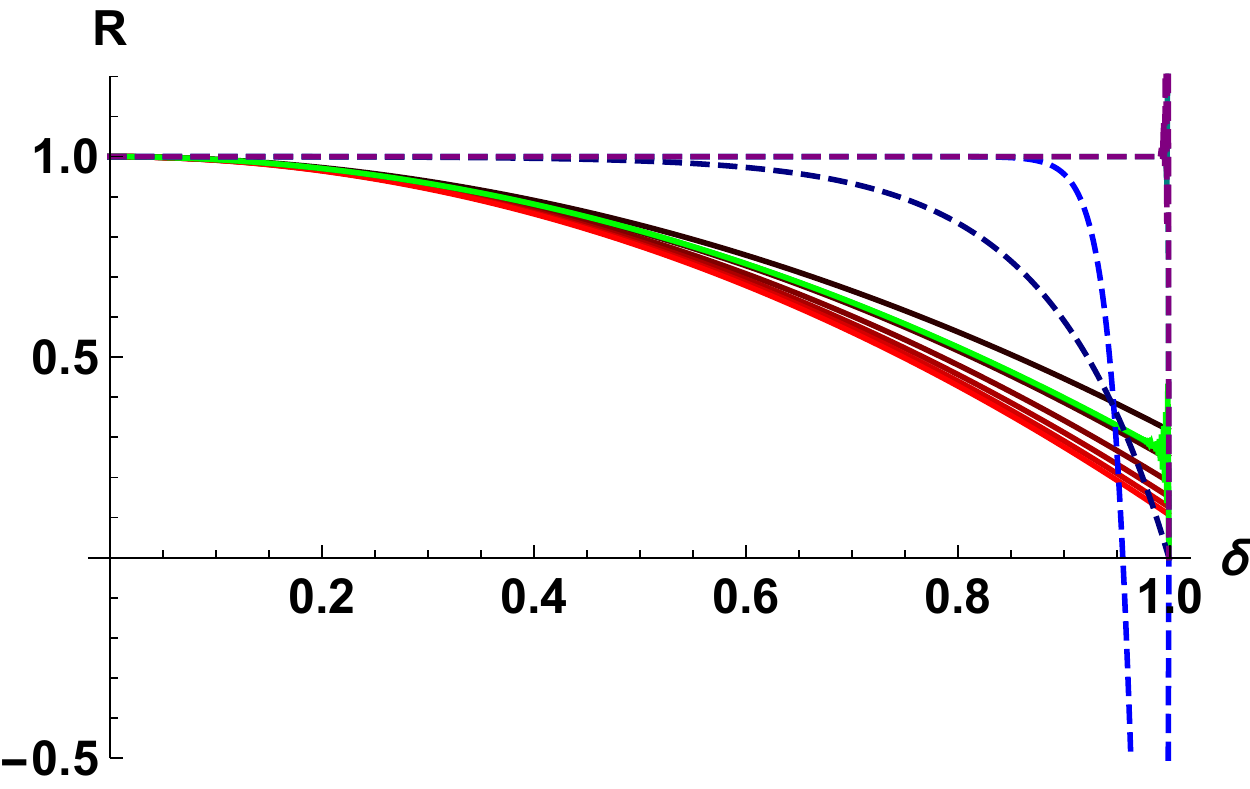}
 \caption{$R^{(M)}_2(L=1000,\delta)$ vs. $\delta$}
\end{subfigure}\hspace*{0.01\textwidth}
\begin{subfigure}{0.48\textwidth}\centering
 \includegraphics[width=\textwidth]{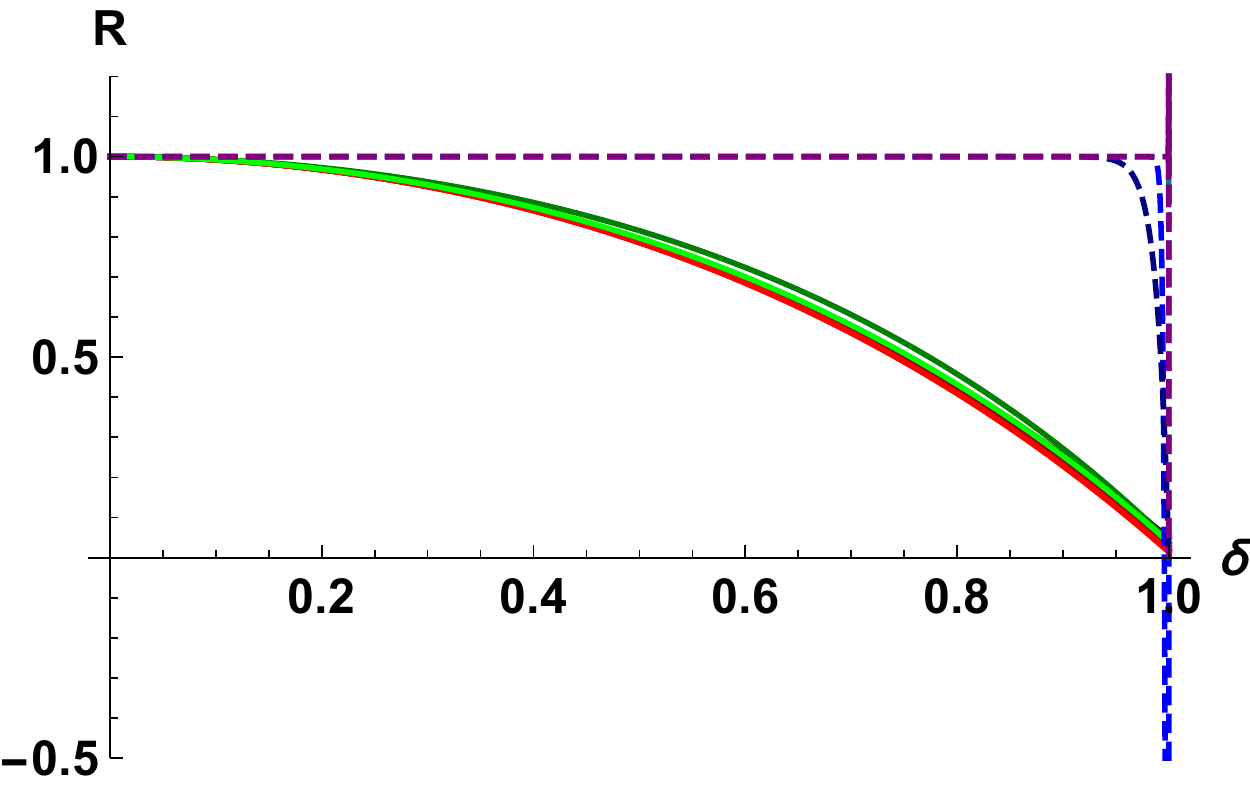}
 \caption{$R^{(M)}_2(L=10000,\delta)$ vs. $\delta$}
\end{subfigure}
 \caption{Mass multipole ratios without $M_0$ in the numerator: $R^{(M)}_2(L,\delta)$.
  For the legend, see fig. \ref{fig:legend}.}
 \label{fig:bubblemassmultipoles-R2}
\end{figure}

\begin{figure}[p]\centering
 \begin{subfigure}{0.45\textwidth}\centering
 \includegraphics[width=\textwidth]{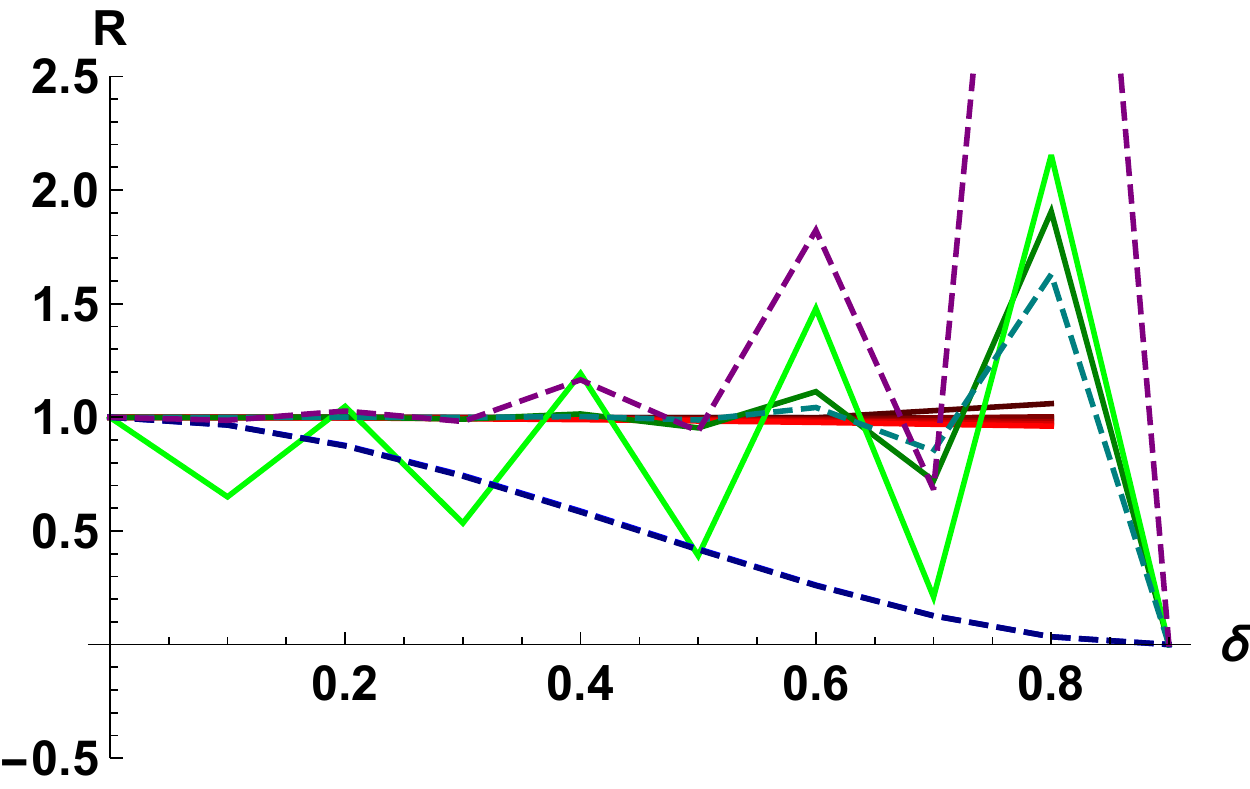}
 \caption{$R^{(M)}_3(L=10,\delta)$ vs. $\delta$}
\end{subfigure}\hspace*{0.01\textwidth}
\begin{subfigure}{0.45\textwidth}\centering
 \includegraphics[width=\textwidth]{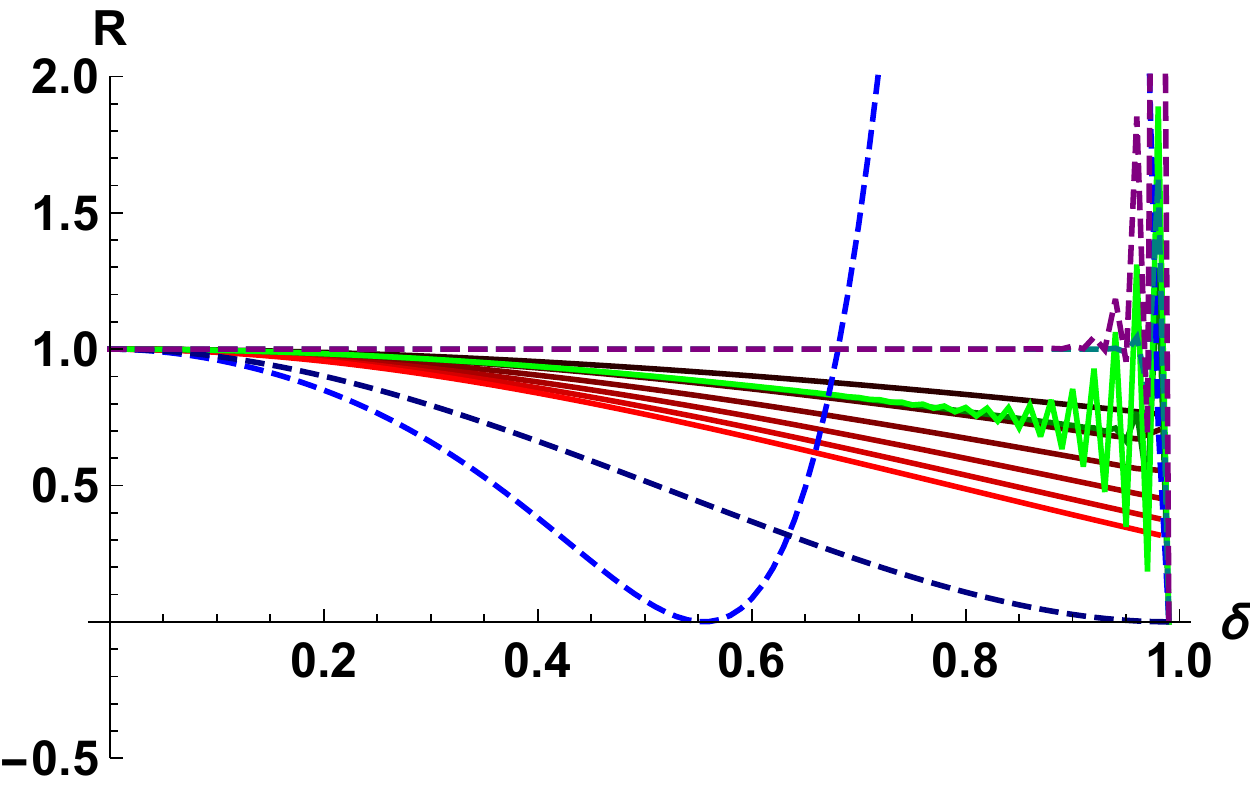}
 \caption{$R^{(M)}_3(L=100,\delta)$ vs. $\delta$}
\end{subfigure}\\
\begin{subfigure}{0.48\textwidth}\centering
 \includegraphics[width=\textwidth]{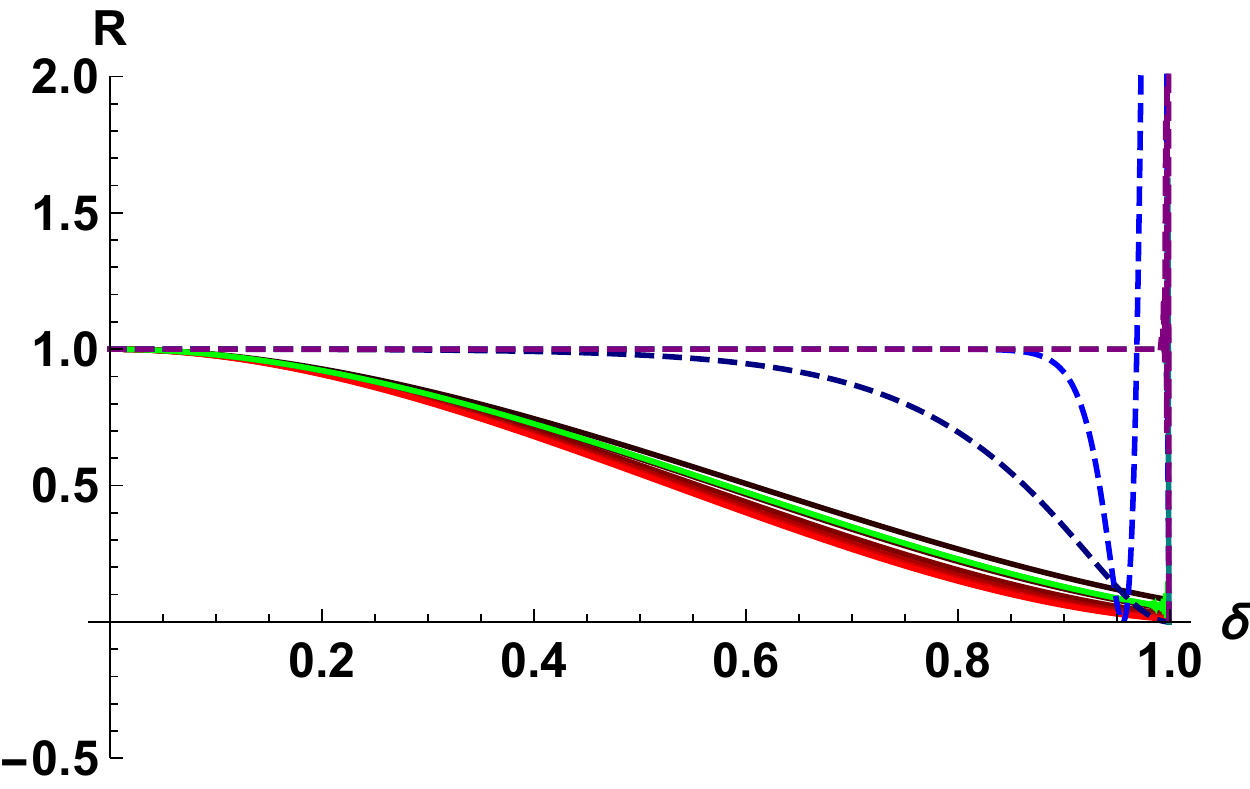}
 \caption{$R^{(M)}_3(L=1000,\delta)$ vs. $\delta$}
\end{subfigure}\hspace*{0.01\textwidth}
\begin{subfigure}{0.48\textwidth}\centering
 \includegraphics[width=\textwidth]{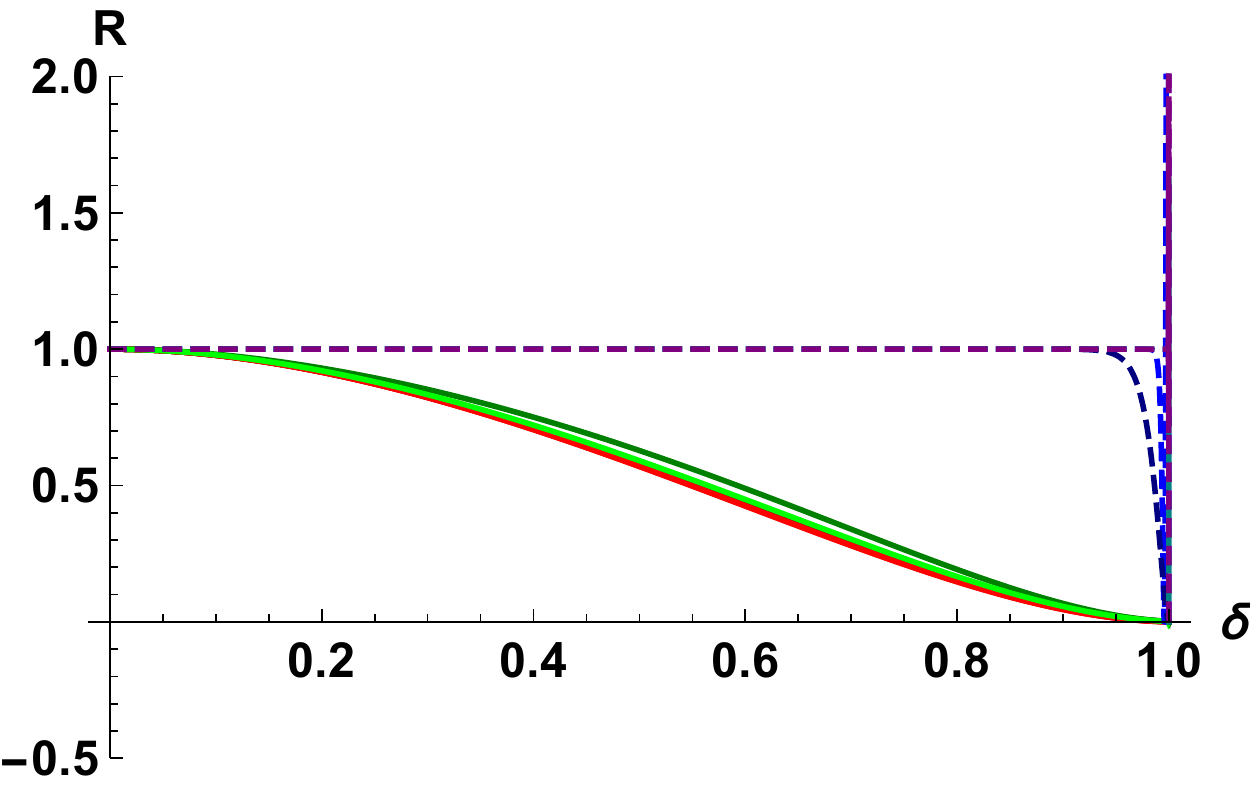}
 \caption{$R^{(M)}_3(L=10000,\delta)$ vs. $\delta$}
\end{subfigure}
 \caption{Mass multipole ratios without $M_0$ in the numerator: $R^{(M)}_3(L,\delta)$. For the legend, see fig. \ref{fig:legend}.}
 \label{fig:bubblemassmultipoles-R3}
\end{figure}

To illustrate this, we can consider $R_2^{(M)}(L,\delta)$ as defined in (\ref{eq:R2def}), as well as $R_3^{(M)}(L,\delta)$ defined by:
\begin{align}
 \label{eq:R3def} R_3^{(M)}(L,\delta)& \equiv \mathcal{M}_{ \{L(1+2\delta),L(1-\delta),L(1-\delta)\}/\{L,L,L\}} = \frac{M_{L(1+2\delta)} M_{L(1-\delta)}M_{L(1-\delta)}}{(M_L)^3}.
 \end{align}
 We plot $R_2^{(M)}(L,\delta)$ and $R_3^{(M)}(L,\delta)$ for various fixed values of $L$ while varying $\delta$ in figs. \ref{fig:bubblemassmultipoles-R2} and \ref{fig:bubblemassmultipoles-R3}.  We see that as the imbalance of the multipoles increases ($\delta\rightarrow 1$), these ratios tend to zero for the pincer $(n_1,n_2)$ geometries while they remain fairly close to 1 for the four-center geometries $A,B,C,D$ (although there appear to be some extra edge effects at $\delta\sim 1$ for these four-center geometries).

\paragraph{Ratios of  mass multipoles where $M_0$  appears in the numerator} 
As we can see from Figure \ref{fig:bubblemassmultipoleswithM0}, when the non-zero multipole orders in the ratio satisfy $a_i/b_j\sim \mathcal{O}(1)$, these multipole ratios are generally (much) bigger than one. Furthermore, it appears that, as the number of centers $N$ gets bigger, these ratios get bigger. 
 
\begin{figure}[p]\centering
\begin{subfigure}{0.48\textwidth}\centering
 \includegraphics[width=\textwidth]{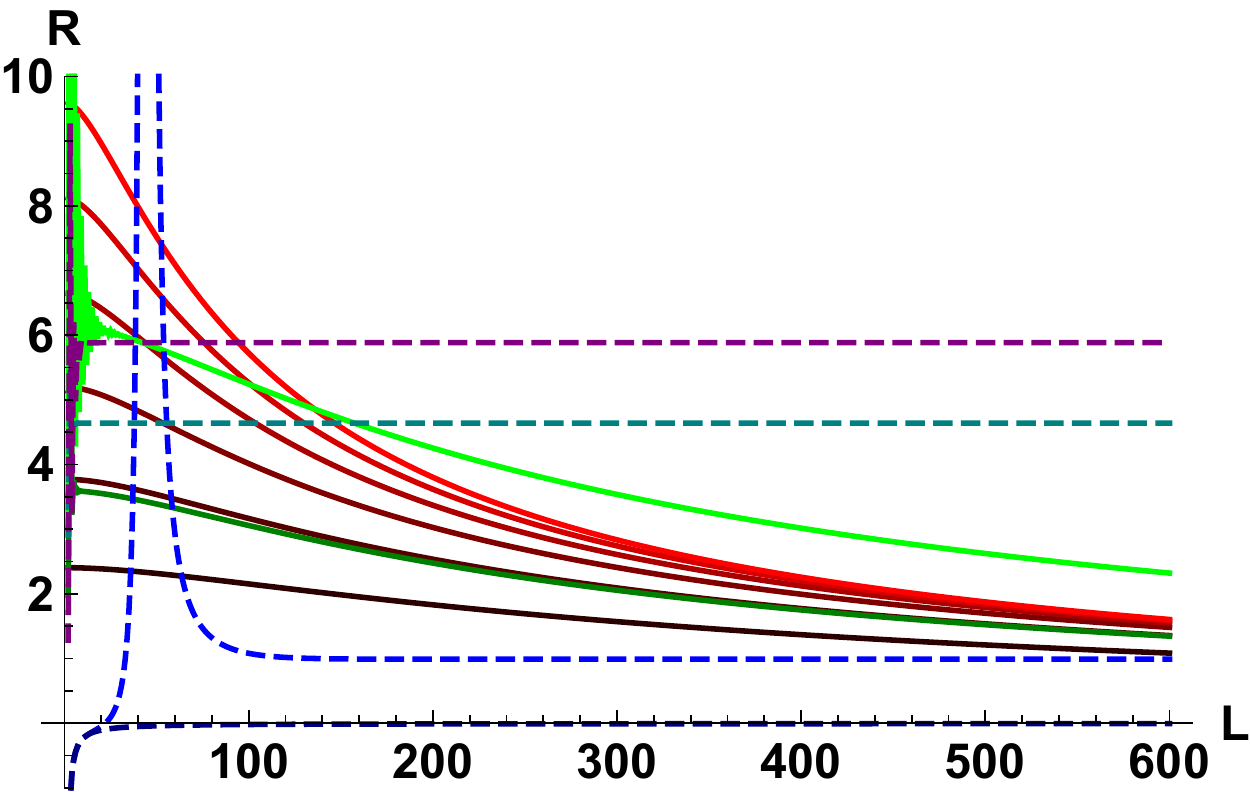}
 \caption{$R^{(M)}_{2,0}(L)$ vs. $L$}
\end{subfigure}\hspace*{0.01\textwidth}
\begin{subfigure}{0.48\textwidth}\centering
 \includegraphics[width=\textwidth]{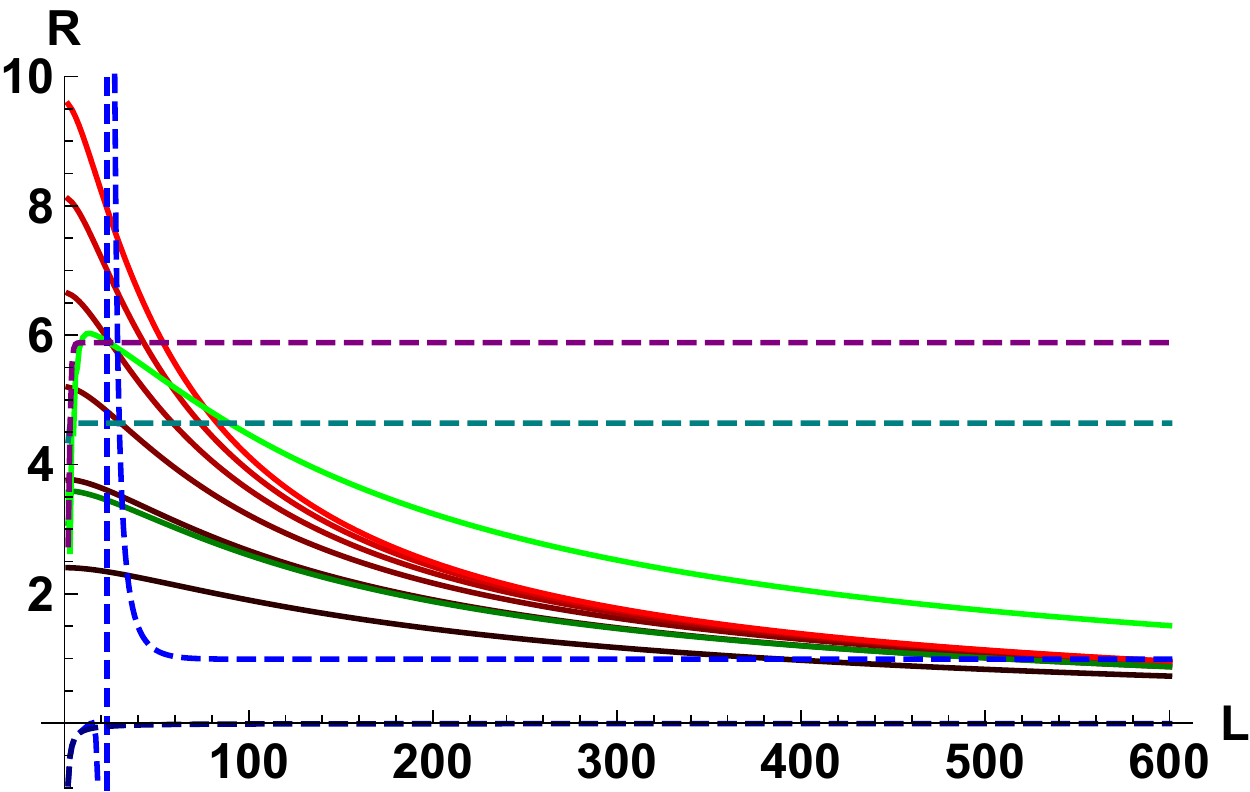}
 \caption{$R^{(M)}_{3,0}(L)$ vs. $L$}
\end{subfigure}
 \caption{Mass multipole ratios with $M_0$ in the numerator: $R^{(M)}_{i,0}(L)$. For the legend, see fig. \ref{fig:legend}.}
 \label{fig:bubblemassmultipoles-Ri0}
\end{figure}

We can also calculate ratios containing $M_0$ where instead $a_i/b_j\nsim \mathcal{O}(1)$. Examples include:
 \begin{align}
 \label{eq:R20def}  R_{2,0}^{(M)}(L) &\equiv \mathcal{M}_{ \{2L,0\}/\{L,L\} } = \frac{ M_{2L} M_0}{(M_L)^2},\\
\label{eq:R30def}  R_{3,0}^{(M)}(L) &\equiv \mathcal{M}_{ \{3L,3L,0\}/\{2L,2L,2L\} } = \frac{ M_{3L}M_{3L} M_0}{(M_{2L})^3}.
 \end{align}
As we can see in fig. \ref{fig:bubblemassmultipoles-Ri0}, such ratios tend to approach  $\mathcal{O}(1)$ numbers for large $L$, for all of the geometries we consider.


\paragraph{Note on the symmetric pincers}

For the symmetric pincers ($n_1=n_2$) of Section \ref{sec:pincers}, the ratios above can only be computed for even mass multipoles since $M_{2n+1}=0$, meaning the number of points evaluated for their graphs in the above figures is typically half of the number of points for the non-${\mathbb Z}_2$-symmetric geometries. Remarkably, the resulting graphs for the symmetric pincers have very similar behaviors to the other geometries, especially the asymmetric pincers. This leads us to believe that the generic behavior of multipole ratios in multi-center geometries is captured by the graphs we give above; the ``accidental'' symmetries leading to certain multipoles vanishing identically are the exception rather than the rule, and do not affect the overall behavior of the multipole ratios.

\subsection{Ratios of current multipoles}\label{sec:subseccurrent}

   \begin{figure}[hp]\centering
\begin{subfigure}{0.48\textwidth}\centering
 \includegraphics[width=\textwidth]{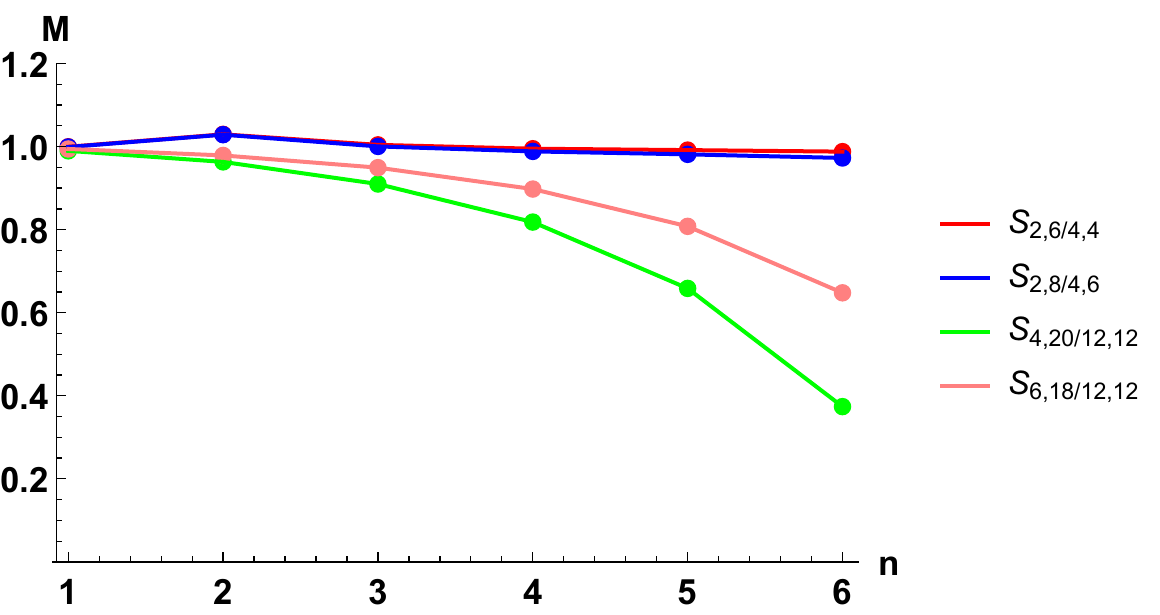}
 \caption{Current multipole ratios without $M_0$}
 \label{fig:bubblecurmultipolesnoM0}
\end{subfigure}\hspace*{0.01\textwidth}
\begin{subfigure}{0.48\textwidth}\centering
  \includegraphics[width=\textwidth]{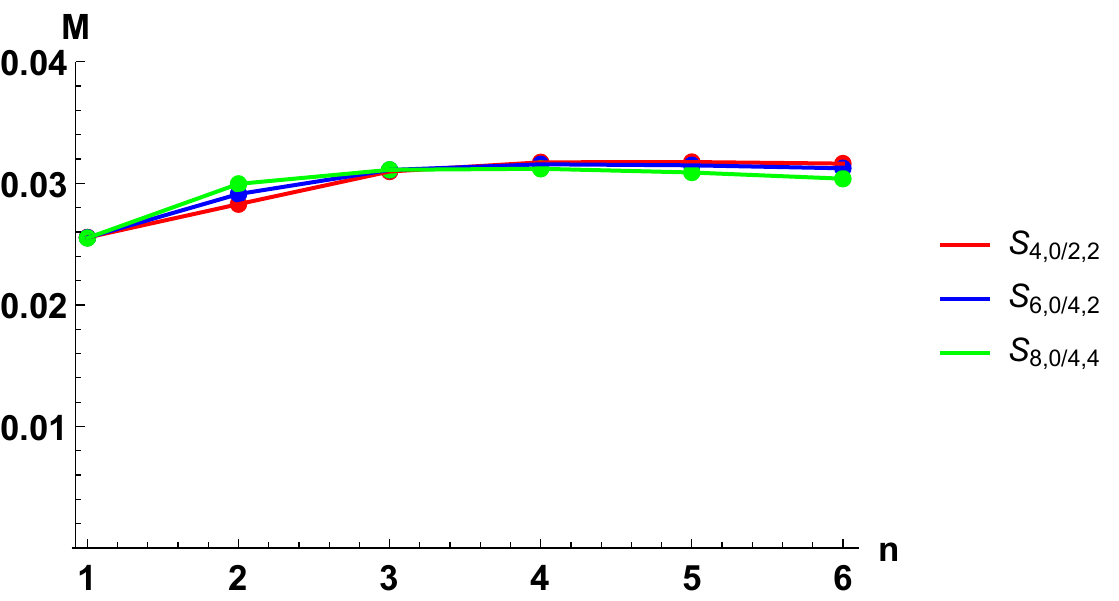}
\caption{Current multipole ratios with $M_0$}
  \label{fig:bubblecurmultipoleswithM0}
\end{subfigure}
 \caption{Current multipole ratios computed from symmetric pincers $(n,n)$ with $n=1,\cdots,6$ (section \ref{sec:pincers}).}
 \label{fig:bubblecurmultipoles}
\end{figure}

Besides mass-multipole ratios, we can also consider current multipoles ratios $\mathcal{S}_{\{a_1,a_2,\cdots\}/\{b_1,b_2,\cdots\}}$, given by (\ref{eq:massmultipoleratiodef}) by replacing $M_l$ by $S_l$, except for $l=0$ where we always leave $M_0$ in the expression (and never $S_0$). We find that these multipole ratios have more or less the same qualitative behavior as the corresponding mass-multipole ratios.

Indeed, current multipole ratios that involve multipoles $a_i,b_i\neq 0$ of the same order ($a_i/b_j\sim \mathcal{O}(1)$) are generically close to 1, see fig. \ref{fig:bubblecurmultipoles}). 

\begin{figure}[p]\centering
 \begin{subfigure}{0.48\textwidth}\centering
 \includegraphics[width=\textwidth]{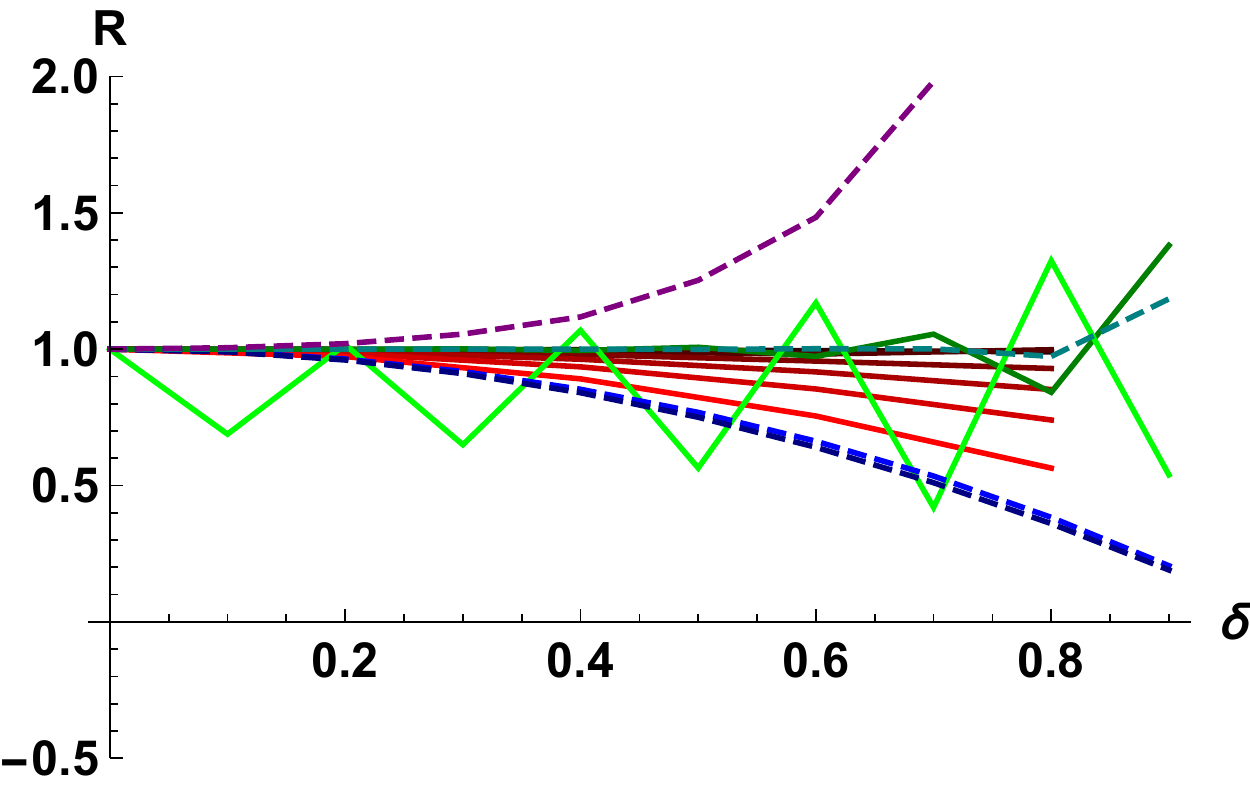}
 \caption{$R^{(S)}_2(L=10,\delta)$ vs. $\delta$}
\end{subfigure}\hspace*{0.01\textwidth}
\begin{subfigure}{0.48\textwidth}\centering
 \includegraphics[width=\textwidth]{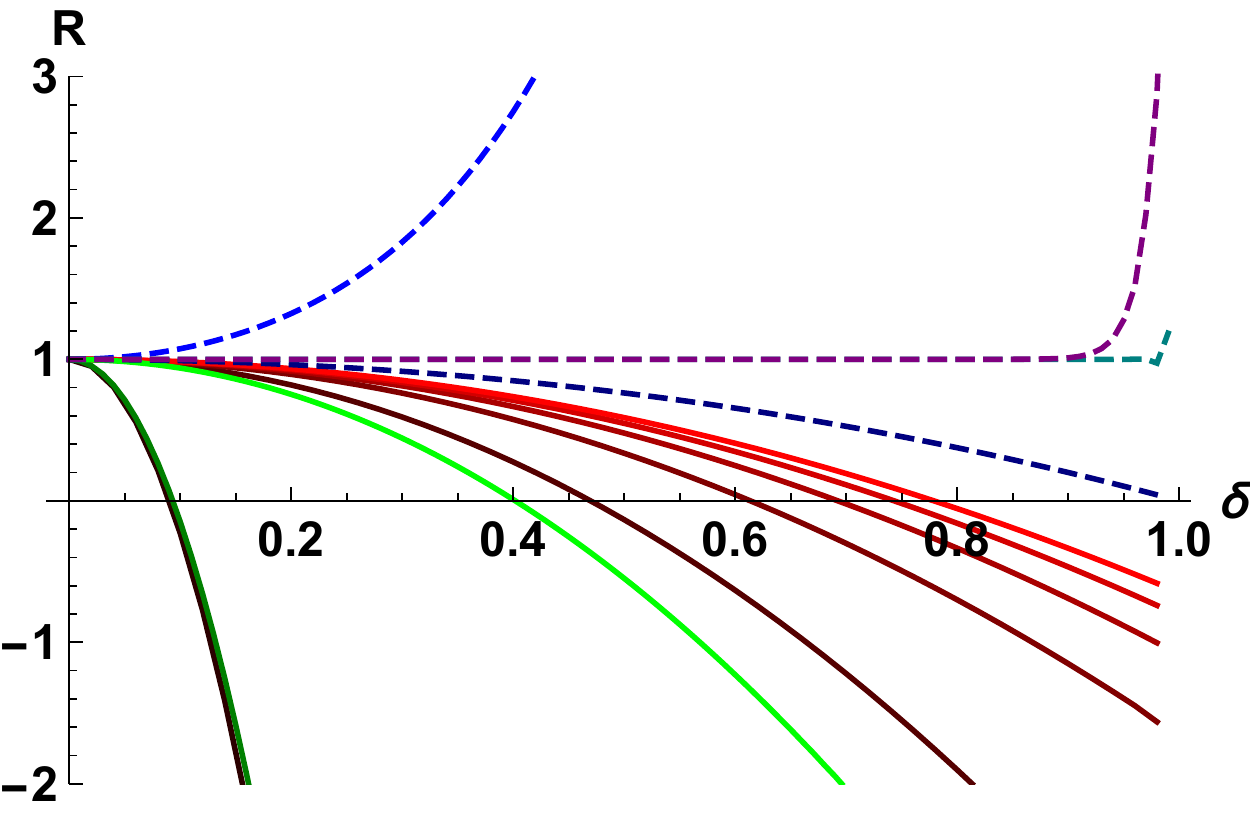}
 \caption{$R^{(S)}_2(L=100,\delta)$ vs. $\delta$}
\end{subfigure}\\
\begin{subfigure}{0.48\textwidth}\centering
 \includegraphics[width=\textwidth]{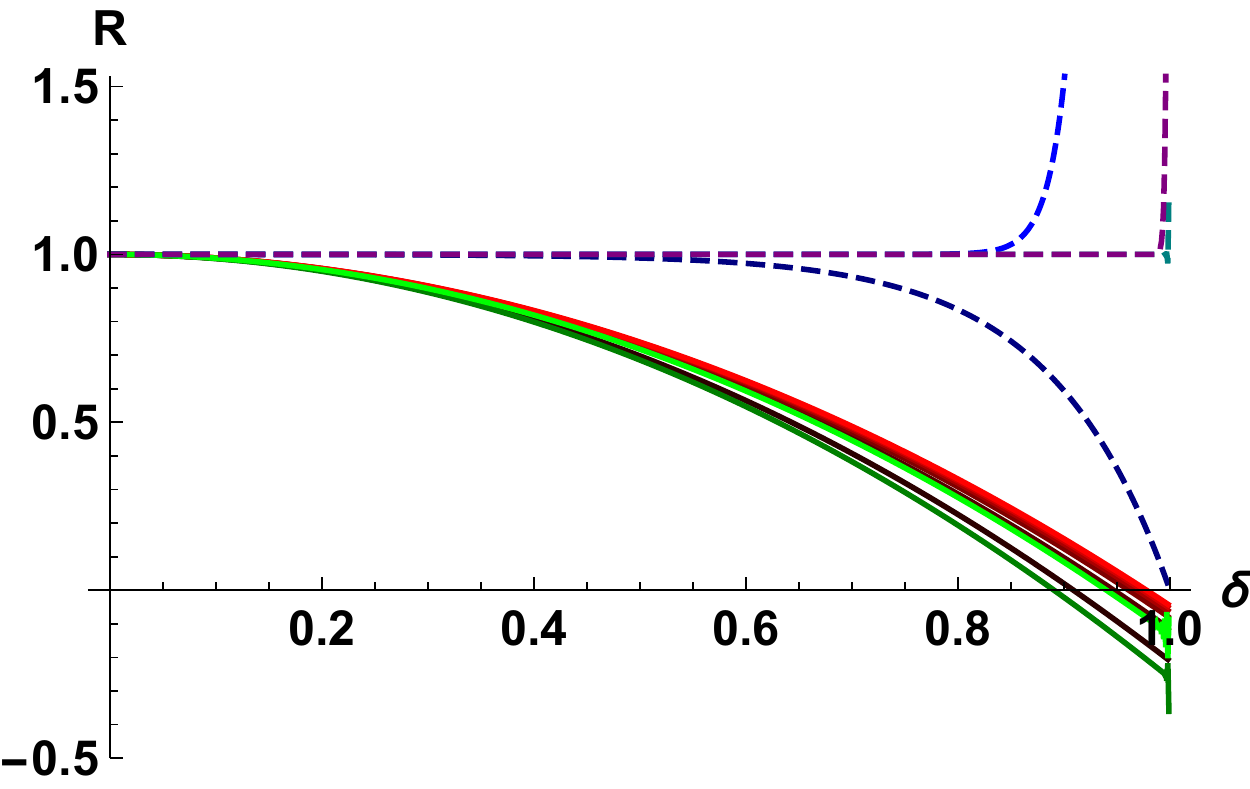}
 \caption{$R^{(S)}_2(L=1000,\delta)$ vs. $\delta$}
\end{subfigure}\hspace*{0.01\textwidth}
\begin{subfigure}{0.48\textwidth}\centering
 \includegraphics[width=\textwidth]{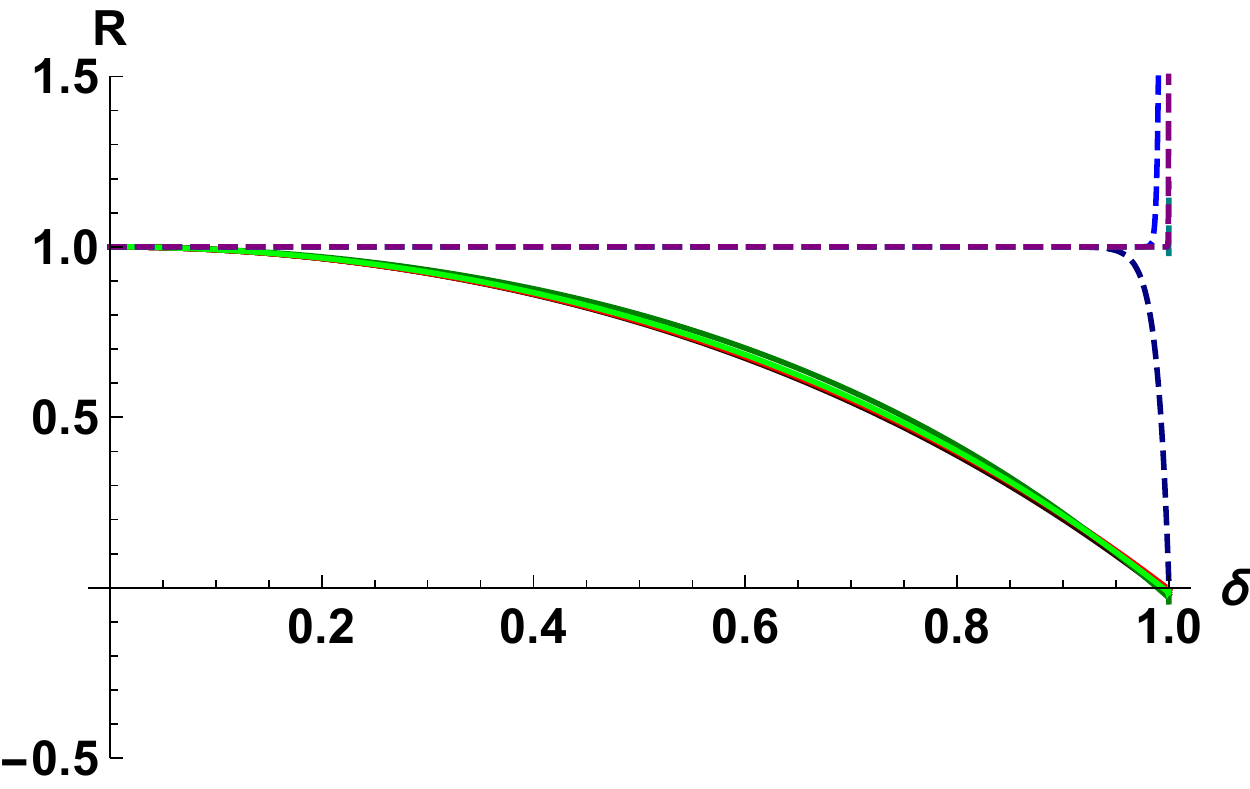}
 \caption{$R^{(S)}_2(L=10000,\delta)$ vs. $\delta$}
\end{subfigure}
 \caption{Current multipole ratios $R^{(S)}_2(L,\delta)$. For the legend, see fig. \ref{fig:legend}.}
 \label{fig:bubblecurmultipoles-R2}
\end{figure}

\begin{figure}[p]\centering
 \begin{subfigure}{0.48\textwidth}\centering
 \includegraphics[width=\textwidth]{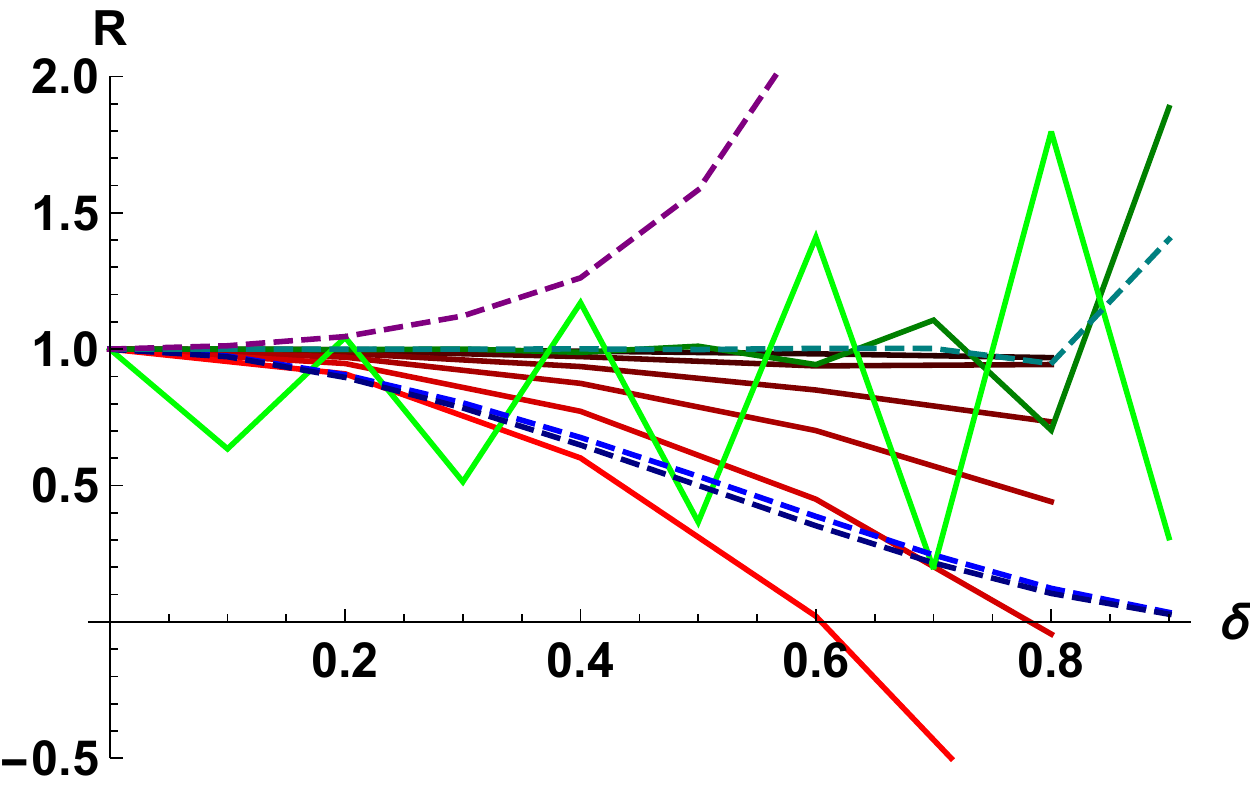}
 \caption{$R^{(S)}_3(L=10,\delta)$ vs. $\delta$}
\end{subfigure}\hspace*{0.01\textwidth}
\begin{subfigure}{0.48\textwidth}\centering
 \includegraphics[width=\textwidth]{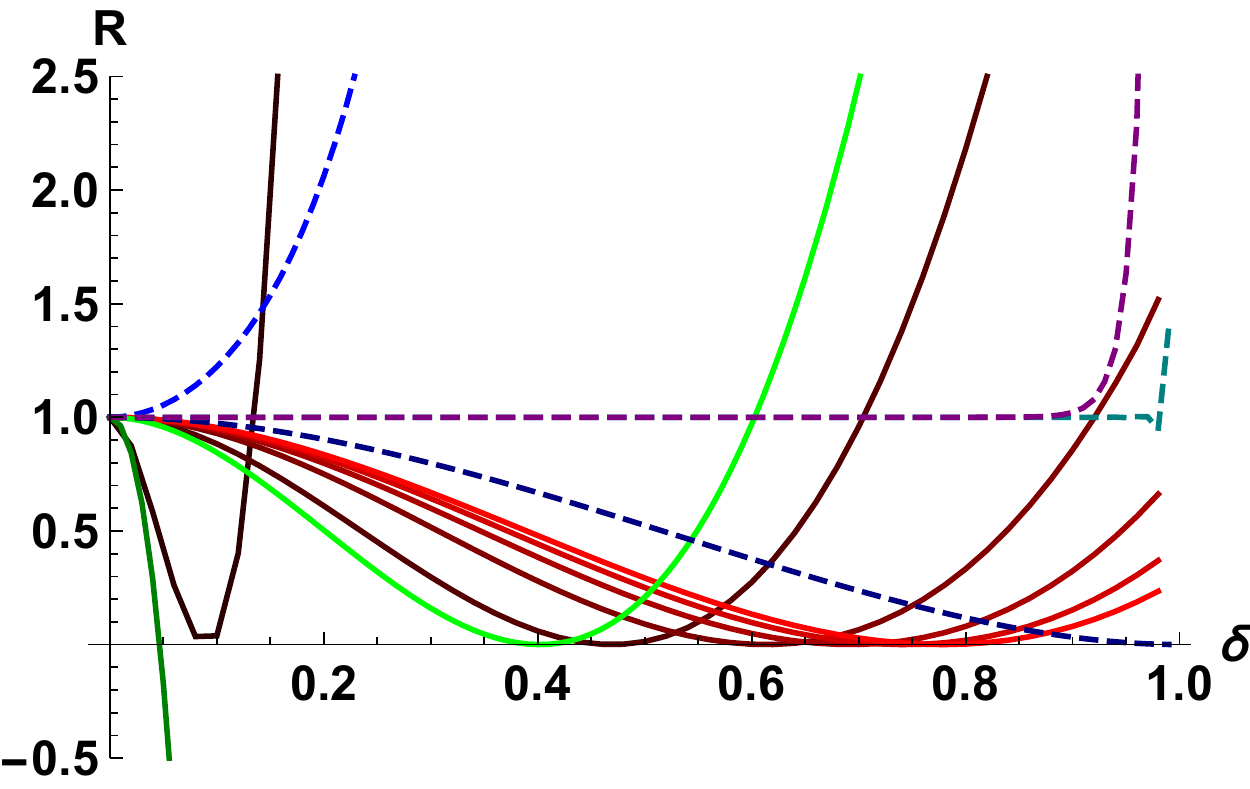}
 \caption{$R^{(S)}_3(L=100,\delta)$ vs. $\delta$}
\end{subfigure}\\
\begin{subfigure}{0.48\textwidth}\centering
 \includegraphics[width=\textwidth]{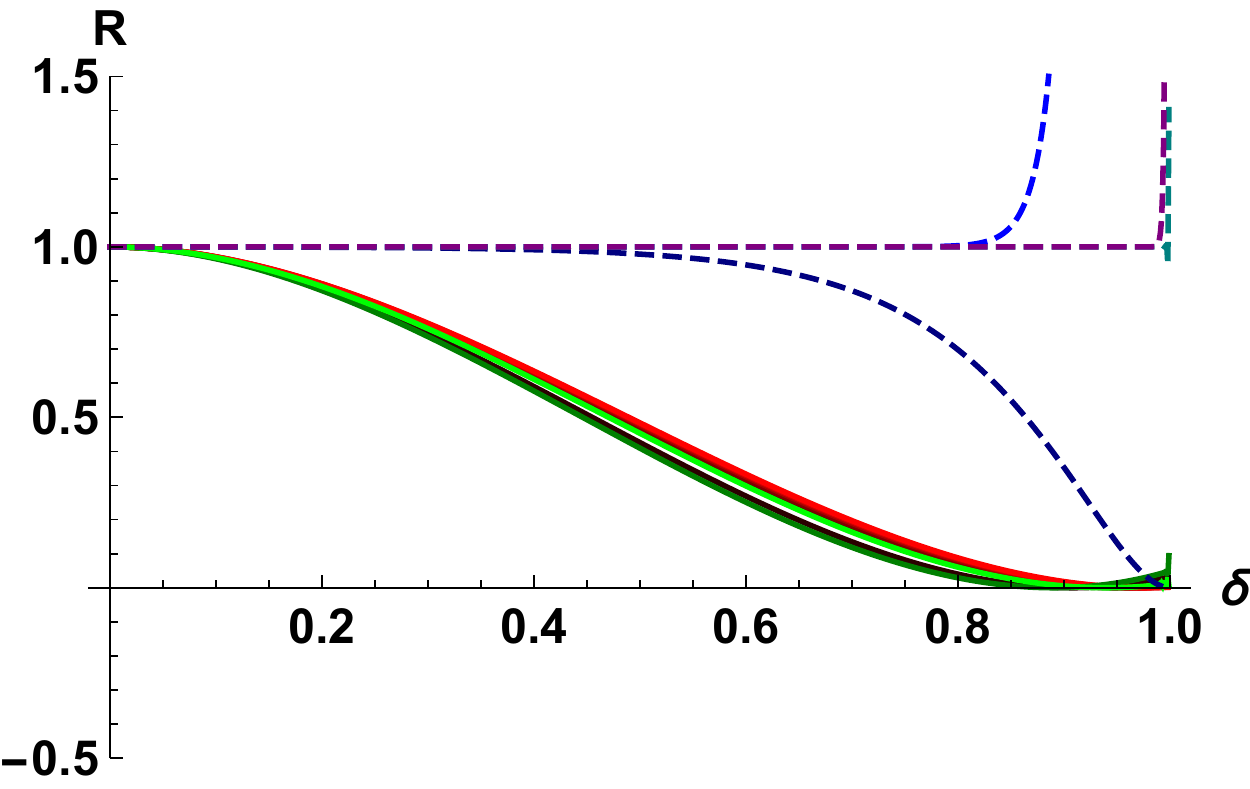}
 \caption{$R^{(S)}_3(L=1000,\delta)$ vs. $\delta$}
\end{subfigure}\hspace*{0.01\textwidth}
\begin{subfigure}{0.48\textwidth}\centering
 \includegraphics[width=\textwidth]{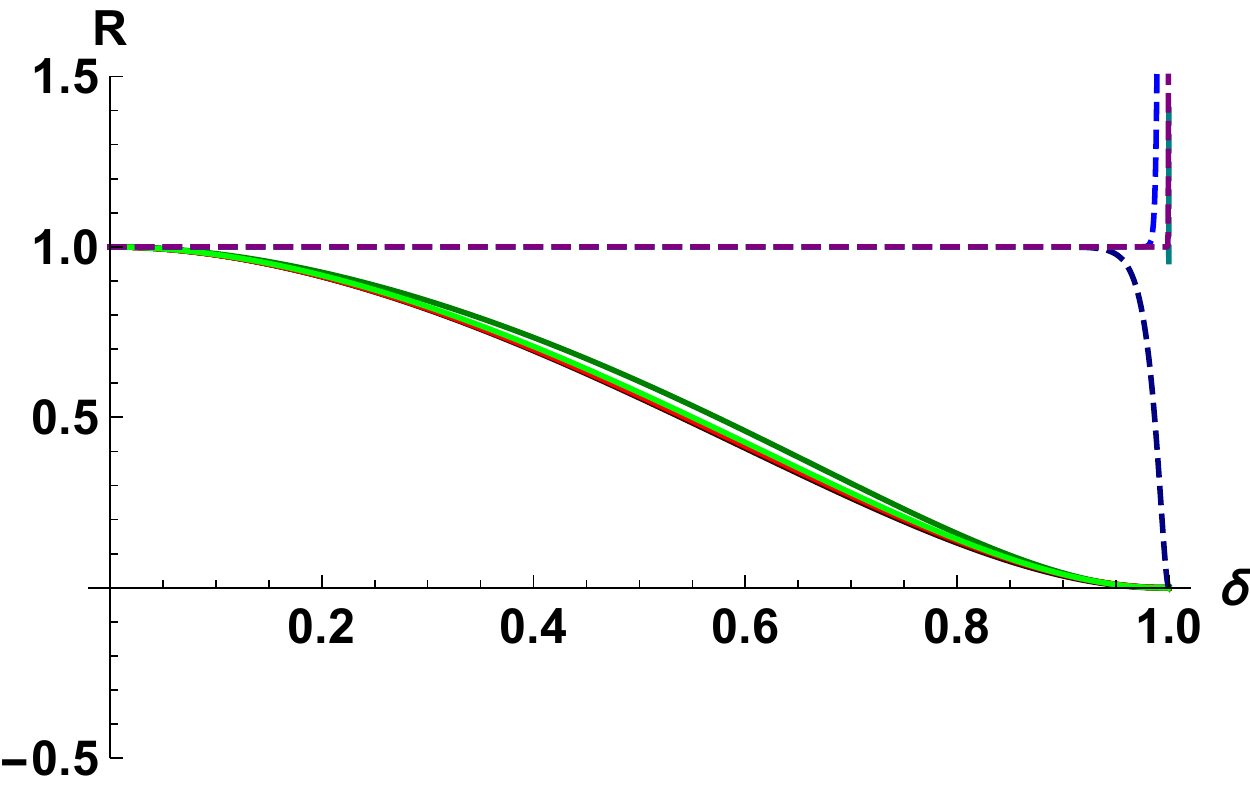}
 \caption{$R^{(S)}_3(L=10000,\delta)$ vs. $\delta$}
\end{subfigure}
 \caption{Current multipole ratios $R^{(S)}_3(L,\delta)$. For the legend, see fig. \ref{fig:legend}.}
 \label{fig:bubblecurmultipoles-R3}
\end{figure}

\begin{figure}[p]\centering
\begin{subfigure}{0.48\textwidth}\centering
 \includegraphics[width=\textwidth]{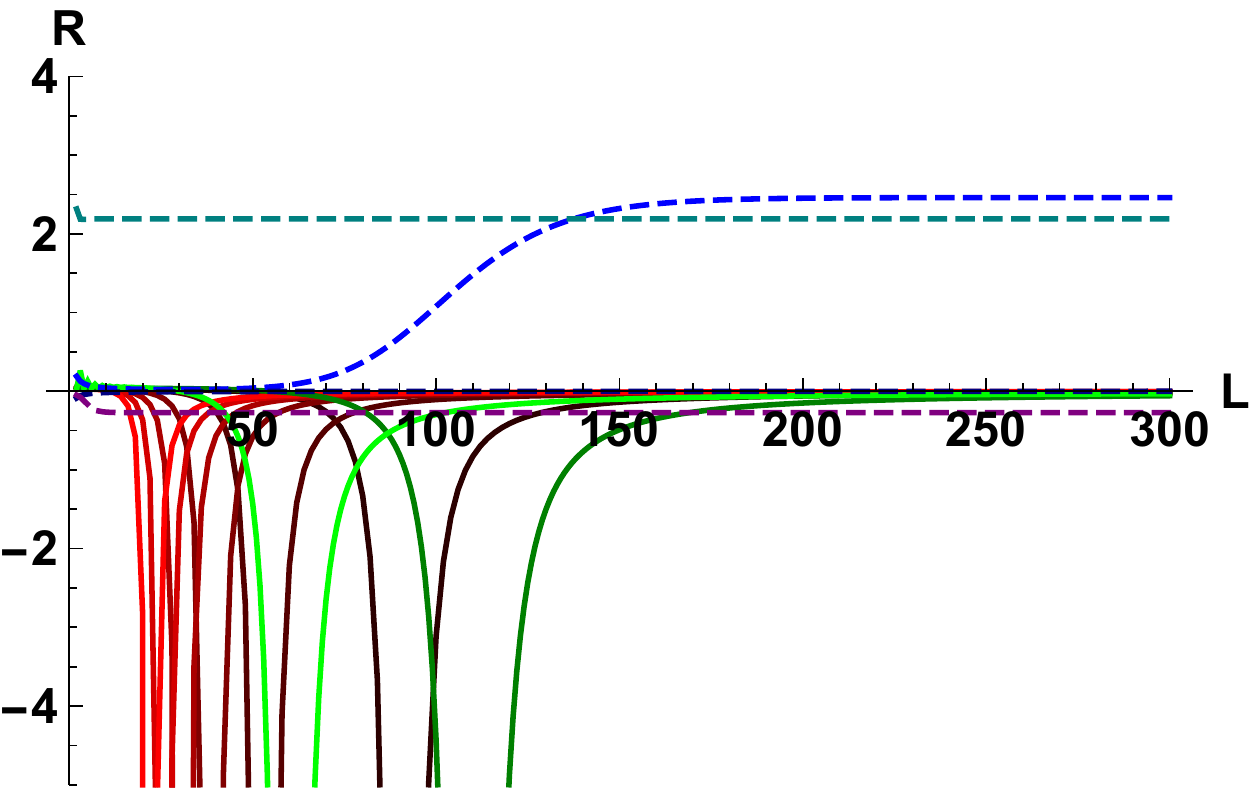}
 \caption{$R^{(S)}_{2,0}(L)$ vs. $L$}
\end{subfigure}\hspace*{0.01\textwidth}
\begin{subfigure}{0.48\textwidth}\centering
 \includegraphics[width=\textwidth]{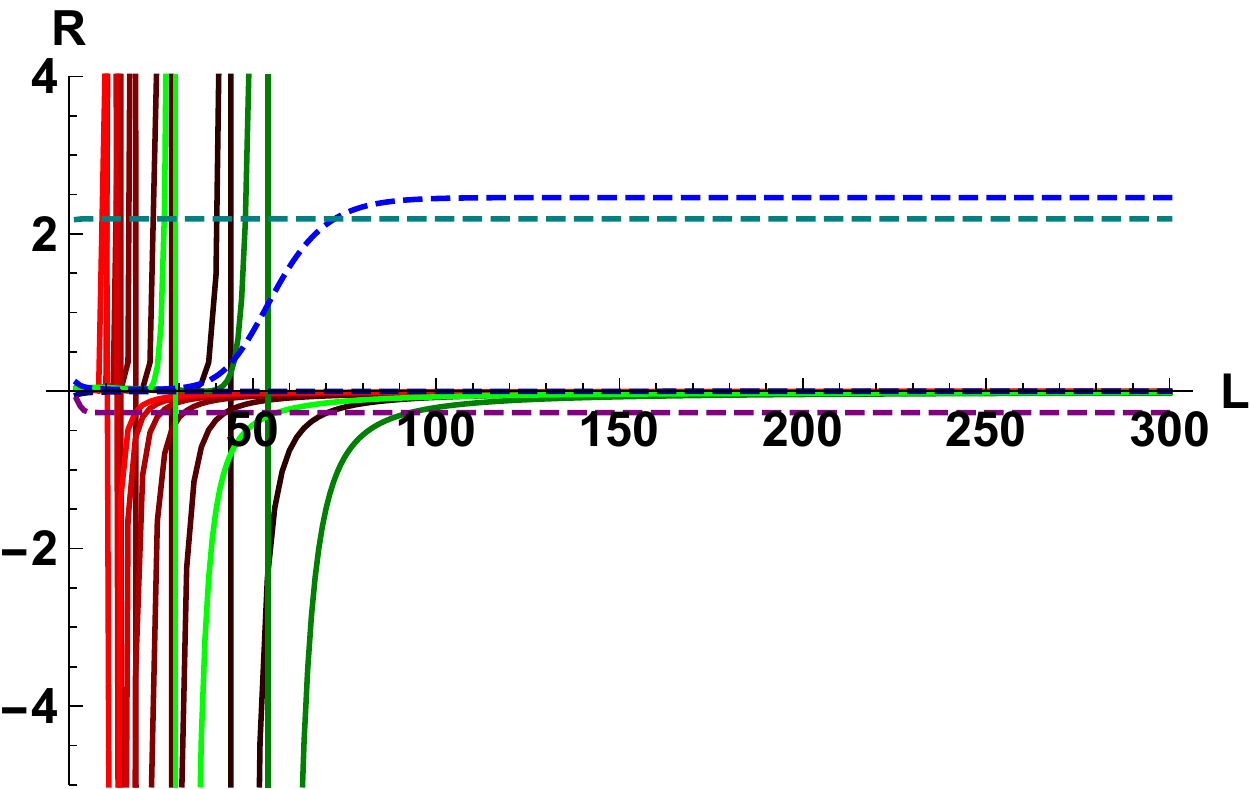}
 \caption{$R^{(S)}_{3,0}(L)$ vs. $L$}
\end{subfigure}
\begin{subfigure}{0.48\textwidth}\centering
 \includegraphics[width=\textwidth]{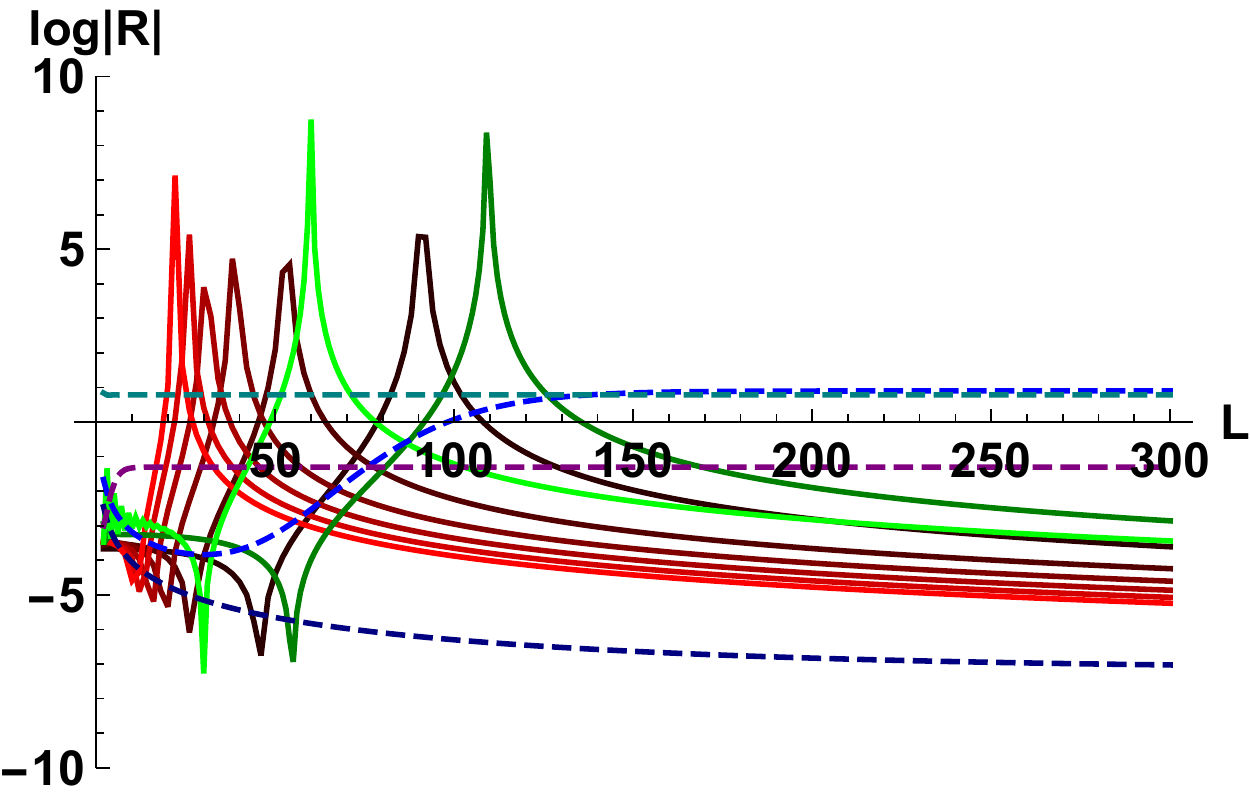}
 \caption{$\log|R^{(S)}_{2,0}(L)|$ vs. $L$}
\end{subfigure}\hspace*{0.01\textwidth}
\begin{subfigure}{0.48\textwidth}\centering
 \includegraphics[width=\textwidth]{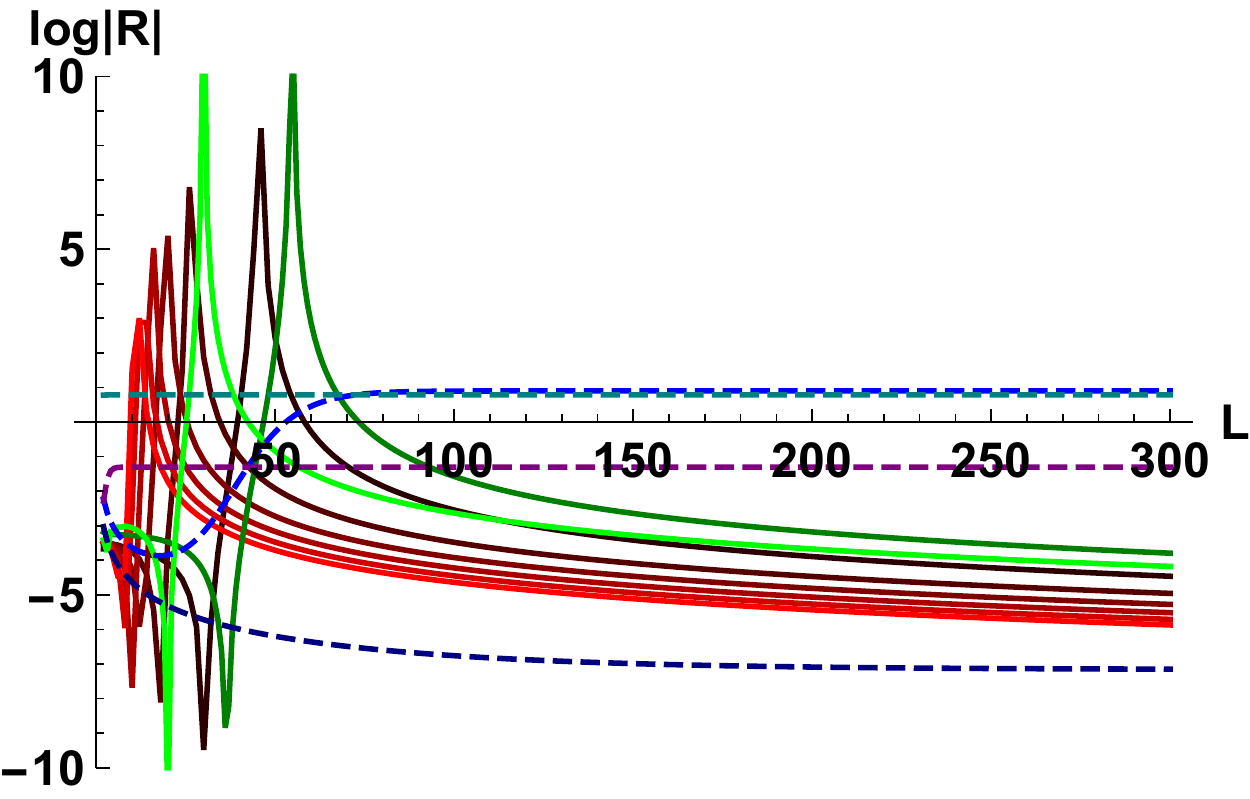}
 \caption{$\log|R^{(S)}_{3,0}(L)|$ vs. $L$}
\end{subfigure}
 \caption{Current multipole ratios $R^{(S)}_{i,0}(L)$.For the legend, see fig. \ref{fig:legend}.\newline The $\log$ plot below shows the ``peak'' structure. The double peak structure is the result of one multipole moment that is exceptionally low. }
 \label{fig:bubblecurmultipoles-Ri0}
\end{figure}

The ratios $R_i^{(S)}(L,\delta)$ and $R_{i,0}^{(S)}(L)$, obtained by replacing every $M_l$ by $S_l$ in (\ref{eq:R2def}), (\ref{eq:R3def}), (\ref{eq:R20def}), and   (\ref{eq:R30def}), also have the same qualitative behavior as the mass-multipole ratios $R_i^{(M)}(L,\delta)$ and $R_{i,0}^{(M)}(L)$, although there seem to be more ``excursions'' away from the general trend for relatively small $L$. See figs. \ref{fig:bubblecurmultipoles-R2}, \ref{fig:bubblecurmultipoles-R3}, and \ref{fig:bubblecurmultipoles-Ri0}.


 \clearpage
 \newpage
\section{General Black Hole Charge Parameters}\label{app:chargeparams}
For reference, we give the relations between the parameters $\mu_{1,2},\nu_{1,2},C,\mathcal{D}$ and $\delta_I,\gamma_I$ for the general black hole of Section \ref{app:generalBH}. The following equations are all taken from (4.18), (4.19), (4.20), and (5.5) in \cite{Chow:2014cca} (with the range of $I$ changed here to $I=0,\cdots, 3$).

First, we define the shorthands:
\begin{align}
 s_{\delta I} &\equiv \sinh \delta_I, & c_{\delta I} &\equiv \cosh \delta_I,
 \end{align}
 and similarly for $s_{\gamma I},c_{\gamma I}$. We also define shorthands for products of these parameters such as:
 \be s_{\delta IJ} \equiv s_{\delta I} s_{\delta J},\ee
 and similar for products of $c_{\delta I},s_{\gamma I},s_{\gamma J}$; we can also specify more than two indices to multiply in such a product, such as $s_{\delta 0123}$.
 
 Then, the charge parameters $\mu_{1,2},\nu_{1,2}$ are given by:
\begin{align}
\mu_1 & = 1 + \sum_I \bigg( \frac{s_{\delta I}^2 + s_{\gamma I}^2}{2} - s_{\delta I}^2 s_{\gamma I}^2 \bigg) + \frac{1}{2} \sum_{I, J} s_{\delta I}^2 s_{\gamma J}^2 ,\\
\mu_2 & = \sum_I s_{\delta I} c_{\delta I} \bigg( \frac{s_{\gamma I}}{c_{\gamma I}} c_{\gamma 0 1 2 3} - \frac{c_{\gamma I}}{s_{\gamma I}} s_{\gamma 0 1 2 3}\bigg) ,\\
\nu_1 & = \sum_I s_{\gamma I} c_{\gamma I} \bigg( \frac{c_{\delta I}}{s_{\delta I}} s_{\delta 0 1 2 3} - \frac{s_{\delta I}}{c_{\delta I}} c_{\delta 0 1 2 3} \bigg) ,\\
\nu_2 &= \iota - \mathcal{D}
\end{align}
where
\begin{align}
\iota &= c_{\delta 0 1 2 3}c_{\gamma 0 1 2 3}+s_{\delta 0 1 2 3} s_{\gamma 0 1 2 3}+ \sum_{I < J} c_{\delta 0 1 2 3} \frac{s_{\delta I J}}{c_{\delta I J}} \frac{c_{\gamma I J}}{s_{\gamma I J}} s_{\gamma 0 1 2 3} , \\
\mathcal{D} &= c_{\delta 0 1 2 3}s_{\gamma 0 1 2 3}+s_{\delta 0 1 2 3}c_{\gamma 0 1 2 3} + \sum_{I < J} c_{\delta 0 1 2 3} \frac{s_{\delta I J}}{c_{\delta I J}} \frac{s_{\gamma I J}}{c_{\gamma I J}} c_{\gamma 0 1 2 3}.
\end{align}
Finally, we have:
\begin{align}
C  &= 1 +   \sum_I (s_{\delta I}^2 c_{\gamma I}^2 + s_{\gamma I}^2 c_{\delta I}^2) +   \sum_{I < J} (s_{\delta I J}^2 + s_{\gamma I J}^2)  +   \sum_{I \neq J} s_{\delta I}^2 s_{\gamma J}^2 +  \sum_I \sum_{J < K} (s_{\delta I}^2 s_{\gamma J K}^2 + s_{\gamma I}^2 s_{\delta J K}^2) \nonumber \\
& \quad + 2  \sum_{I < J} \bigg( s_{\delta 0 1 2 3} c_{\delta 0 1 2 3} \frac{s_{\gamma I J}}{c_{\delta I J}} \frac{c_{\gamma I J}}{s_{\delta I J}}  + s_{\delta 0 1 2 3}^2 \frac{s_{\gamma I J}^2}{s_{\delta I J}^2} + s_{\delta I J} s_{\gamma I J} c_{\delta I J} c_{\gamma I J} + s_{\delta I J}^2 s_{\gamma I J}^2 \bigg) - \nu_1^2 - \nu_2^2 .
\end{align}

\clearpage

\bibliographystyle{toine}
\bibliography{bubble_multipoles}

\end{document}